\documentclass[apj,numberedappendix]{emulateapj}
\usepackage{amsmath,multirow,natbib,placeins}
\shorttitle{IDENTIFYING THE BEST LENSING TELESCOPES}
\shortauthors{WONG ET AL.}

\begin{document}
\title{A NEW APPROACH TO IDENTIFYING THE MOST POWERFUL GRAVITATIONAL LENSING TELESCOPES}
\author{
Kenneth C. Wong\altaffilmark{1}, 
Ann I. Zabludoff\altaffilmark{1},
S. Mark Ammons\altaffilmark{2},
Charles R. Keeton\altaffilmark{3},
David W. Hogg\altaffilmark{4},
and Anthony H. Gonzalez\altaffilmark{5}}
\altaffiltext{1}{Steward Observatory, University of Arizona, 933 North Cherry Avenue, Tucson, AZ 85721}
\altaffiltext{2}{Lawrence Livermore National Laboratory, 7000 East Avenue, Livermore, CA 94550}
\altaffiltext{3}{Department of Physics and Astronomy, Rutgers University, 136 Frelinghuysen Road, Piscataway, NJ 08854}
\altaffiltext{4}{Center for Cosmology and Particle Physics, Department of Physics, New York University, 4 Washington Place, New York, NY 10003}
\altaffiltext{5}{Department of Astronomy, University of Florida, Gainesville, FL 32611}

\begin{abstract}
The best gravitational lenses for detecting distant galaxies are those with the largest mass concentrations and the most advantageous configurations of that mass along the line of sight.  Our new method for finding such gravitational telescopes uses optical data to identify projected concentrations of luminous red galaxies (LRGs).  LRGs are biased tracers of the underlying mass distribution, so lines of sight with the highest total luminosity in LRGs are likely to contain the largest total mass.  We apply this selection technique to the Sloan Digital Sky Survey and identify the 200 fields with the highest total LRG luminosities projected within a $3\farcm5$ radius over the redshift range $0.1 \leq z \leq 0.7$.  The redshift and angular distributions of LRGs in these fields trace the concentrations of non-LRG galaxies.  These fields are diverse; 22.5\% contain one known galaxy cluster and 56.0\% contain multiple known clusters previously identified in the literature.  Thus, our results confirm that these LRGs trace massive structures and that our selection technique identifies fields with large total masses. These fields contain $2-3$ times higher total LRG luminosities than most known strong-lensing clusters and will be among the best gravitational lensing fields for the purpose of detecting the highest redshift galaxies.
\end{abstract}


\section{INTRODUCTION} \label{sec:intro}
Gravitational lensing by galaxy clusters is an important tool for studying the high-redshift universe.  Galaxies at redshifts $1 \lesssim z \lesssim 3$ can be magnified into extended arcs, enabling studies of these sources at spatial resolutions beyond what is feasible in similar unlensed objects \citep[e.g.,][]{brammer2012,frye2012,livermore2012,sharon2012,yuan2012}.   Lensing by foreground galaxy clusters can also magnify very high-redshift ($z \gtrsim 7$) sources into detectability, allowing us to measure their physical properties \citep[e.g.,][]{kneib2004,pello2004,schaerer2005,richard2006,richard2008,stark2007,bradley2008,bradley2012,zheng2009,zheng2012,laporte2011,bouwens2012,hall2012,coe2013} and making them ideal targets for spectroscopic follow-up \citep[e.g.,][]{bradac2012}.  Such studies are particularly important for characterizing objects on the faint end of the galaxy luminosity function at these redshifts, as even the deepest {\it HST} observations in blank fields require too large a time investment to probe to such depths.

The lensing power of foreground clusters depends on a variety of physical properties.  The total mass of the cluster is very important, as the lensing strength depends on the surface mass density of the lens.  \citet{wong2012} found that distributing the mass among multiple cluster-scale halos along the line of sight (LOS) can increase the lensing cross section compared to having the same mass in a single rich cluster.  This effect results from interactions among the multiple lens potentials, boosting the magnification in the field.  Analysis of the Millennium \citep{springel2005} and Millennium XXL \citep{angulo2012} simulations shows that lines of sight with large total masses may contain multiple massive ($\gtrsim 10^{14} M_{\odot}$) halos and produce some of the highest lensing cross sections in the universe (K. D. French et al. 2013, in preparation).  Individual halo properties, including concentration, ellipticity, orientation, and redshift also affect lensing cross sections \citep[e.g.,][]{bartelmann1995,meneghetti2003,wong2012}.

State-of-the-art lensing analyses focus on fields identified by a single massive cluster \citep[e.g.,][]{postman2012}.  Even X-ray surveys \citep[e.g.,][]{bohringer2000,vikhlinin2009} and \citet[SZ;][]{sunyaev1972} effect surveys \citep[e.g.,][]{vanderlinde2010,marriage2011,williamson2011} for great lensing fields are biased toward lines of sight with a dominant cluster-scale halo because the signal is not proportional to projected mass.  In other words, a line of sight with a single massive cluster looks identical in  X-ray or SZ observations to a similar cluster with additional smaller projected halos whose masses may not be sufficient to have a detectible hot X-ray gas component.  In X-ray or SZ observations, the scaling of the signal with halo mass is faster than linear \citep[e.g.,][]{bonamente2008,vikhlinin2009}, so a line of sight with multiple structures will have a lower signal than if the same total mass were concentrated in a single cluster.  Additional lower mass halos, which may not have detectable hot gas components, may be missed entirely in the field.  Thus, these studies do not necessarily select for the largest total mass and/or the most advantageous mass configuration.  {\it The gravitational lenses explored to date may not in fact be the best directions on the sky to look.}

We explore a new optical selection technique to identify the best lines of sight (hereafter referred to as ``beams") for gravitational lensing.  By selecting fields that have the greatest total luminosity in luminous red galaxies \citep[LRGs; e.g.,][]{eisenstein2001}, which are biased tracers of the underlying matter distribution \citep{zehavi2005,li2006,ho2009,white2011} and detectable to high redshifts, we are likely to find beams with the largest single massive halos (galaxy clusters) and with chance alignments of multiple group and cluster-scale halos.  This technique requires a wide-field multi-band photometric dataset with accurate redshifts, photometric or spectroscopic.  In essence, we are using fewer, but more biased, tracers of the mass along the LOS than methods like the Cluster Red Sequence technique \citep[CRS;][]{gladders2000} and the Gaussian Mixture Brightest Cluster Galaxy algorithm \citep[GMBCG;][]{hao2009,hao2010} that exploit the relationship between halo mass and red galaxy counts within the halo \citep{lin2004b}.  \citet{zitrin2012} derived mass models of clusters in the \citet{hao2010} GMBCG sample, including some of the most powerful lenses (Einstein radius $> 30$\arcsec).  Like X-ray and SZ surveys, these approaches may not be sensitive to multiple projected halos, as smaller structures (i.e., poor clusters) may be hard to identify as galaxy overdensities in color-magnitude space.  In contrast, even individual LRGs can be indicative of cluster-scale structures \citep{ho2009}.

The Sloan Digital Sky Survey (SDSS), with its large sky coverage and photometric LRG selection out to $z \sim 0.7$, is ideal for identifying the best lensing beams using LRGs.  Most arc-producing lensing clusters are at intermediate redshift \citep[$0.3 \lesssim z \lesssim 0.8$;][]{bartelmann1998,gladders2003,hennawi2007}, although higher-redshift lensing clusters have been found \citep[e.g.,][]{huang2009,gonzalez2012}.  The SDSS is deep enough to probe a volume-limited sample of very bright LRGs ($M_{i} - 5\mathrm{log}_{10}(h) \lesssim -22.5$) out to $z \sim 0.7$.  The 5-band optical photometry provides LRG selection, luminosities, colors, and photometric redshifts for over $10^{6}$ galaxies \citep[e.g.,][]{padmanabhan2005,padmanabhan2007,ross2011}.

We present our beam selection method and apply it to the SDSS, identifying the 200 beams with the highest LRG luminosity concentrations and therefore likely to contain the largest total masses projected within a radius of $3\farcm5$.  The LRG photometric redshift distributions show that many of these beams have multiple structures along the line of sight.  Follow-up galaxy spectroscopy in the first fields selected using this method has revealed a diversity of structures, including chance alignments of multiple cluster-scale halos and total masses $\gtrsim 2\times10^{15} h^{-1} M_{\odot}$ (S. M. Ammons et al. 2013, in preparation).

We describe our method of selecting massive beams in \S~\ref{sec:selection}.  In \S~\ref{sec:results}, we apply it to the SDSS, list the highest-ranked beams and their properties, and discuss applications of this method to future surveys.  We summarize our main conclusions in \S~\ref{sec:conclusions}.  Throughout this paper, we assume a $\Lambda$CDM cosmology with $\Omega_{m} = 0.274$, $\Omega_{\Lambda} = 0.726$, and $H_{0} = 100~h$ km s$^{-1}$ Mpc$^{-1}$ with $h = 0.71$.  All magnitudes given are on the AB magnitude system \citep{oke1983}.

\section{SELECTION OF MASSIVE BEAMS} \label{sec:selection}
Our approach to selecting lines of sight with large total masses is based on using LRGs as indicators of massive halos.  LRGs are strongly clustered and are biased tracers of massive structure \citep[e.g.,][]{zehavi2005,li2006,ho2009,white2011}.  They are among the most luminous galaxies in optical light \citep[$L \gtrsim L^{*}$;][]{ho2009} and thus are visible to large distances.  LRGs show little variation in their SEDs \citep{eisenstein2003,cool2008}, making them easy to identify through their optical colors in broadband imaging data.  They have been surveyed over large regions of the sky, making them useful probes of the evolution of large scale structure over a cosmologically interesting volume \citep[e.g.][]{eisenstein2001,padmanabhan2005}.  Projected concentrations of LRGs on the sky are therefore indicative of either an extremely rich galaxy cluster or a superposition of multiple group and cluster-scale halos, given that each individual LRG is likely to occupy an overdense region.

Our selection technique makes use of this relationship between LRGs and massive structures, identifying beams that have the highest total LRG luminosity.  The stellar mass-to-light ($M_{*}/L$) ratios of LRGs are strongly correlated with their rest-frame colors \citep{bell2001}, which, given their homogeneous SEDs, implies that they have similar $M_{*}/L$ ratios.  Indeed, \citet{kauffmann2003} find that the $M_{*}/L$ ratio of galaxies flattens at high luminosities with smaller scatter for redder rest-frame optical wavelengths, and that the most luminous galaxies have the highest $M_{*}/L$ ratios \citep[see also][]{zaritsky2006}.  As a result, the optical/near-IR luminosities of LRGs can be used to estimate their stellar masses.

Relating the LRG luminosity to the mass of its host halo is complicated by the relatively flat slope and substantial scatter of the stellar-to-halo mass (SHM) relation for halo masses above $\sim 10^{12} M_{\odot}$ \citep[e.g.,][]{mandelbaum2006,yang2008,conroy2009,more2009,behroozi2010,behroozi2012,moster2010,leauthaud2012}.  While this scatter is smaller than for the luminosity to halo mass relation \citep[e.g.,][]{yang2008,cacciato2009}, stellar mass is still not a precise tracer of halo mass for individual massive galaxies above this threshold.  There is a scaling between the mass and luminosity of a galaxy cluster \citep[e.g.,][]{lin2003,lin2006,tinker2005,cacciato2013a,cacciato2013b}, although central galaxies contribute fractionally less to the stellar mass for larger halo masses \citep{lin2004a,gonzalez2007,leauthaud2012,lidman2012}, further suggesting that individual LRGs may not give a good estimate of halo mass.  On the other hand, this effect should be mitigated when estimating the total mass in a particular field by integrating over all LRGs in the field and in redshift space.  Furthermore, many of these galaxies are satellite galaxies of higher mass clusters, which increases the total LRG luminosity in the most massive halos \citep{white2011,behroozi2012}.  Therefore, lines of sight containing high total LRG luminosities are likely to have large total masses, either distributed in multiple, projected cluster halos or dominated by a single massive cluster.  The former configurations are more challenging to identify through other selection methods.

Simple number counts of LRGs also can be useful, as there is a relationship between number of LRGs and halo mass.  However, the relation has large scatter for individual clusters  \citep[0.21 dex for $M \sim 10^{15} M_{\odot}$ clusters;][]{ho2009}.  Using total luminosity is likely to be a better tracer of total mass due to the SHM relation, despite its shallow slope at high masses.  In addition, the small number of LRGs in clusters leads to large Poisson errors, whereas we are unlikely to miss the brightest galaxies that contribute the most to the total luminosity.  We perform a simple test that demonstrates that using total LRG luminosity provides a better contrast to the field galaxy population than simple number counts (see Appendix~\ref{app:nsort}).  For completeness, we list there the additional beams that would have been selected using number counts instead.

\section{RESULTS \& DISCUSSION} \label{sec:results}
In this section, we apply our massive beam selection technique to the SDSS.  The SDSS is currently the survey that has the best characteristics for our selection technique.  The latest data release includes imaging of roughly a third of the sky in five optical broadband filters.  The depth of the photometric observations is sufficient to detect and classify LRGs within $0.1 \leq z \leq 0.7$ \citep{padmanabhan2005}, where we expect a large number of lensing clusters to lie.  The SDSS also includes spectroscopic redshifts for roughly a third of the LRGs.  We examine the LRG redshift distributions in a comparison sample of known lensing clusters, comparing these fields to the 200 best beams in the SDSS as ranked by their integrated LRG luminosity.  Our top beams have higher total LRG luminosity and potentially more mass than even these known lensing clusters.  Roughly $75\%$ of our beams contain known galaxy clusters, confirming the power of our selection technique.  We also discuss possible applications of this technique to current and future surveys.

\subsection{Defining the LRG Sample} \label{subsec:lrg}
We select our sample of LRGs from the SDSS Data Release 9 \citep[DR9;][]{ahn2012}.  We identify LRGs using a modified version of the photometric selection criteria of \citet{padmanabhan2005,padmanabhan2007}.  The criteria consist of two separate cuts, denoted ``Cut I" and ``Cut II", which are designed to select LRGs at $z \lesssim 0.4$ and $z \gtrsim 0.4$, respectively.  The details of the LRG selection are given in Appendix~\ref{app:selection}.

The photometric redshifts in the DR9 catalog are computed using the method in \citet{csabai2003}.  We limit our sample to LRGs at redshifts $0.1 \leq z \leq 0.7$.  At $z \lesssim 0.1$, the Cut I criteria are too permissive, resulting in a large fraction of interlopers and causing biases in the photometric redshifts when compared to spectroscopic redshifts.  At $z \gtrsim 0.7$, most objects are at the faint edge of our sample, resulting in larger photometric errors.  We do not have enough spectroscopically observed objects to assess the quality of the photometric redshifts beyond this point.

\begin{figure}
\centering
\plotone{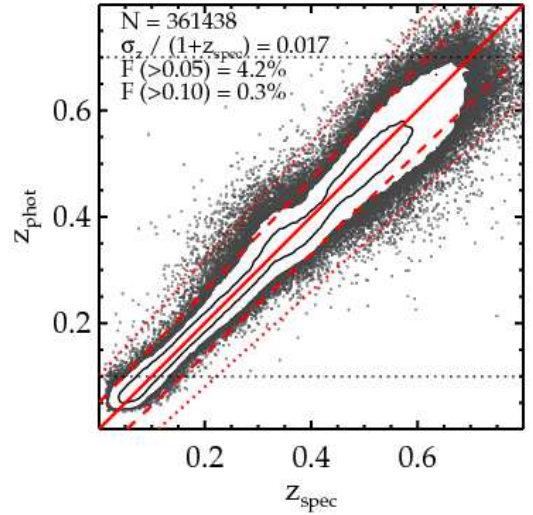}
\caption{Comparison of SDSS DR9 photometric redshifts with SDSS DR9 spectroscopic redshifts.  The contours enclose 68\% and 95\%, respectively, of the galaxies that appear in both samples, with the remaining objects plotted individually (grey points).  The red lines represent the 1-to-1 line (solid) and the 5\% (dashed) and 10\% (dotted) deviations in $\Delta z / (1+z)$.  We plot all galaxies, but only calculate statistics for objects with $0.1 \leq z_{phot} \leq 0.7$, represented by the black dotted lines.  We expect the most powerful lenses to lie within this redshift range.  We exclude objects at redshifts $z \leq 0.1$ due to biases in the photometric redshifts and the unfavorable lensing geometry.  We exclude objects at $z \geq 0.7$ due to large photometric errors and a lack of spectroscopically observed objects to assess directly the quality of the photometric redshifts. The number of objects, redshift accuracy, and fraction of outliers greater than 5\% and 10\% in $|\Delta z| / (1+z)$ are given in the top left corner.  We find good agreement with the spectroscopic redshifts between $0.1 \leq z_{phot} \leq 0.7$, suggesting that our chosen redshift range is reasonable.} \label{fig:zvz}
\end{figure}

We find good agreement of the photometric redshifts with the DR9 spectroscopic redshifts between $0.1 \leq z \leq 0.7$ (Figure~\ref{fig:zvz}).  We define the photometric redshift accuracy to be the normalized median absolute deviation, $\sigma_{z}/(1+z) \equiv 1.48 \times \mathrm{median}(|\Delta z| / (1+z))$.  For objects with $0.1 \leq z_{phot} \leq 0.7$, we calculate $\sigma_{z} / (1+z) = 0.017$, with catastrophic outlier rates of 4.2\% with $|\Delta z| / (1+z) > 0.05$ and 0.3\% with $|\Delta z| / (1+z) > 0.1$.  The photometric redshifts are unbiased to within $\overline{\Delta z} / (1+z) \leq 0.01$ for most of the redshift range probed. There is a slight bias at the $\overline{\Delta z} / (1+z) \leq 0.02$ level at $0.6 \leq z \leq 0.7$, which affects less than 10\% of our sample.  The subsample of objects at the faint end of our magnitude range show the same behavior as the full sample and do not contain additional biases.

To improve the redshift accuracy of our sample, we replace the LRG photometric redshifts and errors with SDSS spectroscopic redshifts and errors where available, which is roughly for one-third of the LRG sample.  The redshift uncertainties and outlier rates given in Figure~\ref{fig:zvz} are thus upper limits.  Hereafter, when referring to an LRG's redshift and its uncertainty, we mean the spectroscopic redshift when available and the photometric redshift otherwise.

In deriving the LRG luminosities, we account for K-corrections and luminosity evolution using an elliptical galaxy template generated by evolving a \citet{bruzual2003} stellar population synthesis model.  \citet{delucia2006} find that massive elliptical galaxies in dense environments have roughly solar metallicities and stellar populations with a median formation redshift of $z \sim 2.5$ for $M_{*} \gtrsim 10^{11} M_{\odot}$, with higher formation redshifts for more massive systems.  Similar formation redshifts for massive ellipticals are supported by observational studies \citep[e.g.,][]{vandokkum2003,treu2005a,treu2005b,perezgonzalez2008}.  There is evidence that massive early-type galaxies are well-characterized by a \citet{salpeter1955} initial mass function \citep[e.g.,][]{auger2010,treu2010} or even more bottom-heavy IMFs \citep[e.g.,][]{cappellari2012,conroy2012,spiniello2012,vandokkum2012}.  Therefore, we generate the template SED assuming an instantaneous burst of star formation at $z = 3$ with a Salpeter IMF and solar metallicity.

We perform K-corrections using the template SED at the age of the galaxy at its observed redshift, and the model is then passively evolved to $z = 0$.  All galaxy luminosities are normalized to $z = 0$ quantities to ensure a fair comparison of their luminosities.  We use a single model instead of fitting templates to the observed photometry \citep[e.g.,][]{eisenstein2001}, as LRGs are typically red, quiescent galaxies with little recent star formation and very homogeneous SEDs \citep{eisenstein2003,cool2008}.  Furthermore, we are interested in corrections to the rest-frame $i$-band, which for the template fitting method can only be determined beyond $z \sim 0.2$ by extrapolating the fits to redder rest-frame wavelengths than are covered by the SDSS photometry.  We choose the $i$-band because it is less affected by extinction and tends to trace stellar mass, and thus total mass, better than bluer filters as a result of being less sensitive to recent star formation.

We also only include objects with derived absolute magnitudes within the broad range $-24.7 < M_{i} - 5\mathrm{log}_{10}(h) < -21$ to eliminate objects with spurious photometric redshifts or aberrant inferred luminosities, while retaining the most luminous LRGs.  We visually inspect objects at the bright end of this range to ensure that we include the brightest LRGs in our sample.  Our final sample contains 1,151,117 LRGs, of which 361,438 have spectroscopic redshifts.

\subsection{Characteristics of the LRG Sample} \label{subsec:lrgchar}
\begin{figure*}
\centering
\plotone{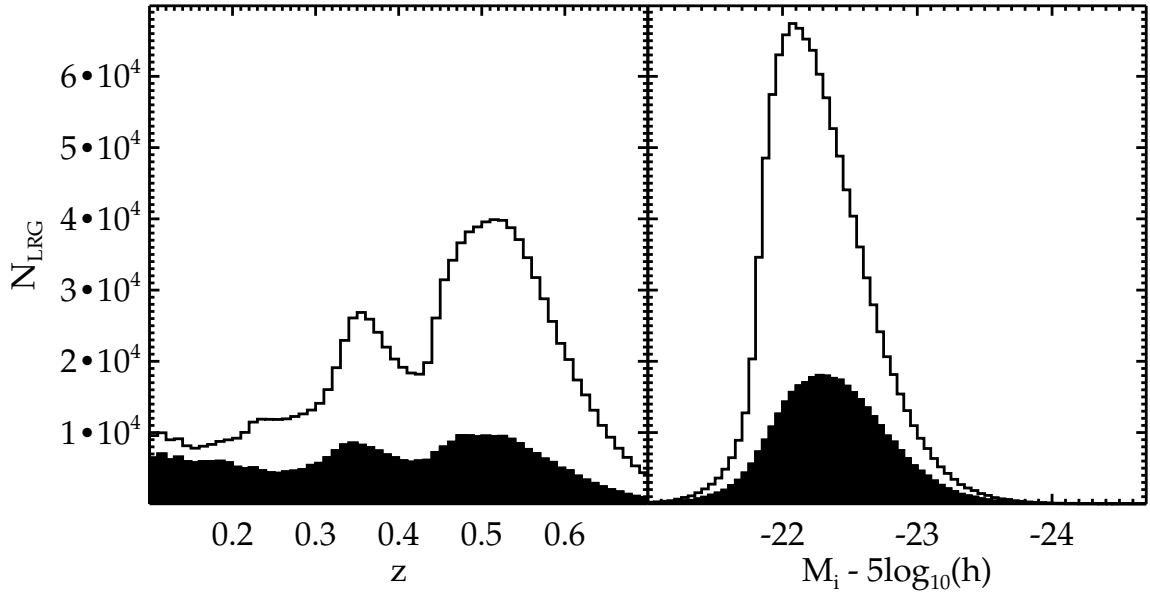}
\caption{Distribution of redshift (left) and $i$-band absolute magnitude (right) for our LRG sample.  The open histogram shows the distribution for all LRGs, while the shaded histogram shows the LRGs with spectroscopic redshifts.  The spike in the redshift histogram at $z \sim 0.35$ and trough at $z \sim 0.4$ arise from the combination of several effects, including misidentification of the 4000 \AA~break in the photometric selection criteria, the transition between the Cut I and Cut II criteria, and photometric redshift biases.  \label{fig:samplehist}}
\end{figure*}

Here, we investigate the properties of our final SDSS LRG sample.  The redshift and $i$-band luminosity (after accounting for K-corrections and luminosity evolution) distributions of our sample are shown in Figure~\ref{fig:samplehist}.

In the left panel of Figure~\ref{fig:samplehist}, the spike in the LRG redshift histogram at $z \sim 0.35$ and trough at $z \sim 0.4$ arise from the combination of several effects.  $z \sim 0.35$ is roughly the redshift at which the 4304 \AA~G-band absorption feature is redshifted into the SDSS $r$-band, which can masquerade as the 4000 \AA~break in color-redshift space.  As a result, the photometric selection criteria select an excess of galaxies around this redshift.  This bias also affects our spectroscopic subsample and in spectroscopic LRG samples with similar photometric selection criteria \citep[e.g.,][]{zehavi2005}.  Secondly, the transition between galaxies selected by the Cut I and Cut II criteria is roughly at $z \sim 0.4$.  Cut I has an apparent magnitude cut at $r < 19.7$, whereas Cut II has a cut at $i < 20$.  This results in a sharp decrease in the number of Cut I-selected LRGs around $z \sim 0.4$ because we hit the $r < 19.7$ magnitude limit.  Meanwhile, Cut II, while probing fainter objects, is optimized to select LRGs at $z \sim 0.5$ and is less efficient at $z \sim 0.4$.  This results in a deficit of LRGs around $z \sim 0.4$, accentuating the $z \sim 0.35$ peak.

In addition to these selection effects, there is a small photometric redshift bias in the range $0.3 \lesssim z \lesssim 0.4$ at the $|\Delta z| / (1+z) < 0.01$ level that pulls galaxies toward $z \sim 0.35$, despite the fact that the photometric redshifts remain unbiased overall.  The spectroscopic LRG subsample is not affected by this problem.  Around this redshift, LRGs transition nearly orthogonally in color-color space, leading to a degeneracy that makes the photometric redshifts less precise, as was also noted by \citet{padmanabhan2005}.  This bias, while still small, conspires with the other effects to ``sharpen" the spike at $z \sim 0.35$, which was already present due to the selection effects discussed above.  This feature is not in the photometric redshift distribution of the \citet{padmanabhan2005} LRG sample due to known biases in their photometric redshifts that drive some objects with true redshifts near $z \sim 0.35$ to photometric redshifts of $z \sim 0.4$, smoothing out the feature.  For $\sim62\%$ of the galaxies with photometric redshifts in the range $0.3 \leq z \leq 0.4$ and a measured spectroscopic redshift, the absolute difference between the photometric and spectroscopic redshift is smaller than the photometric redshift error given in the SDSS catalog.

\begin{figure*}
\centering
\plotone{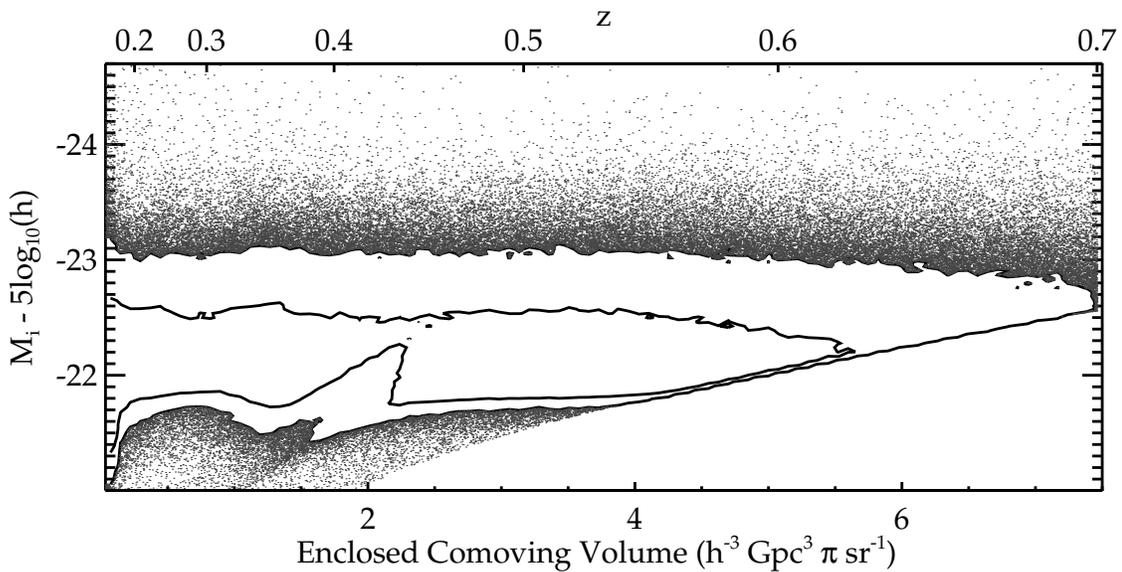}
\caption{Absolute $i$-band magnitude as a function of comoving volume enclosed within a survey of $\pi$ steradians at the redshift of each LRG.  The contours include 68\% and 95\% of the LRGs, respectively, with the remaining LRGs plotted individually.  The redshift is indicated by the upper x-axis.  This visualization shows that our LRG selection is approximately volume-limited out to $z \approx 0.4$ for $M_{i} - 5\mathrm{log}_{10}(h) \leq -21$ and to our upper redshift limit of $z = 0.7$ for $M_{i} - 5\mathrm{log}_{10}(h) \lesssim -22.5$, with the caveat that our selection is less efficient around $z \sim 0.4$. \label{fig:magvol}}
\end{figure*}

In Figure~\ref{fig:magvol}, we plot the distribution of LRG absolute $i$-band magnitudes as a function of redshift.  The redshift axis has been rescaled so that it is linear in the enclosed comoving volume within a survey area of $\pi$ steradians.  The sharp edge at the lower right where LRG selection is truncated represents the apparent magnitude cut at $i < 20$ for the Cut II objects.  The sharpness of this cutoff is due to our single-model method of handling K-corrections and luminosity evolution.  Applying a template-fitting method to the observed photometry \citep[e.g.,][]{eisenstein2001} would result in scatter about this cutoff.  The ``sawtooth" feature at $z \sim 0.4$, which is also in the LRG redshift-luminosity distribution of \citet{padmanabhan2007}, results from the $r < 19.7$ apparent magnitude cut for the Cut I objects and corresponds to the trough in the redshift distribution (see Figure~\ref{fig:samplehist}).  This visualization shows that our LRG selection is approximately volume-limited out to $z = 0.4$ for $M_{i} - 5\mathrm{log}_{10}(h) \lesssim -21$ and to our upper redshift limit of $z = 0.7$ for $M_{i} - 5\mathrm{log}_{10}(h) \lesssim -22.5$, with the caveat that our selection is less efficient around $z \sim 0.4$.

\subsection{LRG Properties of Known Lensing Cluster Fields} \label{subsec:known}
We examine the LRG properties of a sample of known strong lensing clusters from \citet{hennawi2008} and the Cluster Lensing and Supernova Survey with Hubble \citep[CLASH;][]{postman2012} that lie within our chosen redshift range and overlap the sky coverage of our SDSS LRG sample.  Our goal is to identify the best lensing fields using our new selection technique, and this set of known lensing clusters provides a calibration to which we can compare the LRG properties of our new beams (\S~\ref{subsec:beams}).  If our beams have more total LRG luminosity ($\sim$mass) than known strong lenses, as well as multiple lensing planes in some cases, it is likely that they comprise a better sample of strong lenses.  \citet{hennawi2008} select their sample from the SDSS using the CRS selection method to identify clusters between $0.1 \lesssim z \lesssim 0.6$.  We only use those systems labeled by \citet{hennawi2008} as ``definite" or ``tentative" lensing clusters from visual identification of lensed arcs.  The CLASH sample is a mostly X-ray selected sample of 20 massive clusters with an additional five known lensing clusters.  Abell 2261 is in both samples, but we treat it as a part of the CLASH sample here.

\begin{figure*}
\centering
\plotone{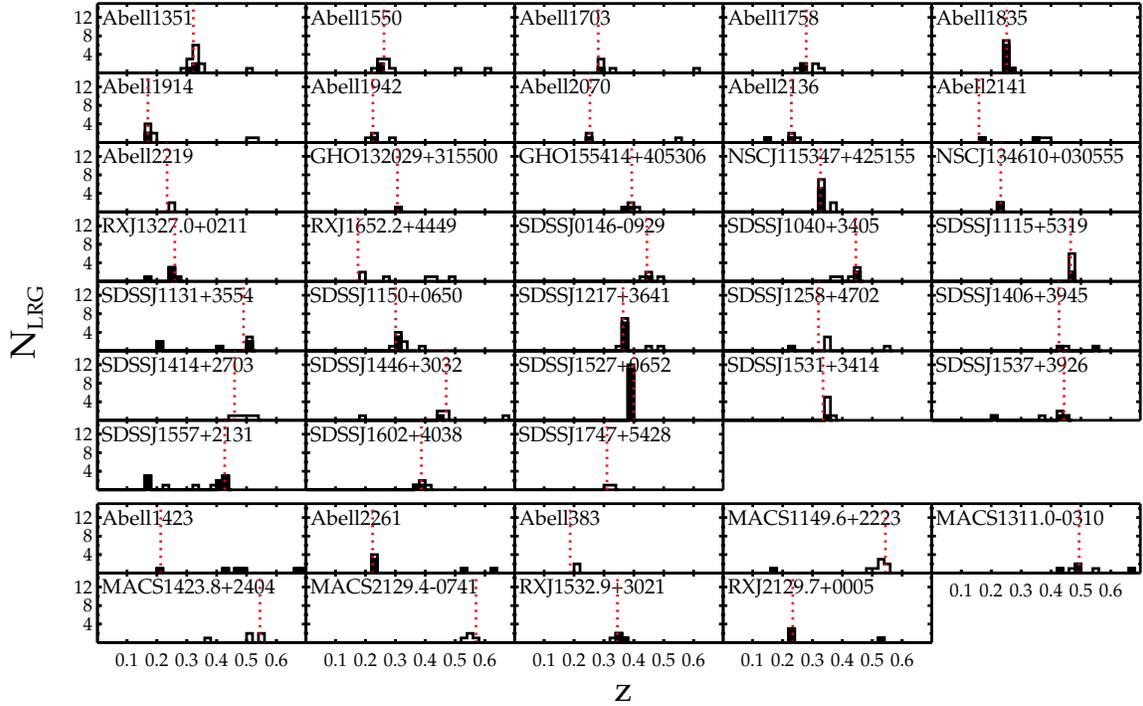}
\caption{Redshift histograms for the LRGs in 33 \citet{hennawi2008} (top group) and nine CLASH (bottom group) known lensing clusters.  The bin size is $\Delta z = 0.02$, roughly the median photometric redshift uncertainty of the LRG sample.  The open histograms show the distribution for all LRGs in the beam, while the shaded histograms show those LRGs with spectroscopic redshifts.  The redshift of the known cluster is indicated by the dotted red line.  Most fields are dominated by a single massive structure, although several have multiple structures distributed throughout redshift space.  In most fields, even the massive lensing cluster is marked by only a few LRGs.  \label{fig:compzhist}}
\end{figure*}

We identify LRGs in these fields within an aperture of $3\farcm5$ (as we do for our SDSS beams in \S~\ref{subsec:beams}) and show their redshift distributions in Figure~\ref{fig:compzhist}.  In many fields, even the massive lensing cluster is marked by only a few LRGs.  By selecting beams from SDSS that have a larger total luminosity in LRGs than these fields, we maximize the chance of finding mass concentrations that can act as powerful lenses.  While these comparison fields tend to be single mass concentrations, several show non-negligible mass concentrations projected along the line of sight that are unassociated with the main cluster.  Thus, the lens modeling of these known clusters should account for LOS mass concentrations.  Such effects have not been explicitly treated in most past analyses, but can influence the inferred mass model \citep[e.g.,][]{hoekstra2011}.

\subsection{Selection of SDSS Beams} \label{subsec:beams}
To find beams with high total LRG luminosity, we search a fixed angular radius around each LRG in our sample.  In principle, this selection can be performed with an arbitrary search radius.  For our search, we choose an aperture of radius $3\farcm5$.  Chance projections of the most massive clusters ($\sim 10^{15} M_{\odot}$) tend to benefit from interactions among their lens potentials \citep{wong2012}, even in regions beyond their Einstein radii.  In particular, the boost at intermediate magnifications ($\mu \sim 3-10$) due to lensing interactions at larger radii is critical in increasing the detectability of very high-redshift lensed galaxies.   Beyond 3.5', the strength of these interactions fall off to the point where the halos can be treated as independent lensing fields.  Furthermore, typical ground-based near-infrared detectors that are well suited for follow-up observations of lensed high-redshift galaxies have fields of view roughly this size or larger.

For each beam centered on an LRG, we tabulate the total number of LRGs within the aperture, as well as the integrated rest-frame $i$-band luminosity of those LRGs as a proxy for total mass in the beam.  We rank all the beams centered on an LRG in descending order by the total LRG luminosity in the beam.  Overlapping beams containing the same LRGs but that are centered on different LRGs are further ranked by the luminosity of the LRG at the beam center.  This choice makes it more likely that in beams with single dominant clusters, we select the central galaxy, which is often the most luminous \citep{lin2004a}, though not always \citep[e.g.,][]{vonderlinden2007,coziol2009,skibba2011,hikage2012}.  Starting from the beam with the highest total luminosity and moving down this ranked list, we construct our list of top beams.  If a beam is centered on an LRG that has already been included as part of a previous beam, we skip over that beam, but allow LRGs within it to be counted in beams further down the list.  This method makes it possible for an LRG to be part of multiple beams, but minimizes overlap among beams in dense regions.  There will always be some beams adjacent to or overlapping one another whose centers are separated by slightly more than the selection radius.  While these beams could be counted as a single field for follow-up purposes, we count them as separate beams for consistency.  In our final catalog of the top 200 beams, there are 28 that overlap another beam in the top 200.

This method can find beams containing dense concentrations of LRGs (e.g., clusters), but may not necessarily be centered on a cluster center.  This can occur if an LRG in the outer parts of a cluster has other LRGs within $3\farcm5$ of it that are not within $3\farcm5$ of the more central LRGs of that cluster.  This is not a flaw in the methodology as we are not specifically looking for fields centered on a single dense cluster.  Rather, our selection is more likely to find lines of sight containing multiple mass concentrations.  We do not account for the boundaries of the survey region when performing our beam selection.  This is conservative because we can only underestimate the total luminosity of LRGs in a given field by ignoring these edge effects.

\begin{figure*}
\centering
\plotone{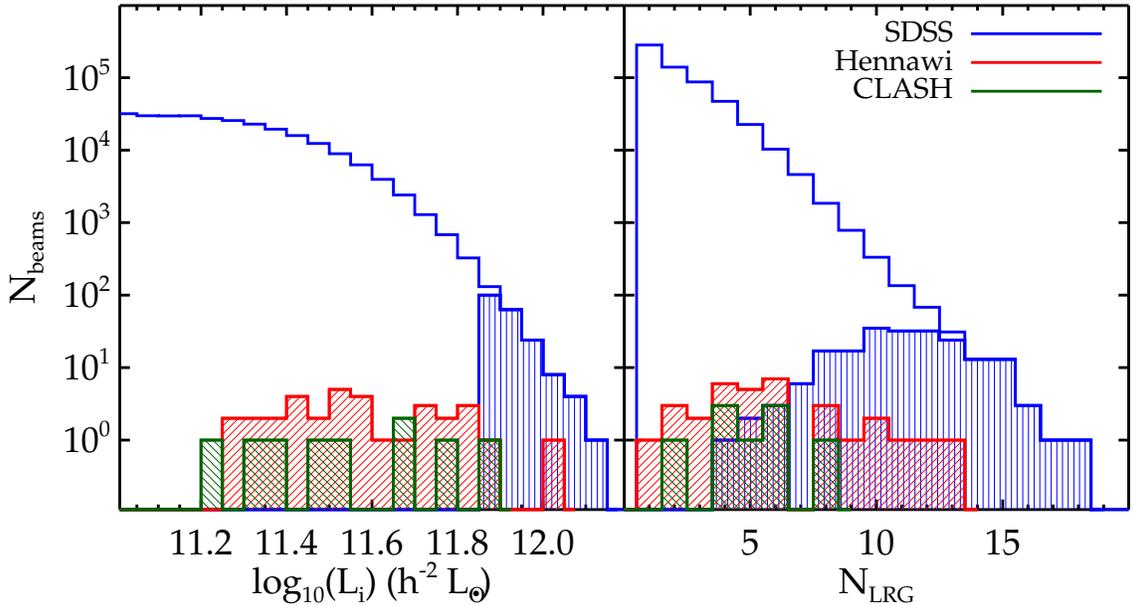}
\caption{{\it Left:} Histogram of total $i$-band luminosity of LRGs in each of our SDSS beams (blue).  For comparison, we show the same quantity for similar size beams centered on the coordinates of 33 definite and tentative lensing clusters from \citet{hennawi2008} (red) and nine X-ray and lensing-selected clusters from the CLASH \citep{postman2012} survey (green).  The shaded blue histogram shows the distribution of total LRG luminosity for our top 200 beams as ranked by total LRG luminosity.   {\it Right:} Histogram of total number of LRGs in each of our selected beams compared to the Hennawi and CLASH beams.  The shaded blue histogram represents the LRG number count distribution of the same top 200 beams ranked by total LRG luminosity.  Our top 200 beams generally contain larger total LRG luminosity and number counts than even these known lensing clusters.  Most of the high-luminosity beams are also the beams with the largest number of LRGs.  \label{fig:nlhist}}
\end{figure*}

We compare the total luminosity and total number of LRGs in our full sample of beams to that of the comparison sample of lensing clusters from \citet{hennawi2008} and CLASH in Figure~\ref{fig:nlhist}.  Both the \citet{hennawi2008} and CLASH samples tend to have lower total LRG luminosity and number counts compared to the SDSS beams, suggesting that our beams at the extreme tail of these distributions contain larger total masses.

We present a list of the 200 best beams as ranked by their total LRG luminosity in Table~\ref{tab:beams}.  We choose a sample size of 200 because this is roughly the beam rank above which our beams exceed the total LRG luminosities of massive lensing clusters.  These top 200 beams have total LRG luminosities $2-3$ times greater than the average total LRG luminosity of the comparison sample.  We do recover five of the Hennawi/CLASH clusters in our sample, although our beams may be centered on different coordinates that include more LRGs in the field.  We find that $\sim60\%$ of our selected beams overlap with the top 200 beams selected by LRG number counts.  For completeness, we provide a separate list of beams in Appendix~\ref{app:nsort} that would have been in the top 200 (or had an equal number of LRGs to beams in our top 200) if we had chosen to sort by LRG number counts instead.

\begin{table*}
\caption{List of SDSS LRG Beams \label{tab:beams}}
\begin{ruledtabular}
\begin{tabular}{ccccccccc}
\multirow{2}{*}{Rank} &
\multirow{2}{*}{RA} &
\multirow{2}{*}{Dec} &
\multirow{2}{*}{N$_{LRG}$} &
log$_{10}$(L$_{i}$)\tablenotemark{a} &
MC log$_{10}$(L$_{i}$)\tablenotemark{a,b} &
\multirow{2}{*}{MC Rank}\tablenotemark{c} &
\multirow{2}{*}{Comments}\tablenotemark{d} &
\multirow{2}{*}{Known Clusters}\tablenotemark{e} \\
& & & &
($h^{-2}$ L$_{\odot}$) &
($h^{-2}$ L$_{\odot}$) &
& &
\\
\tableline
1 &
11:23:53.231 &
+50:52:53.458 &
15 &
12.10 &
$12.08^{+0.02}_{-0.04}$ &
1 &
... &
3 \\
2 &
12:44:02.713 &
+16:51:53.185 &
16 &
12.09 &
$11.99^{+0.05}_{-0.05}$ &
6 &
... &
4 \\
3 &
21:28:25.929 &
+01:33:15.508 &
13 &
12.07 &
$12.06^{+0.03}_{-0.03}$ &
3 &
overlaps with 2128+0137 &
... \\
4 &
11:42:24.777 &
+58:32:05.333 &
14 &
12.06 &
$12.07^{+0.02}_{-0.02}$ &
2 &
... &
3 \\
5 &
00:53:29.221 &
+16:40:29.335 &
13 &
12.05 &
$11.90^{+0.06}_{-0.07}$ &
52 &
overlaps with 0053+1637 &
... \\
6 &
23:22:36.940 &
+09:24:08.046 &
18 &
12.04 &
$11.93^{+0.08}_{-0.08}$ &
25 &
overlaps with 2322+0922 &
... \\
7 &
09:49:58.602 &
+17:08:24.240 &
14 &
12.04 &
$11.99^{+0.03}_{-0.04}$ &
7 &
overlaps with 0950+1704 &
5 \\
8 &
15:26:13.832 &
+04:40:30.109 &
15 &
12.04 &
$11.96^{+0.05}_{-0.06}$ &
17 &
overlaps with 1526+0439 &
2 \\
9 &
10:53:30.330 &
+56:41:21.327 &
17 &
12.02 &
$11.99^{+0.03}_{-0.03}$ &
5 &
... &
3 \\
10 &
02:18:39.886 &
--00:12:31.752 &
15 &
12.02 &
$11.99^{+0.03}_{-0.05}$ &
10 &
... &
3 \\
11 &
10:06:11.170 &
+08:03:00.487 &
11 &
12.01 &
$11.99^{+0.04}_{-0.04}$ &
9 &
... &
2 \\
12 &
11:05:34.246 &
+17:35:28.209 &
12 &
12.00 &
$11.95^{+0.05}_{-0.06}$ &
22 &
overlaps with 1105+1737 &
2 \\
13 &
15:10:15.493 &
+51:45:25.444 &
13 &
12.00 &
$11.97^{+0.04}_{-0.04}$ &
13 &
... &
... \\
14 &
16:16:11.888 &
+06:57:44.695 &
14 &
12.00 &
$11.99^{+0.02}_{-0.02}$ &
8 &
overlaps with 1615+0655 &
5 \\
15 &
14:17:45.068 &
+21:18:27.860 &
12 &
11.99 &
$11.98^{+0.03}_{-0.03}$ &
11 &
... &
3 \\
16 &
14:01:09.485 &
--07:50:57.943 &
14 &
11.99 &
$11.97^{+0.04}_{-0.04}$ &
14 &
... &
... \\
17 &
02:39:53.126 &
--01:34:55.980 &
12 &
11.99 &
$12.05^{+0.03}_{-0.03}$ &
4 &
... &
1 \\
18 &
08:16:58.695 &
+49:33:42.953 &
13 &
11.99 &
$11.96^{+0.03}_{-0.03}$ &
16 &
... &
2 \\
19 &
01:04:33.196 &
+12:51:14.258 &
16 &
11.99 &
$11.92^{+0.05}_{-0.06}$ &
32 &
... &
... \\
20 &
08:19:53.827 &
+31:59:57.660 &
15 &
11.98 &
$11.91^{+0.04}_{-0.05}$ &
37 &
... &
3 \\
21 &
11:52:08.844 &
+31:42:35.278 &
12 &
11.98 &
$11.95^{+0.03}_{-0.04}$ &
20 &
... &
3 \\
22 &
00:28:32.205 &
+09:01:02.414 &
15 &
11.98 &
$11.88^{+0.05}_{-0.06}$ &
71 &
... &
... \\
23 &
10:50:03.460 &
+28:29:58.080 &
12 &
11.97 &
$11.98^{+0.03}_{-0.03}$ &
12 &
overlaps with 1050+2828 &
4 \\
24 &
12:09:15.992 &
+26:43:33.237 &
14 &
11.97 &
$11.92^{+0.04}_{-0.04}$ &
33 &
... &
1 \\
25 &
13:40:46.678 &
--02:51:50.541 &
11 &
11.97 &
$11.94^{+0.03}_{-0.05}$ &
24 &
... &
1 \\
26 &
11:39:26.649 &
+47:04:27.608 &
13 &
11.96 &
$11.94^{+0.02}_{-0.03}$ &
23 &
overlaps with 1139+4704b &
3 \\
27 &
13:51:32.691 &
+52:03:33.346 &
14 &
11.96 &
$11.85^{+0.07}_{-0.07}$ &
123 &
... &
2 \\
28 &
14:37:40.295 &
+30:12:00.275 &
12 &
11.96 &
$11.95^{+0.03}_{-0.03}$ &
18 &
... &
3 \\
29 &
12:43:08.915 &
+20:22:51.768 &
14 &
11.96 &
$11.88^{+0.05}_{-0.05}$ &
69 &
... &
3 \\
30 &
17:43:22.152 &
+63:42:57.575 &
11 &
11.96 &
$11.90^{+0.05}_{-0.06}$ &
55 &
... &
3 \\
31 &
02:02:01.333 &
--08:29:02.252 &
12 &
11.96 &
$11.89^{+0.07}_{-0.08}$ &
62 &
... &
4 \\
32 &
09:10:41.968 &
+38:50:33.710 &
13 &
11.96 &
$11.91^{+0.03}_{-0.04}$ &
42 &
... &
3 \\
33 &
09:42:55.372 &
+14:27:20.611 &
13 &
11.95 &
$11.95^{+0.03}_{-0.03}$ &
19 &
... &
2 \\
34 &
14:33:25.780 &
+29:27:45.979 &
15 &
11.95 &
$11.93^{+0.02}_{-0.02}$ &
26 &
... &
4 \\
35 &
13:06:54.628 &
+46:30:36.691 &
11 &
11.95 &
$11.95^{+0.03}_{-0.03}$ &
21 &
overlaps with 1307+4633 &
5 \\
36 &
10:51:34.352 &
+42:23:29.626 &
11 &
11.95 &
$11.97^{+0.02}_{-0.02}$ &
15 &
... &
3 \\
37 &
01:37:26.699 &
+07:52:09.305 &
15 &
11.95 &
$11.90^{+0.03}_{-0.03}$ &
54 &
overlaps with 0137+0755 &
3 \\
38 &
17:22:13.049 &
+32:06:51.773 &
8 &
11.95 &
$11.93^{+0.02}_{-0.03}$ &
27 &
... &
3 \\
39 &
12:52:58.597 &
+23:42:00.034 &
11 &
11.95 &
$11.92^{+0.04}_{-0.04}$ &
31 &
... &
4 \\
40 &
01:19:56.788 &
+12:18:34.735 &
15 &
11.95 &
$11.91^{+0.03}_{-0.04}$ &
41 &
... &
1 \\
41 &
00:36:44.118 &
--21:03:56.989 &
10 &
11.95 &
$11.92^{+0.04}_{-0.05}$ &
34 &
... &
... \\
42 &
10:35:35.605 &
+31:17:47.478 &
10 &
11.95 &
$11.86^{+0.07}_{-0.11}$ &
92 &
... &
5 \\
43 &
23:34:23.927 &
--00:25:00.606 &
13 &
11.95 &
$11.91^{+0.03}_{-0.04}$ &
36 &
... &
3 \\
44 &
23:22:53.938 &
+09:22:51.440 &
16 &
11.95 &
$11.82^{+0.07}_{-0.06}$ &
155 &
overlaps with 2322+0924 &
... \\
45 &
15:26:26.690 &
+04:39:04.707 &
11 &
11.95 &
$11.90^{+0.04}_{-0.05}$ &
56 &
overlaps with 1526+0440 &
1 \\
46 &
00:01:58.481 &
+12:03:58.021 &
7 &
11.94 &
$11.72^{+0.04}_{-0.04}$ &
195 &
... &
2 \\
47 &
14:39:56.246 &
+54:51:14.221 &
15 &
11.94 &
$11.86^{+0.05}_{-0.06}$ &
106 &
... &
3 \\
48 &
09:43:29.460 &
+33:18:49.403 &
13 &
11.94 &
$11.91^{+0.03}_{-0.05}$ &
43 &
... &
5 \\
49 &
13:22:06.243 &
+53:53:26.158 &
10 &
11.94 &
$11.91^{+0.05}_{-0.06}$ &
40 &
... &
... \\
50 &
22:43:28.058 &
--00:25:58.808 &
13 &
11.94 &
$11.89^{+0.04}_{-0.03}$ &
65 &
... &
3 \\
\end{tabular}
\end{ruledtabular}
\tablenotetext{1}{Total rest-frame $i$-band luminosity in LRGs.}
\tablenotetext{2}{Median of Monte Carlo total luminosity distribution.  The error bars represent the difference between the median and the 16/84\% quantiles of the distribution.}
\tablenotetext{3}{Rank when ordered by median of Monte Carlo total luminosity distribution.}
\tablenotetext{4}{``Overlap" with another beam means that the beam centers are separated by $< 7\arcmin$ and can have LRGs in common.}
\tablenotetext{5}{See Appendix~\ref{app:clusters} for details of known clusters in each beam.}
\end{table*}

\addtocounter{table}{-1}
\begin{table*}
\caption{Continued.}
\begin{ruledtabular}
\begin{tabular}{ccccccccc}
\multirow{2}{*}{Rank} &
\multirow{2}{*}{RA} &
\multirow{2}{*}{Dec} &
\multirow{2}{*}{N$_{LRG}$} &
log$_{10}$(L$_{i}$)\tablenotemark{a} &
MC log$_{10}$(L$_{i}$)\tablenotemark{a,b} &
\multirow{2}{*}{MC Rank}\tablenotemark{c} &
\multirow{2}{*}{Comments}\tablenotemark{d} &
\multirow{2}{*}{Known Clusters}\tablenotemark{e} \\
& & & &
($h^{-2}$ L$_{\odot}$) &
($h^{-2}$ L$_{\odot}$) &
& &
\\
\tableline
51 &
15:01:56.456 &
+33:20:41.059 &
12 &
11.94 &
$11.90^{+0.04}_{-0.05}$ &
47 &
... &
1 \\
52 &
02:20:56.845 &
+06:52:09.157 &
10 &
11.94 &
$11.93^{+0.05}_{-0.05}$ &
28 &
... &
1 \\
53 &
01:48:08.231 &
+00:00:59.692 &
14 &
11.93 &
$11.81^{+0.06}_{-0.07}$ &
165 &
... &
2 \\
54 &
09:55:05.022 &
+28:57:41.137 &
13 &
11.93 &
$11.83^{+0.08}_{-0.08}$ &
146 &
... &
... \\
55 &
23:32:19.553 &
+09:08:21.028 &
10 &
11.93 &
$11.92^{+0.05}_{-0.05}$ &
35 &
... &
... \\
56 &
10:39:51.501 &
+15:27:25.227 &
10 &
11.93 &
$11.90^{+0.04}_{-0.06}$ &
49 &
... &
3 \\
57 &
16:15:59.965 &
+06:55:18.520 &
15 &
11.93 &
$11.91^{+0.03}_{-0.03}$ &
45 &
overlaps with 1616+0657 &
1 \\
58 &
22:58:30.947 &
+09:13:49.117 &
11 &
11.93 &
$11.80^{+0.07}_{-0.10}$ &
167 &
overlaps with 2258+0915 &
... \\
59 &
01:59:59.711 &
--08:49:39.704 &
12 &
11.93 &
$11.89^{+0.04}_{-0.05}$ &
68 &
... &
3 \\
60 &
10:56:14.771 &
+28:22:23.064 &
13 &
11.93 &
$11.87^{+0.04}_{-0.06}$ &
80 &
... &
2 \\
61 &
09:14:23.778 &
+21:24:52.512 &
12 &
11.93 &
$11.90^{+0.03}_{-0.03}$ &
48 &
... &
2 \\
62 &
09:43:34.462 &
+03:45:19.979 &
7 &
11.93 &
$11.77^{+0.13}_{-0.21}$ &
186 &
... &
... \\
63 &
09:02:16.490 &
+38:07:07.073 &
15 &
11.92 &
$11.84^{+0.04}_{-0.06}$ &
127 &
... &
2 \\
64 &
21:28:19.511 &
+01:37:42.682 &
10 &
11.92 &
$11.92^{+0.04}_{-0.04}$ &
30 &
overlaps with 2128+0133 &
... \\
65 &
16:54:24.482 &
+44:42:10.793 &
11 &
11.92 &
$11.86^{+0.04}_{-0.04}$ &
104 &
... &
1 \\
66 &
09:26:35.472 &
+29:34:22.128 &
14 &
11.92 &
$11.86^{+0.05}_{-0.05}$ &
94 &
... &
5 \\
67 &
15:03:01.311 &
+27:57:48.781 &
10 &
11.92 &
$11.91^{+0.03}_{-0.04}$ &
38 &
... &
1 \\
68 &
00:59:42.039 &
+13:10:48.304 &
13 &
11.92 &
$11.85^{+0.04}_{-0.05}$ &
115 &
overlaps with 0059+1315 &
... \\
69 &
14:33:54.319 &
+50:40:45.173 &
11 &
11.92 &
$11.87^{+0.04}_{-0.05}$ &
83 &
... &
2 \\
70 &
08:53:00.399 &
+26:22:13.805 &
11 &
11.92 &
$11.88^{+0.03}_{-0.04}$ &
72 &
... &
2 \\
71 &
01:25:00.069 &
--05:31:22.841 &
15 &
11.92 &
$11.89^{+0.04}_{-0.05}$ &
64 &
... &
... \\
72 &
11:00:10.270 &
+19:16:17.104 &
12 &
11.92 &
$11.90^{+0.03}_{-0.04}$ &
57 &
... &
2 \\
73 &
23:26:37.180 &
+11:57:48.578 &
7 &
11.92 &
$11.63^{+0.21}_{-0.12}$ &
198 &
... &
... \\
74 &
13:26:36.060 &
+53:53:57.959 &
13 &
11.91 &
$11.87^{+0.04}_{-0.04}$ &
76 &
... &
2 \\
75 &
01:19:34.439 &
+14:52:08.957 &
12 &
11.91 &
$11.80^{+0.09}_{-0.22}$ &
169 &
... &
3 \\
76 &
20:54:38.816 &
--16:48:57.240 &
8 &
11.91 &
$11.81^{+0.08}_{-0.11}$ &
164 &
... &
... \\
77 &
11:56:12.252 &
--00:21:02.814 &
11 &
11.91 &
$11.90^{+0.03}_{-0.04}$ &
53 &
... &
2 \\
78 &
23:17:33.137 &
+11:51:58.264 &
12 &
11.91 &
$11.80^{+0.07}_{-0.09}$ &
173 &
... &
... \\
79 &
12:12:08.759 &
+27:34:06.919 &
8 &
11.91 &
$11.92^{+0.03}_{-0.03}$ &
29 &
... &
3 \\
80 &
16:16:27.616 &
+58:12:38.798 &
9 &
11.91 &
$11.89^{+0.03}_{-0.03}$ &
67 &
... &
1 \\
81 &
12:58:32.002 &
+43:59:47.314 &
8 &
11.91 &
$11.90^{+0.03}_{-0.04}$ &
46 &
... &
1 \\
82 &
08:07:56.920 &
+65:25:07.350 &
6 &
11.91 &
$11.56^{+0.30}_{-0.07}$ &
199 &
... &
1 \\
83 &
10:50:20.408 &
+28:28:04.966 &
10 &
11.91 &
$11.90^{+0.03}_{-0.03}$ &
50 &
overlaps with 1050+2829 &
2 \\
84 &
23:19:33.487 &
--01:19:26.377 &
10 &
11.91 &
$11.90^{+0.04}_{-0.04}$ &
51 &
... &
2 \\
85 &
09:21:11.999 &
+30:29:24.946 &
14 &
11.91 &
$11.86^{+0.04}_{-0.04}$ &
97 &
... &
2 \\
86 &
13:04:10.776 &
+46:37:48.615 &
8 &
11.91 &
$11.85^{+0.04}_{-0.09}$ &
110 &
... &
... \\
87 &
12:06:57.409 &
+30:29:22.828 &
15 &
11.91 &
$11.86^{+0.04}_{-0.05}$ &
95 &
... &
3 \\
88 &
15:48:35.149 &
+17:02:22.535 &
10 &
11.91 &
$11.77^{+0.12}_{-0.12}$ &
184 &
... &
1 \\
89 &
09:11:06.757 &
+61:08:18.085 &
6 &
11.91 &
$11.91^{+0.04}_{-0.05}$ &
39 &
... &
3 \\
90 &
02:09:44.322 &
+27:17:09.282 &
11 &
11.91 &
$11.86^{+0.04}_{-0.05}$ &
99 &
... &
... \\
91 &
08:22:49.875 &
+41:28:12.007 &
11 &
11.91 &
$11.86^{+0.05}_{-0.06}$ &
103 &
... &
2 \\
92 &
23:03:44.474 &
+00:09:38.406 &
11 &
11.91 &
$11.85^{+0.05}_{-0.06}$ &
124 &
... &
1 \\
93 &
09:16:14.956 &
--00:25:31.237 &
14 &
11.91 &
$11.89^{+0.04}_{-0.04}$ &
63 &
... &
5 \\
94 &
12:01:25.380 &
+23:50:58.316 &
7 &
11.90 &
$11.86^{+0.06}_{-0.16}$ &
93 &
... &
4 \\
95 &
12:20:34.594 &
+23:01:06.765 &
8 &
11.90 &
$11.88^{+0.04}_{-0.05}$ &
70 &
... &
... \\
96 &
12:42:19.077 &
+40:23:40.425 &
10 &
11.90 &
$11.87^{+0.03}_{-0.04}$ &
85 &
... &
2 \\
97 &
11:59:04.900 &
+51:11:15.803 &
13 &
11.90 &
$11.85^{+0.03}_{-0.04}$ &
116 &
... &
1 \\
98 &
11:11:23.230 &
+26:01:58.357 &
12 &
11.90 &
$11.90^{+0.03}_{-0.03}$ &
59 &
... &
3 \\
99 &
11:23:08.271 &
+54:01:58.928 &
10 &
11.90 &
$11.67^{+0.05}_{-0.06}$ &
197 &
... &
1 \\
100 &
14:32:40.352 &
+31:41:36.116 &
7 &
11.90 &
$11.76^{+0.15}_{-0.09}$ &
187 &
... &
2 \\
\end{tabular}
\end{ruledtabular}
\end{table*}

\addtocounter{table}{-1}
\begin{table*}
\caption{Continued.}
\begin{ruledtabular}
\begin{tabular}{ccccccccc}
\multirow{2}{*}{Rank} &
\multirow{2}{*}{RA} &
\multirow{2}{*}{Dec} &
\multirow{2}{*}{N$_{LRG}$} &
log$_{10}$(L$_{i}$)\tablenotemark{a} &
MC log$_{10}$(L$_{i}$)\tablenotemark{a,b} &
\multirow{2}{*}{MC Rank}\tablenotemark{c} &
\multirow{2}{*}{Comments}\tablenotemark{d} &
\multirow{2}{*}{Known Clusters}\tablenotemark{e} \\
& & & &
($h^{-2}$ L$_{\odot}$) &
($h^{-2}$ L$_{\odot}$) &
& &
\\
\tableline
101 &
12:45:04.700 &
+02:29:08.618 &
10 &
11.90 &
$11.90^{+0.03}_{-0.04}$ &
58 &
... &
2 \\
102 &
02:22:47.925 &
+06:28:11.294 &
4 &
11.90 &
$11.85^{+0.07}_{-0.22}$ &
113 &
... &
... \\
103 &
09:15:50.686 &
+42:57:08.567 &
13 &
11.90 &
$11.84^{+0.03}_{-0.03}$ &
134 &
... &
2 \\
104 &
13:41:08.628 &
+12:33:45.316 &
9 &
11.90 &
$11.81^{+0.07}_{-0.11}$ &
161 &
... &
1 \\
105 &
12:25:49.069 &
+08:24:48.700 &
11 &
11.89 &
$11.88^{+0.03}_{-0.03}$ &
74 &
... &
1 \\
106 &
08:31:34.886 &
+26:52:25.307 &
13 &
11.89 &
$11.86^{+0.04}_{-0.05}$ &
100 &
... &
3 \\
107 &
10:14:11.602 &
+22:31:53.131 &
12 &
11.89 &
$11.73^{+0.03}_{-0.04}$ &
193 &
... &
3 \\
108 &
15:36:44.604 &
+02:46:50.702 &
12 &
11.89 &
$11.87^{+0.04}_{-0.05}$ &
87 &
... &
2 \\
109 &
14:54:16.517 &
+04:34:40.491 &
8 &
11.89 &
$11.86^{+0.04}_{-0.07}$ &
96 &
... &
1 \\
110 &
15:27:45.828 &
+06:52:33.629 &
12 &
11.89 &
$11.89^{+0.03}_{-0.04}$ &
60 &
... &
3 \\
111 &
00:06:11.544 &
--10:28:19.512 &
13 &
11.89 &
$11.87^{+0.05}_{-0.05}$ &
84 &
... &
4 \\
112 &
00:46:43.567 &
--01:52:23.434 &
13 &
11.89 &
$11.81^{+0.05}_{-0.06}$ &
158 &
... &
... \\
113 &
00:53:37.578 &
+16:37:29.457 &
8 &
11.89 &
$11.74^{+0.09}_{-0.08}$ &
191 &
overlaps with 0053+1640 &
... \\
114 &
00:15:23.386 &
--09:18:51.103 &
12 &
11.89 &
$11.86^{+0.05}_{-0.06}$ &
108 &
... &
2 \\
115 &
11:16:01.248 &
+18:24:23.300 &
9 &
11.89 &
$11.91^{+0.04}_{-0.04}$ &
44 &
... &
1 \\
116 &
02:04:17.610 &
--12:34:03.800 &
10 &
11.89 &
$11.83^{+0.05}_{-0.06}$ &
144 &
... &
... \\
117 &
00:59:28.085 &
+13:15:48.171 &
10 &
11.89 &
$11.86^{+0.04}_{-0.04}$ &
91 &
overlaps with 0059+1310 &
... \\
118 &
10:42:47.206 &
+33:12:17.845 &
11 &
11.89 &
$11.73^{+0.07}_{-0.08}$ &
192 &
... &
1 \\
119 &
15:33:49.313 &
+02:38:36.105 &
11 &
11.89 &
$11.81^{+0.05}_{-0.06}$ &
163 &
... &
2 \\
120 &
17:52:27.692 &
+60:10:12.774 &
10 &
11.89 &
$11.83^{+0.05}_{-0.05}$ &
143 &
... &
1 \\
121 &
12:08:19.794 &
+61:22:03.732 &
10 &
11.89 &
$11.84^{+0.05}_{-0.09}$ &
132 &
... &
1 \\
122 &
01:19:07.658 &
--09:34:02.693 &
13 &
11.89 &
$11.82^{+0.05}_{-0.05}$ &
153 &
... &
2 \\
123 &
09:38:11.613 &
+27:35:43.705 &
8 &
11.89 &
$11.89^{+0.03}_{-0.03}$ &
66 &
... &
3 \\
124 &
23:02:14.123 &
+06:49:30.820 &
11 &
11.89 &
$11.80^{+0.05}_{-0.05}$ &
175 &
... &
... \\
125 &
14:52:00.837 &
+01:06:56.447 &
12 &
11.88 &
$11.86^{+0.03}_{-0.05}$ &
107 &
... &
4 \\
126 &
12:24:45.458 &
--00:39:14.796 &
10 &
11.88 &
$11.85^{+0.06}_{-0.06}$ &
122 &
... &
3 \\
127 &
15:54:59.348 &
+51:37:23.214 &
10 &
11.88 &
$11.80^{+0.05}_{-0.07}$ &
176 &
... &
2 \\
128 &
22:43:27.197 &
+20:39:48.807 &
11 &
11.88 &
$11.79^{+0.05}_{-0.05}$ &
181 &
... &
... \\
129 &
12:19:21.841 &
+50:53:28.236 &
11 &
11.88 &
$11.84^{+0.05}_{-0.05}$ &
139 &
... &
3 \\
130 &
09:51:40.088 &
--00:14:20.218 &
6 &
11.88 &
$11.86^{+0.03}_{-0.04}$ &
90 &
... &
3 \\
131 &
00:51:24.585 &
--10:49:09.758 &
12 &
11.88 &
$11.79^{+0.05}_{-0.08}$ &
179 &
... &
1 \\
132 &
22:26:27.277 &
+00:53:29.136 &
12 &
11.88 &
$11.87^{+0.02}_{-0.02}$ &
86 &
... &
1 \\
133 &
15:12:31.251 &
+17:12:15.057 &
8 &
11.88 &
$11.80^{+0.06}_{-0.09}$ &
166 &
... &
... \\
134 &
15:50:36.108 &
+39:48:56.718 &
9 &
11.88 &
$11.84^{+0.04}_{-0.04}$ &
133 &
... &
3 \\
135 &
11:33:41.604 &
+39:52:25.291 &
9 &
11.88 &
$11.85^{+0.03}_{-0.03}$ &
114 &
... &
5 \\
136 &
13:01:02.878 &
+05:35:29.711 &
10 &
11.88 &
$11.84^{+0.04}_{-0.05}$ &
131 &
... &
2 \\
137 &
21:55:56.718 &
+05:49:22.752 &
12 &
11.88 &
$11.84^{+0.03}_{-0.04}$ &
130 &
... &
... \\
138 &
11:33:37.447 &
+66:24:44.842 &
11 &
11.88 &
$11.81^{+0.05}_{-0.06}$ &
162 &
... &
2 \\
139 &
01:57:54.644 &
--00:57:11.347 &
12 &
11.88 &
$11.79^{+0.07}_{-0.10}$ &
180 &
... &
1 \\
140 &
11:13:46.352 &
+56:40:34.462 &
14 &
11.88 &
$11.80^{+0.07}_{-0.07}$ &
168 &
... &
3 \\
141 &
14:44:19.496 &
+16:20:12.303 &
10 &
11.88 &
$11.84^{+0.04}_{-0.05}$ &
135 &
... &
3 \\
142 &
01:54:36.729 &
--19:31:19.437 &
10 &
11.88 &
$11.80^{+0.03}_{-0.04}$ &
170 &
$\sim$20\% outside survey edge &
... \\
143 &
22:24:24.708 &
--02:39:33.138 &
11 &
11.88 &
$11.85^{+0.04}_{-0.06}$ &
118 &
... &
... \\
144 &
10:27:02.072 &
+09:16:40.107 &
8 &
11.88 &
$11.79^{+0.05}_{-0.07}$ &
177 &
... &
1 \\
145 &
16:48:00.199 &
+33:40:03.887 &
11 &
11.88 &
$11.83^{+0.03}_{-0.04}$ &
145 &
... &
2 \\
146 &
10:22:32.057 &
+50:07:07.870 &
8 &
11.88 &
$11.87^{+0.03}_{-0.03}$ &
88 &
... &
2 \\
147 &
09:26:51.422 &
+04:58:17.559 &
12 &
11.88 &
$11.88^{+0.04}_{-0.05}$ &
75 &
... &
6 \\
148 &
12:17:31.158 &
+36:41:11.240 &
11 &
11.88 &
$11.85^{+0.03}_{-0.06}$ &
117 &
... &
3 \\
149 &
11:53:05.648 &
+41:45:20.510 &
9 &
11.87 &
$11.82^{+0.04}_{-0.08}$ &
147 &
... &
2 \\
150 &
14:33:05.416 &
+51:03:16.905 &
12 &
11.87 &
$11.75^{+0.07}_{-0.08}$ &
190 &
... &
1 \\
\end{tabular}
\end{ruledtabular}
\end{table*}

\addtocounter{table}{-1}
\begin{table*}
\caption{Continued.}
\begin{ruledtabular}
\begin{tabular}{ccccccccc}
\multirow{2}{*}{Rank} &
\multirow{2}{*}{RA} &
\multirow{2}{*}{Dec} &
\multirow{2}{*}{N$_{LRG}$} &
log$_{10}$(L$_{i}$)\tablenotemark{a} &
MC log$_{10}$(L$_{i}$)\tablenotemark{a,b} &
\multirow{2}{*}{MC Rank}\tablenotemark{c} &
\multirow{2}{*}{Comments}\tablenotemark{d} &
\multirow{2}{*}{Known Clusters}\tablenotemark{e} \\
& & & &
($h^{-2}$ L$_{\odot}$) &
($h^{-2}$ L$_{\odot}$) &
& &
\\
\tableline
151 &
09:01:04.594 &
+39:54:49.063 &
10 &
11.87 &
$11.84^{+0.04}_{-0.05}$ &
136 &
... &
1 \\
152 &
10:50:38.567 &
+35:49:12.425 &
13 &
11.87 &
$11.80^{+0.05}_{-0.05}$ &
174 &
... &
1 \\
153 &
20:55:07.146 &
--11:44:38.837 &
10 &
11.87 &
$11.87^{+0.03}_{-0.03}$ &
78 &
... &
... \\
154 &
08:50:07.915 &
+36:04:13.650 &
12 &
11.87 &
$11.87^{+0.02}_{-0.03}$ &
79 &
... &
3 \\
155 &
17:24:47.122 &
+32:02:10.676 &
10 &
11.87 &
$11.83^{+0.03}_{-0.04}$ &
141 &
... &
... \\
156 &
08:45:43.999 &
+30:10:07.090 &
9 &
11.87 &
$11.84^{+0.04}_{-0.05}$ &
140 &
... &
4 \\
157 &
14:55:07.993 &
+38:36:04.879 &
11 &
11.87 &
$11.82^{+0.04}_{-0.05}$ &
150 &
... &
4 \\
158 &
02:27:57.609 &
+03:09:16.489 &
14 &
11.87 &
$11.76^{+0.05}_{-0.06}$ &
188 &
... &
... \\
159 &
11:07:19.334 &
+53:04:17.938 &
9 &
11.87 &
$11.86^{+0.04}_{-0.06}$ &
98 &
... &
2 \\
160 &
15:38:02.025 &
+39:27:39.159 &
8 &
11.87 &
$11.87^{+0.03}_{-0.03}$ &
82 &
overlaps with 1538+3922 &
4 \\
161 &
01:03:24.248 &
+00:55:37.011 &
9 &
11.87 &
$11.84^{+0.05}_{-0.05}$ &
126 &
... &
2 \\
162 &
10:40:17.611 &
+54:37:08.607 &
9 &
11.87 &
$11.86^{+0.03}_{-0.04}$ &
109 &
... &
3 \\
163 &
12:28:58.786 &
+53:37:27.671 &
11 &
11.87 &
$11.84^{+0.03}_{-0.04}$ &
125 &
... &
1 \\
164 &
15:38:04.005 &
+39:22:32.253 &
8 &
11.87 &
$11.86^{+0.03}_{-0.03}$ &
101 &
overlaps with 1538+3927 &
1 \\
165 &
01:39:10.116 &
+07:03:09.893 &
8 &
11.87 &
$11.82^{+0.05}_{-0.04}$ &
151 &
... &
... \\
166 &
09:50:00.059 &
+17:04:27.060 &
12 &
11.87 &
$11.82^{+0.03}_{-0.03}$ &
154 &
overlaps with 0949+1708 &
4 \\
167 &
10:29:10.561 &
+33:22:35.618 &
5 &
11.87 &
$11.40^{+0.30}_{-0.24}$ &
200 &
... &
... \\
168 &
12:34:49.804 &
+23:03:42.109 &
11 &
11.87 &
$11.85^{+0.04}_{-0.05}$ &
111 &
... &
3 \\
169 &
01:27:10.589 &
+23:14:19.797 &
11 &
11.87 &
$11.86^{+0.03}_{-0.03}$ &
105 &
... &
1 \\
170 &
13:15:23.033 &
--02:50:35.192 &
11 &
11.87 &
$11.77^{+0.06}_{-0.07}$ &
185 &
... &
1 \\
171 &
11:39:02.869 &
+47:04:43.290 &
12 &
11.87 &
$11.81^{+0.04}_{-0.05}$ &
159 &
overlaps with 1139+4704a &
2 \\
172 &
14:31:48.010 &
+09:00:15.869 &
13 &
11.86 &
$11.72^{+0.07}_{-0.09}$ &
194 &
... &
1 \\
173 &
09:47:14.189 &
+38:10:22.088 &
11 &
11.86 &
$11.85^{+0.04}_{-0.05}$ &
112 &
... &
3 \\
174 &
12:17:05.124 &
+26:05:18.445 &
13 &
11.86 &
$11.84^{+0.04}_{-0.05}$ &
129 &
... &
1 \\
175 &
13:26:25.383 &
+53:24:58.472 &
12 &
11.86 &
$11.82^{+0.04}_{-0.05}$ &
156 &
... &
1 \\
176 &
01:53:42.190 &
+05:35:44.062 &
10 &
11.86 &
$11.87^{+0.04}_{-0.03}$ &
77 &
... &
1 \\
177 &
10:54:40.435 &
+55:23:56.307 &
7 &
11.86 &
$11.85^{+0.02}_{-0.03}$ &
119 &
... &
2 \\
178 &
09:58:25.077 &
+42:39:44.430 &
9 &
11.86 &
$11.82^{+0.07}_{-0.03}$ &
152 &
... &
... \\
179 &
00:36:09.793 &
+23:37:17.923 &
9 &
11.86 &
$11.75^{+0.05}_{-0.06}$ &
189 &
... &
... \\
180 &
23:33:42.330 &
+24:41:06.662 &
8 &
11.86 &
$11.79^{+0.09}_{-0.10}$ &
178 &
... &
... \\
181 &
00:40:36.695 &
+25:29:12.961 &
10 &
11.86 &
$11.80^{+0.04}_{-0.05}$ &
171 &
... &
... \\
182 &
14:37:17.666 &
+34:18:22.187 &
12 &
11.86 &
$11.85^{+0.04}_{-0.04}$ &
121 &
... &
5 \\
183 &
14:15:08.392 &
--00:29:35.680 &
10 &
11.86 &
$11.82^{+0.03}_{-0.02}$ &
148 &
... &
3 \\
184 &
20:53:55.128 &
--06:34:51.054 &
8 &
11.86 &
$11.81^{+0.04}_{-0.06}$ &
160 &
... &
1 \\
185 &
12:35:44.353 &
+35:32:47.968 &
12 &
11.86 &
$11.83^{+0.05}_{-0.06}$ &
142 &
... &
1 \\
186 &
11:52:35.385 &
+37:15:43.111 &
9 &
11.86 &
$11.86^{+0.00}_{-0.03}$ &
102 &
... &
2 \\
187 &
14:45:34.036 &
+48:00:12.417 &
10 &
11.86 &
$11.84^{+0.04}_{-0.05}$ &
128 &
... &
2 \\
188 &
08:40:08.745 &
+21:56:03.214 &
10 &
11.86 &
$11.85^{+0.04}_{-0.05}$ &
120 &
... &
1 \\
189 &
13:48:53.073 &
+57:23:46.617 &
11 &
11.86 &
$11.82^{+0.04}_{-0.04}$ &
149 &
... &
3 \\
190 &
13:07:03.631 &
+46:33:47.849 &
9 &
11.86 &
$11.87^{+0.03}_{-0.03}$ &
81 &
overlaps with 1306+4630 &
6 \\
191 &
11:40:40.199 &
+44:07:40.291 &
9 &
11.86 &
$11.84^{+0.04}_{-0.03}$ &
137 &
... &
5 \\
192 &
14:48:20.246 &
+20:43:31.168 &
10 &
11.86 &
$11.82^{+0.04}_{-0.05}$ &
157 &
... &
2 \\
193 &
12:41:56.529 &
+03:43:59.760 &
5 &
11.86 &
$11.86^{+0.02}_{-0.02}$ &
89 &
... &
2 \\
194 &
08:41:23.880 &
+25:13:05.204 &
9 &
11.86 &
$11.84^{+0.03}_{-0.04}$ &
138 &
... &
2 \\
195 &
22:58:17.244 &
+09:15:12.899 &
9 &
11.86 &
$11.78^{+0.07}_{-0.07}$ &
182 &
overlaps with 2258+0913 &
... \\
196 &
00:24:59.715 &
+08:26:16.778 &
12 &
11.86 &
$11.71^{+0.06}_{-0.06}$ &
196 &
... &
1 \\
197 &
15:50:16.987 &
+34:18:33.901 &
10 &
11.86 &
$11.89^{+0.03}_{-0.03}$ &
61 &
... &
3 \\
198 &
01:37:18.176 &
+07:55:44.482 &
13 &
11.86 &
$11.88^{+0.03}_{-0.04}$ &
73 &
overlaps with 0137+0752 &
1 \\
199 &
11:05:20.978 &
+17:37:16.830 &
10 &
11.86 &
$11.78^{+0.05}_{-0.06}$ &
183 &
overlaps with 1105+1735 &
3 \\
200 &
03:33:12.198 &
--06:52:24.614 &
10 &
11.86 &
$11.80^{+0.06}_{-0.07}$ &
172 &
... &
1 \\
\end{tabular}
\end{ruledtabular}
\end{table*}

\clearpage

The redshift distributions of the LRGs in the top 200 beams are shown in Figure~\ref{fig:zhist}.  The beams show a wide diversity of configurations, including beams dominated by a single massive peak, as well as beams with multiple structures along the line of sight.  The latter configurations indicate chance alignments of galaxy clusters or groups in these fields, which can lead to advantageous lensing configurations for magnifying very high-redshift ($z \gtrsim 7$) galaxies \citep{wong2012}.  Even single LRGs can trace massive lensing clusters (see Figure~\ref{fig:compzhist}).

\begin{figure*}
\centering
\plotone{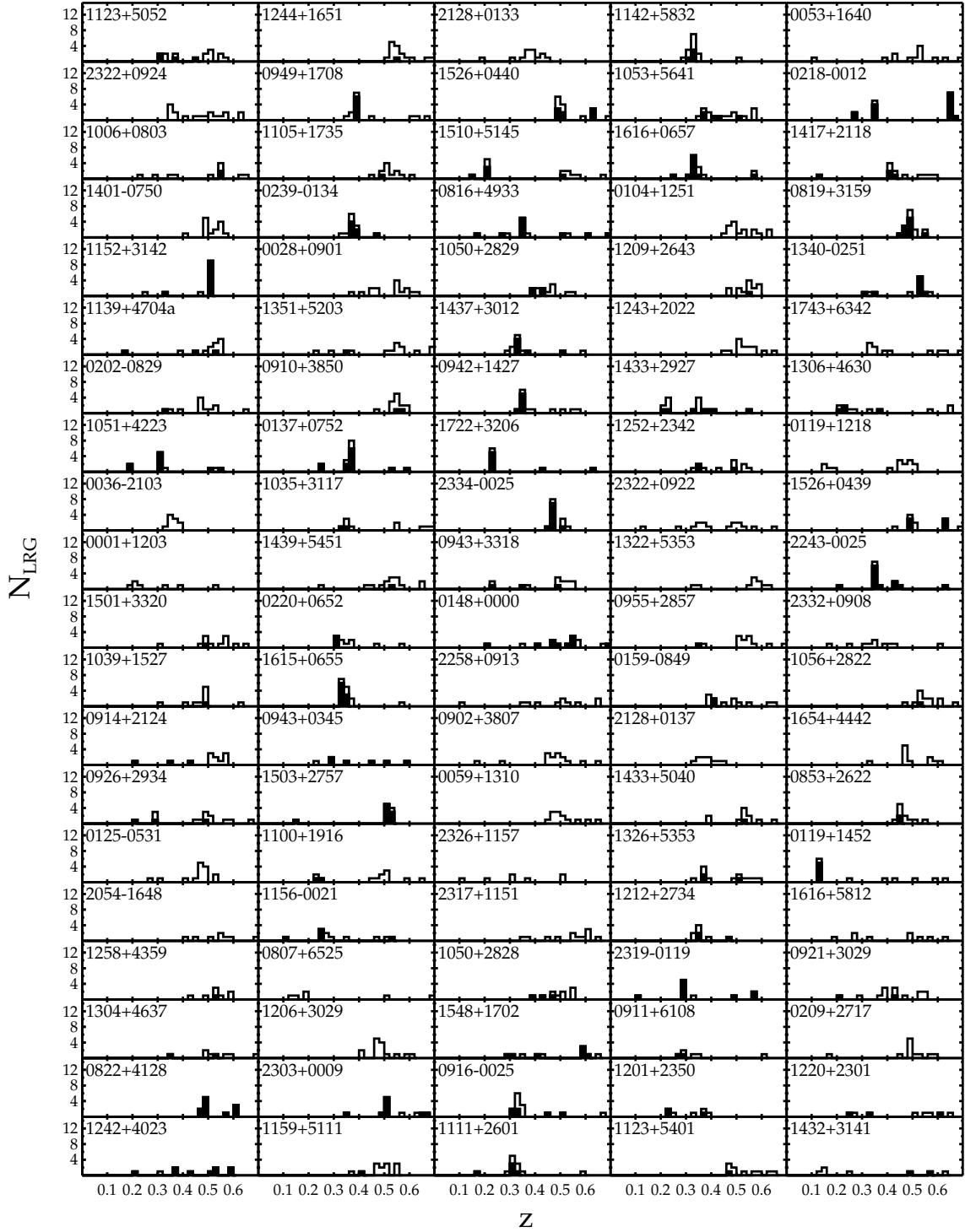}
\caption{Same as Figure~\ref{fig:compzhist}, but for our top 200 beams as ranked by total LRG luminosity.  Our beams contain many more LRGs than the majority of the comparison sample of Hennawi and CLASH lensing clusters.  Five of the Hennawi/CLASH clusters are recovered here (although our beams may be centered at different coordinates that include more LRGs).  Our beams show a diversity of configurations, with some beams dominated by a single massive structure and others having multiple structures distributed throughout redshift space.  Even single LRGs can trace massive lensing clusters (see Figure~\ref{fig:compzhist}). \label{fig:zhist}}
\end{figure*}

\addtocounter{figure}{-1}
\begin{figure*}
\centering
\plotone{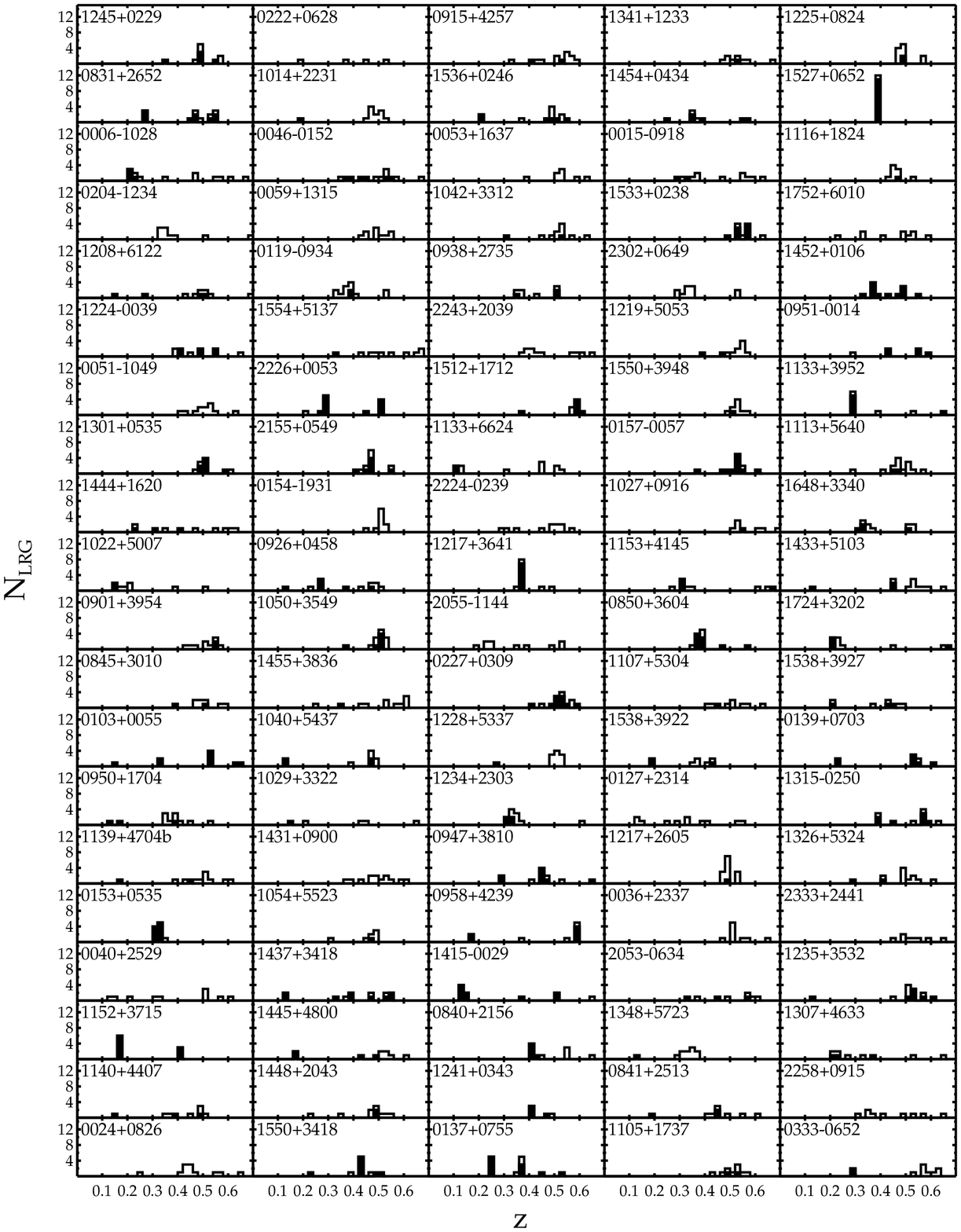}
\caption{(Continued.)}
\end{figure*}

We show color images of our top 20 beams in Figure~\ref{fig:images} to give a sense of the nature of these fields.  The images are taken from the SDSS DR9 SkyServer\footnote{http://skyserver.sdss3.org/dr9/} and show 7\arcmin$\times$7\arcmin~fields centered on the coordinates of each beam to roughly match our circular selection aperture.  The contours overplotted on the images trace the projected galaxy overdensities as determined from the full catalog of primary SDSS photometric galaxies brighter than $r = 22$.  The contours generally trace the LRGs, confirming that the LRGs mark the densest structures in these fields.  Like the redshift distributions of these beams, the angular distributions show a large diversity.  Some beams are dominated by a single concentration of galaxies, whereas others have multiple clumps projected on the sky.

\begin{figure*}
\centering
\plotone{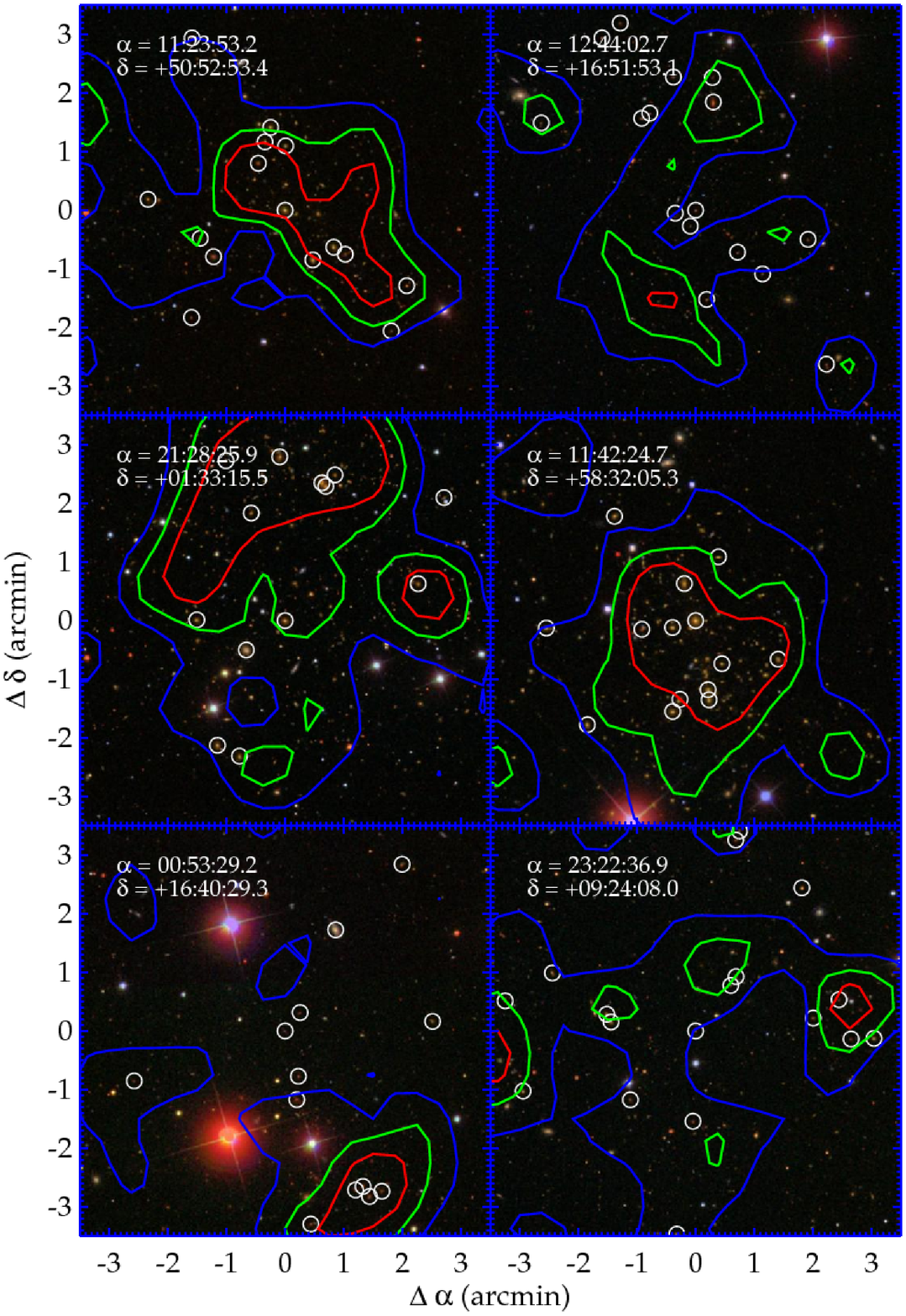}
\caption{SDSS DR9 SkyServer images of the top 20 beams selected from the SDSS.  Each panel is 7\arcmin$\times$7\arcmin~on a side.  Each LRG in the field is indicated by a white circle.  The contours enclose the projected overdensities of galaxies from the SDSS primary photometric galaxy catalog from the highest (red) to lowest (blue) projected overdensities.  The coordinates of the beam center are given in the upper left corner of each panel.  LRGs generally trace the overdensities.  Like the redshift histograms, the projected galaxy densities in these fields show a variety of configurations, with some beams dominated by a single concentration of galaxies and others comprising multiple structures in projection.  \label{fig:images}}
\end{figure*}

\addtocounter{figure}{-1}
\begin{figure*}
\centering
\plotone{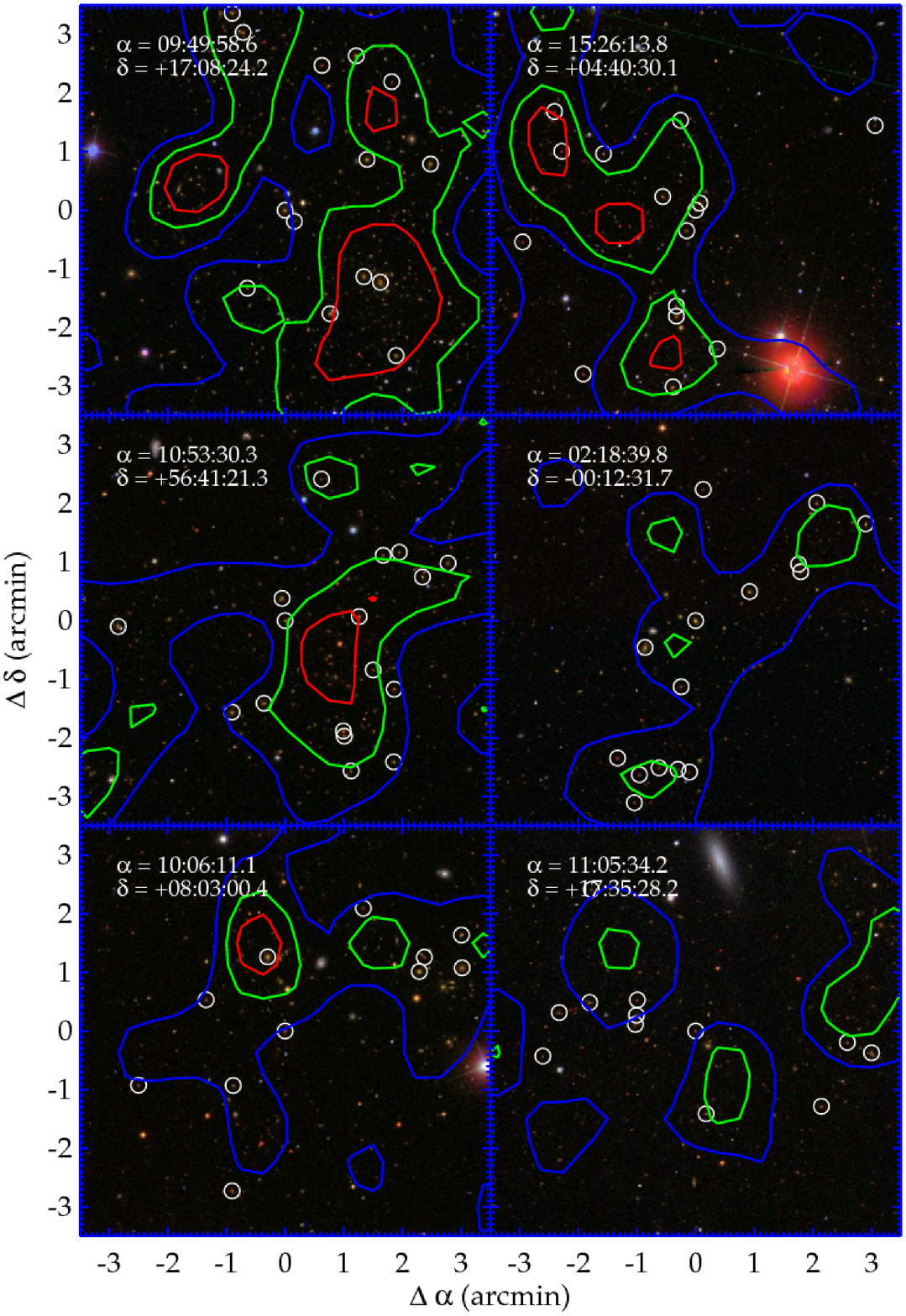}
\caption{(Continued.)}
\end{figure*}

\addtocounter{figure}{-1}
\begin{figure*}
\centering
\plotone{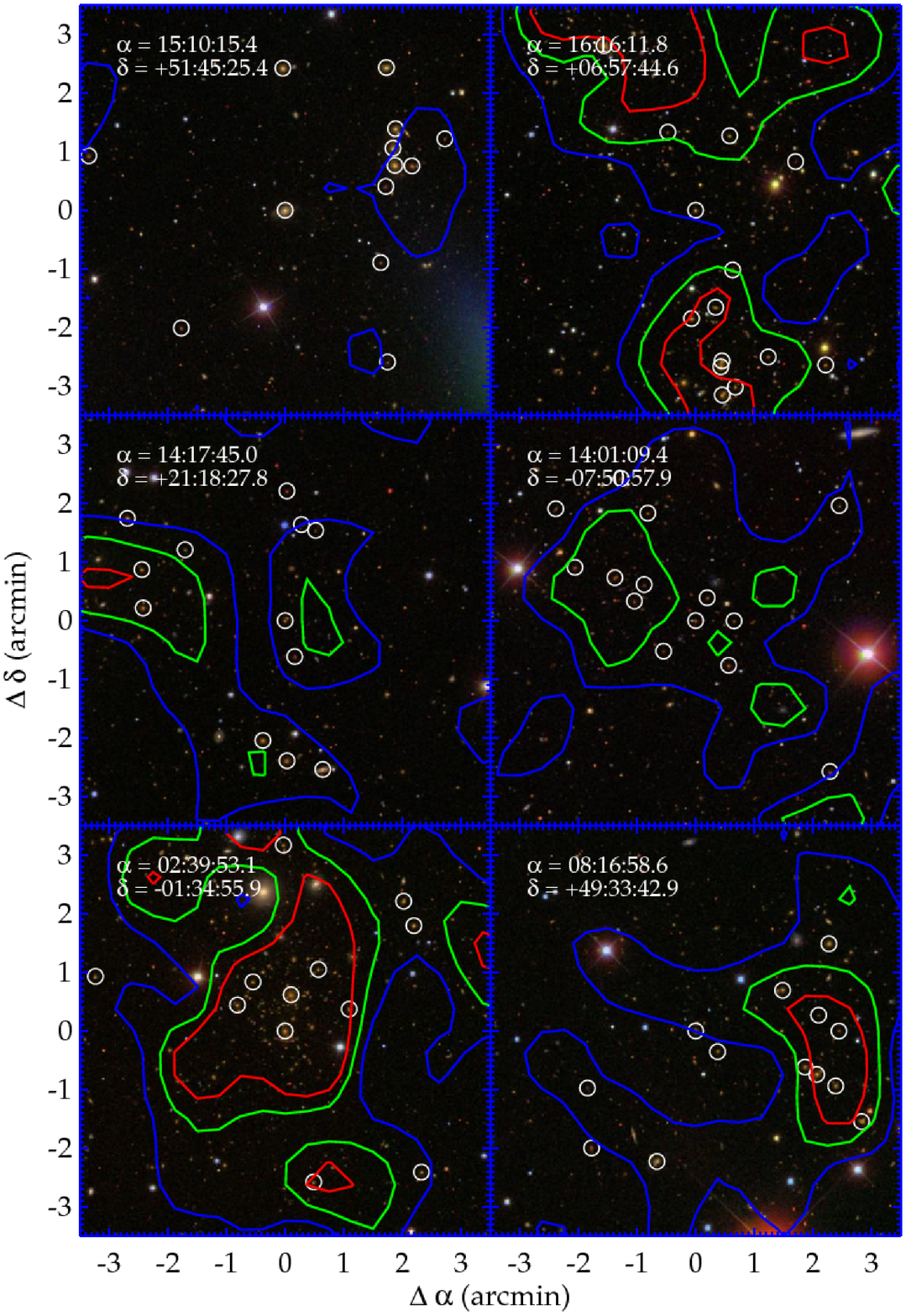}
\caption{(Continued.)}
\end{figure*}

\addtocounter{figure}{-1}
\begin{figure*}
\centering
\plotone{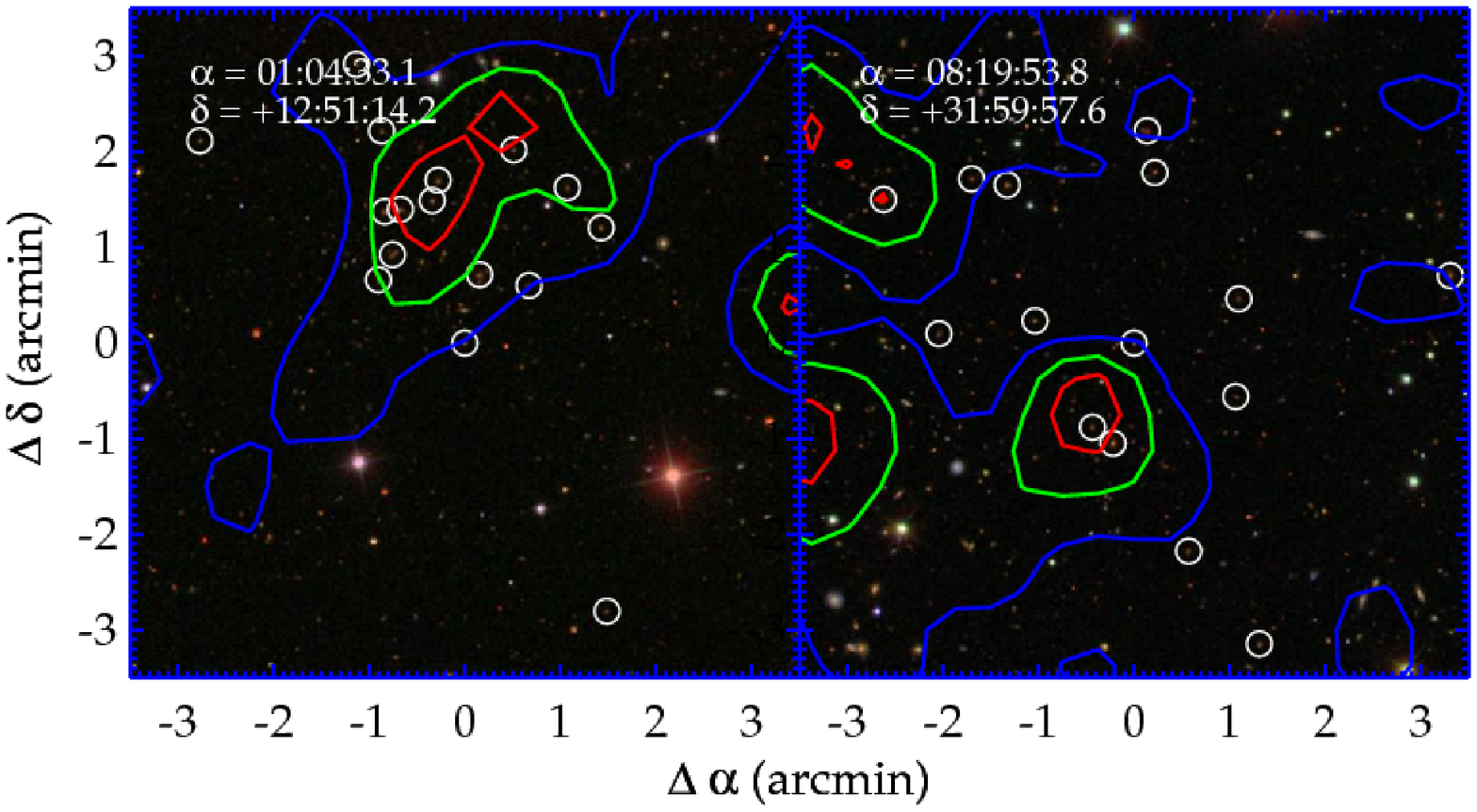}
\caption{(Continued.)}
\end{figure*}

We perform a search in the NASA/IPAC Extragalactic Database (NED) for galaxy clusters that have been identified by past studies within our selected beams.  We find that 22.5\% of our beams contain one known cluster, and 56.0\% contain multiple clusters, confirming the high mass concentrations projected along these lines of sight.  We list the number of known clusters in each beam in Table~\ref{tab:beams} and present a list of these clusters in Appendix~\ref{app:clusters}.

We visually inspect the SDSS SkyServer images of each of our top 200 fields.  The majority show large concentrations of spheroidal galaxies indicative of cluster-scale structures.  The other beams could be chance alignments of smaller group-scale halos, or could contain a number of misclassified objects masquerading as LRGs.  While we have limited the number of misclassifications with our selection cuts (see Appendix~\ref{app:selection}), there are inevitably a small fraction of objects that contaminate our LRG catalog.  We will describe the first results from our spectroscopic follow-up survey of several of these beams in S. M. Ammons et al. (2013, in preparation).

\subsection{Possible Errors in Beam Selection} \label{subsec:errors}
We use a Monte Carlo method to calculate the uncertainty on the total LRG luminosity of each of our top beams.  We begin by searching the SDSS DR9 for all primary photometric objects classified as galaxies within our beams.  Objects fainter than $r = 22$ are removed, as this is roughly where the morphological star/galaxy separation breaks down\footnote{See www.sdss3.org/dr9/imaging/other\_info.php\#stargalaxy}.  We replace the photometric redshift and its associated error with a spectroscopic redshift and error where available.

\begin{figure*}
\centering
\plotone{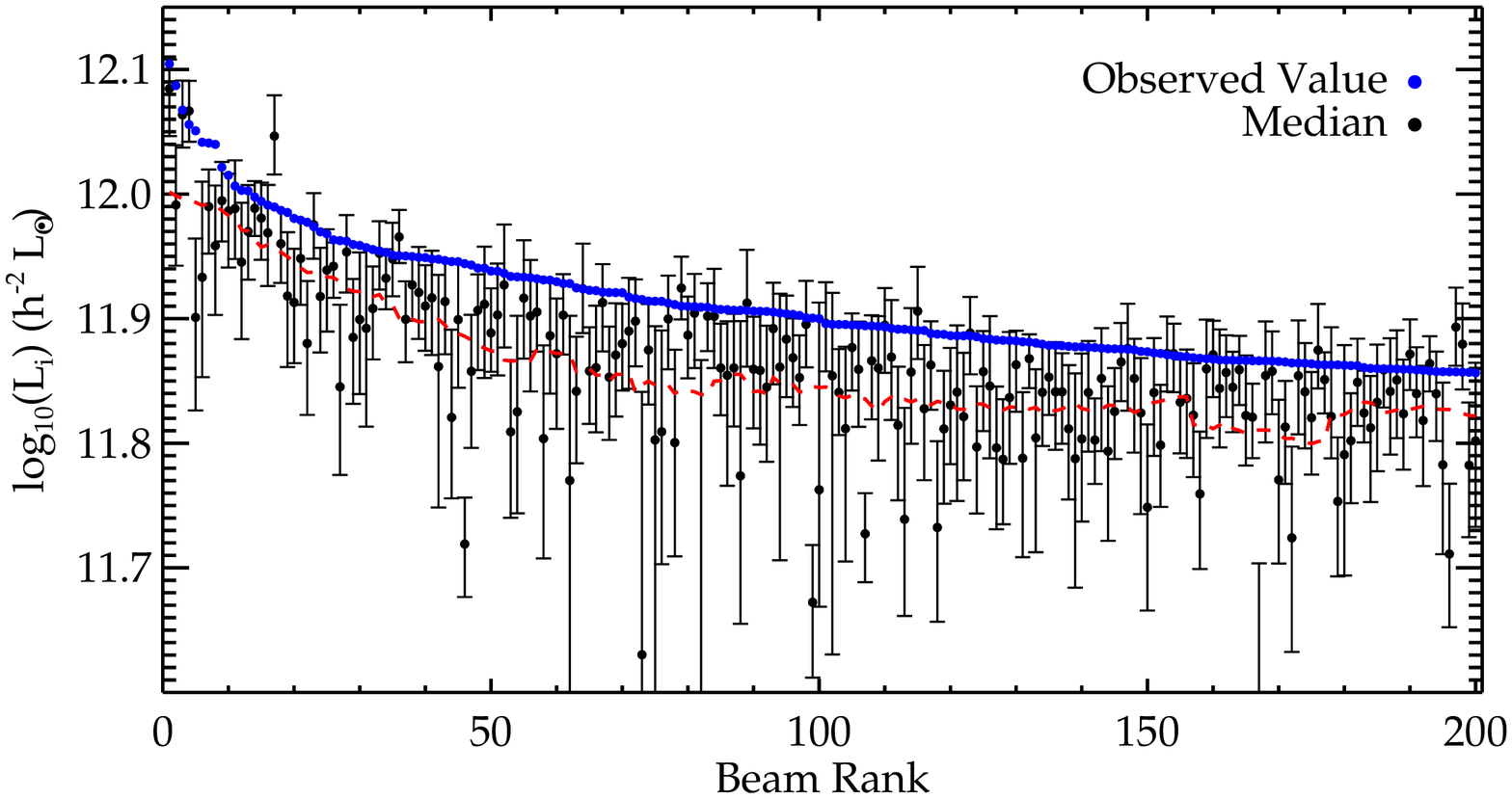}
\caption{Integrated LRG luminosity as a function of beam rank for the top 200 beams.  The observed value is given by the blue points.  The black points represent the median total LRG luminosity over 1000 Monte Carlo realizations of each beam.  The error bars represent the 16\% and 84\% quantiles of the luminosity distributions. The red dashed line is the 20-beam central moving average of the median Monte Carlo luminosities.  The median luminosities are generally lower than the measured values due to expected biases when selecting the top-ranked beams, although the values still noticeably trend downward with increasing beam rank.  When the top 1000 beams are re-ranked by the median integrated LRG luminosities over 200 trials per beam, $\sim70 - 75$\% of the beams within the top $N$ beams, where $N \leq 200$, remain within the top $N$ beams, showing that our beam selection is stable over this range. \label{fig:beamstats}}
\end{figure*}

For each beam, we apply Gaussian errors to the quantities defined in our LRG selection criteria (e.g. apparent magnitude, redshift; see Appendix~\ref{app:selection}) over 1000 Monte Carlo trials.  For a given trial, we re-determine which galaxies in the beam satisfy the selection criteria and compute the total LRG luminosity.  The final distribution of 1000 total LRG luminosities for each beam therefore accounts for both the intrinsic errors in the galaxy properties themselves, as well as galaxies falling into or out of the LRG selection cuts.  Because the resulting luminosity distributions are non-Gaussian in general, we report median values in Table~\ref{tab:beams}.  The error bars associated with these median values correspond to the 16\% and 84\% quantiles of the distribution for each beam.  In Figure~\ref{fig:beamstats}, we plot the observed total luminosity for each beam, along with the median of the Monte Carlo trials with the associated error range.

In general, the medians of the luminosity distributions generated through the Monte Carlo trials tend to be lower than the observed values.  This is expected, because these beams were selected to be the top-ranked beams by total LRG luminosity.  Given the scatter in the total luminosity, the selected beams are likely those that are among the best and also happen to scatter upwards in observed total LRG luminosity.  Also, while this scatter affects the detailed beam rankings relative to the ``true" rankings, Figure~\ref{fig:beamstats} shows that even among the top 200, there is a noticeable decrease in the median integrated LRG luminosity when ordered by the observed values, indicating that the observed values reliably rank the beams.  One may also suspect that our top beams are biased toward low mass-to-light (M/L) ratios.  For the initial beams that we have followed-up spectroscopically (S. M. Ammons et al. 2013, in preparation), we compare their M/L ratios to those of a sample of our comparison lensing clusters (\S~\ref{subsec:known}) using virial masses from the literature \citep{mantz2010,zitrin2011,coe2012}.  We find that our beams have roughly comparable M/L ratios, suggesting that a potential bias in M/L is not a dominant effect.

We show the 20-beam central moving average of the median total LRG luminosities of the Monte Carlo trials in Figure~\ref{fig:beamstats}.  While the values trend downward with beam rank as expected, the steepness of the trend noticeably decreases after the first $20-30$ beams.  To test the robustness of our selected beams, we calculate statistics for a larger sample of the top 1000 beams.  When re-ranked by the median total LRG luminosities over 200 trials per beam, $\sim70-75$\% of the beams within the top $N$ beams, where $N \leq 200$, remain within the top $N$ beams.  Only five of the top 200 beams fall out of the top 400 when re-ranked in this manner.  Thus, most of the top 200 beams are robust to these measurement uncertainties.

As a result of the flux-limited nature of our sample (see Figure~\ref{fig:magvol}), the reported LRG luminosities in our beams are probably biased low given that we are not as sensitive to lower-luminosity LRGs at higher redshifts ($z \gtrsim 0.4$).  Furthermore, galaxies at higher redshifts that could otherwise scatter into the LRG selection cuts are likely to have larger magnitude errors due to their faintness, which can exclude them from being classified as LRGs based on the $r$-band magnitude error cut (see Appendix~\ref{app:selection}).  Thus, these biases potentially affect beams with high-redshift structures more than ones with the bulk of LRGs at lower redshift.

\subsection{Potential Applications to Other Surveys} \label{subsec:future}
Our selection method can be applied not just to current surveys like the SDSS, but also to other ongoing and planned wide-area photometric and spectroscopic surveys.  Ideally, we want a deeper survey than SDSS to probe a volume-limited sample of LRGs to higher redshift ($z \sim 1$) so as to include any cluster-scale lenses at $0.7 \lesssim z \lesssim 1$.  However, any further observing time would be better utilized by expanding the survey area rather than going much deeper in redshift, as the number density of massive cluster-scale halos is decreasing at higher redshift \citep[e.g.,][]{tinker2008} and the lensing geometry for magnifying $z \geq 7$ galaxies is becoming unfavorable.

Surveys with goals that include detection of weak gravitational lensing, with their wide field of view, deep accurate multi-band photometry \citep[particularly for cosmic magnification/convergence studies, e.g.,][]{vanwaerbeke2010,hildebrandt2011,ford2012}, and accurate photometric redshifts, are promising for our selection method.  Current examples include the Kilo-Degree Survey \citep[KiDS;][]{dejong2013}  and the complementary VISTA Kilo-degree Infrared Galaxy Survey (VIKING), as well as the $3\pi$ survey with the Panoramic Survey Telescope and Rapid Response System \citep[Pan-STARRS;][]{kaiser2004}.  The ongoing Dark Energy Survey (DES) will provide deep coverage in the southern sky, complementary to the SDSS coverage.  Planned wide-field instruments and observatories, such as the Hyper Suprime-Cam on Subaru and the Large Synoptic Survey Telescope (LSST), will probe large areas of the sky to unprecedented depths and thus produce excellent datasets for our method.  Space-based missions such as Euclid \citep{laureijs2011} and the Wide-Field Infrared Survey Telescope (WFIRST) will generate deep high-resolution photometry over a large fraction of the sky.

Spectroscopic surveys have the advantage of more robust LRG selection and more accurate redshifts (and therefore luminosities).  A spectroscopic survey with sufficient resolution can also provide information on the physical clustering of the LRGs, which can lead to higher-order estimates of the underlying matter distribution.  Because LRGs are ideal targets for studies of the baryon acoustic oscillation (BAO) feature \citep[e.g.,][]{eisenstein2005}, the goals of BAO surveys match up well with searches for massive beams using our selection method.  Spectra from the ongoing Baryon Oscillation Spectroscopic Survey \citep[BOSS;][]{dawson2013}, a fraction of which are included in the DR9 sample used here, will provide deep spectroscopic data across large areas of the sky and specifically target LRGs.  The upcoming Big Baryon Oscillation Spectroscopic Survey (BigBOSS) will also achieve these goals and probe to deeper redshifts.

\section{CONCLUSIONS} \label{sec:conclusions}
We present a new method of selecting lines of sight (``beams") that contain large total masses and are likely to be the most powerful gravitational lenses for magnifying very high-redshift galaxies.  These fields are important for studying the first galaxies, as their extreme lensing strengths can magnify the background source population into detectability.  We select fields based on the total luminosity in luminous red galaxies \citep[LRGs; e.g.,][]{eisenstein2001} along the line of sight.

We identify the 200 lines of sight in the SDSS DR9 that have the highest total LRG luminosities projected within a radius of $3\farcm5$ and within $0.1 \leq z \leq 0.7$, a key redshift range for lensing high-redshift galaxies.  The total luminosities of LRGs in these fields ($\sim 10^{11.85} - 10^{12.1} h^{-2} L_{\odot}$), which can include $4-18$ total LRGs, are $2-3$ times larger than those of most known strong lensing galaxy clusters from the \citet{hennawi2008} and CLASH \citep{postman2012} surveys, suggesting that they contain larger total masses.  In those beams with multiple LRG peaks in redshift space, those peaks can be individually as rich as known lensing clusters, which may only be traced by a few LRGs.

The distribution of LRGs in these fields, both along the line-of-sight and in projection on the sky, show a diversity of structure.  The LRGs trace both the redshift and angular concentrations of non-LRG galaxies.  Some beams are dominated by a single mass peak, while others contain multiple mass peaks distributed along the line of sight, which can maximize the source plane area that is highly magnified for the purpose of detecting high-redshift galaxies \citep{wong2012}.  Visual inspection of the fields reveals many beams with obvious galaxy clusters, confirming that the LRGs do trace dense structures.  22.5\% of the top 200 beams contain one known cluster and 56.0\% contain multiple known clusters previously identified in the literature.  The rest of the beams may contain new groups and clusters.

Our analysis of the uncertainties in the integrated LRG luminosities of these beams shows that while the detailed rankings are susceptible to fluctuations, the top 200 beams generally comprise fields that have large concentrations of massive galaxies.  $70-75\%$ of the beams remain in the top 200 when an extended sample of the top 1000 beams is sorted by the median LRG luminosity derived by our Monte Carlo error analysis, which accounts for uncertainties in the individual LRG luminosities and for galaxies falling into or out of our selection criteria.

Our follow-up galaxy spectroscopy in a subset of these beams has revealed multiple massive halos and large total masses ($\gtrsim 2\times10^{15} h^{-1} M_{\odot}$), confirming the power of using LRGs as tracers of massive structure (S. M. Ammons et al. 2013, in preparation).  We are modeling the mass distributions in these beams from the spectroscopy alone, but will eventually combine that analysis with a strong lensing analysis of the arcs detected in these fields.  The large total masses and multiple projected massive clusters make these beams likely to be among the best gravitational lenses known.  Future science applications include the detection of faint lensed $z \geq 7$ galaxies that can be followed-up spectroscopically, high spatial resolution studies of strongly lensed galaxies at $z \sim 1-3$, weak lensing studies, and improved detections of $\gamma-$ray sources and supernovae at cosmological distances.

\acknowledgments
We thank Decker French for her contributions to this project.  We are particularly indebted to Daniel Eisenstein and Nikhil Padmanabhan for useful discussions regarding the LRG selection criteria.  We also thank Leon Baruah, Michael Blanton, Marcello Cacciato, Shirley Ho, Daniel Marrone, Jeremiah Ostriker, Ashley Ross, Jeremy Tinker, and Adi Zitrin for helpful discussions and input.  This work was supported by NSF grants AST-0908280 and AST-1211385 and NASA grants ADAP-NNX10AD476 and ADAP-NNX10AE88G.  This work performed in part under the auspices of the U.S. Department of Energy by Lawrence Livermore National Laboratory under Contract DE-AC52-07NA27344.  A.I.Z. thanks the Max Planck Institute for Astronomy and the Center for Cosmology and Particle Physics at New York University for their hospitality and support during her stays there.  This research has made use of the NASA/IPAC Extragalactic Database (NED), which is operated by the Jet Propulsion Laboratory, California Institute of Technology, under contract with the National Aeronautics and Space Administration.

\appendix
\section{SORTING BY NUMBER OF LRGS VS TOTAL LRG LUMINOSITY} \label{app:nsort}
In selecting beams likely to contain large total masses as traced by LRGs, we choose to use total LRG luminosity rather than simple number counts, despite the fact that there are additional uncertainties associated with this method.  This choice arises from the expectation that LRG luminosity traces stellar mass, which is related to halo mass.  While there have not been quantitative studies of the relationship between halo mass and total LRG luminosity at the halo masses we are interested in, we run a simple test here to see whether total LRG luminosity or number counts in galaxy clusters provides a better contrast relative to the field population, i.e., a stronger measure of overdensity.  We can then invoke the known relationship between halo mass and LRG number counts \citep{ho2009} to suggest that either total LRG luminosity or number counts is a better tracer of halo mass.  We use early-type galaxy luminosity functions to compare the expected galaxy luminosity and number counts in beams containing a massive cluster and field galaxies to beams with field galaxies alone.

Adopting the $i$-band luminosity function (LF) of \citet{bernardi2003} to represent the field population of LRGs, we calculate the number of galaxies within $0.1 \leq z \leq 0.7$ and brighter than $M_{i} - 5\mathrm{log}_{10}(h) \leq -21$ expected in a circular field of view of radius $3\farcm5$.  We then take the cluster galaxy $i$-band luminosity function of \citet{popesso2005} to represent the population of LRGs in clusters.  We use the \citet{popesso2005} LF constraints for a cluster-centric radius $r \leq 2.0~h^{-1}$ Mpc with local background subtraction, although our results are qualitatively similar using their results for different cluster-centric radii.  We combine the \citet{popesso2005} bright-end and faint-end components into a single LF, although in practice, the faint-end component contributes very little at the luminosities we are interested in.  Both the field and cluster samples are based on $z < 0.3$ LFs from SDSS data, so evolution in the LF is not taken into account, but both the field and cluster galaxy LFs will have smaller normalizations at higher redshifts.

We integrate the cluster LF down to the chosen limiting magnitude to determine a normalization factor, then normalize the cluster LF for a range of cluster richness, defined as the number of LRGs in the cluster.  Since the \citet{popesso2005} analysis ignored the contribution of the brightest cluster galaxy (BCG) when computing the LF, we add its contribution separately.  We identify the luminosity at which the integrated LF is equal to one galaxy and double the differential number counts in the LF bins brighter than that luminosity.  This effectively modifies the integral of the LF such that there is one additional galaxy in the cluster with the same average luminosity as its brightest (non-BCG) member.  This is conservative in the sense that we will underestimate the luminosity of the BCG since there is a known magnitude gap between the two brightest galaxies in clusters \citep[e.g.,][]{more2012}.

For a range of cluster richness, we compare number counts and total luminosity for the field plus cluster to those for the field alone.  We define the number of LRGs in the field to be 
\begin{equation} \label{eq:nfield}
N_{field} = \left( \frac{\pi R^{2}}{4\pi~\mathrm{ster}} \right) \\
  \int_{z_{min}}^{z_{max}} (c/H(z)) (1+z)^{2} D_{A}(z)^2 dz \\
  \int_{L_{min}}^{\infty} \phi_{field}(L) dL,
\end{equation}
where $R$ is the beam radius, $c/H(z)$ is the Hubble distance at redshift $z$, $D_{A}$ is angular diameter distance, and $\phi_{field}(L)$ is the luminosity function for the field.  The total luminosity for galaxies in the field is then given by 
\begin{equation} \label{eq:lfield}
L_{field} = \left( \frac{\pi R^{2}}{4\pi~\mathrm{ster}} \right) \\
  \int_{z_{min}}^{z_{max}} (c/H(z)) (1+z)^{2} D_{A}(z)^2 dz \\
  \int_{L_{min}}^{\infty} \phi_{field}(L) L dL.
\end{equation}
For the cluster galaxies, we define $N_{cluster}$ to be the number of non-BCG galaxies in the cluster.  The total cluster richness is therefore $N_{cluster} + 1$, and the total luminosity of the cluster galaxies as a function of richness is
\begin{equation} \label{eq:lcluster}
L_{cluster}(N_{cluster} + 1) = \\
  \frac{N_{cluster} \int_{L_{min}}^{\infty} (1 + \mathcal{H}(L - L_{BCG})) \phi_{cluster}(L) L dL}{\\
  \int_{L_{min}}^{\infty} \phi_{cluster}(L) dL},
\end{equation}
where $\phi_{cluster}(L)$ is the luminosity function for the cluster with arbitrary normalization.  The factor of $N_{cluster} / \int_{L_{min}}^{\infty} \phi_{cluster}(L) dL$ serves to normalize the total luminosity for the given richness.  The $1 + \mathcal{H}(L - L_{BCG})$ term, where $\mathcal{H}$ is the Heaviside step function, accounts for the luminosity of the BCG.  $L_{BCG}$ is determined by solving the equation
\begin{equation} \label{eq:lbcg}
\int_{L_{BCG}}^{\infty} \phi_{cluster}(L) dL  = \\
  \frac{\int_{L_{min}}^{\infty} \phi_{cluster}(L) dL}{N_{cluster}}
\end{equation}
for $L_{BCG}$.  Note that the arbitrary normalization of the cluster LF cancels on both sides of Equation~\ref{eq:lbcg}.

We assume $R = 3\farcm5$, $z_{min} = 0.1$, $z_{max} = 0.7$, and test a range of luminosity cuts from $-22.5 \leq M_{min} -5\mathrm{log}_{10}(h) \leq -21$, where $M_{min} = -2.5\mathrm{log}_{10}(L_{min}/L_{\odot})$.  We then define the ``richness contrast" to be
\begin{equation} \label{eq:ncontrast}
C_{N}(N_{cluster}+1) = \frac{N_{field} + N_{cluster}+1}{N_{field}},
\end{equation}
and the ``luminosity contrast" to be 
\begin{equation} \label{eq:lcontrast}
C_{L}(N_{cluster}+1) = \frac{L_{field}+L_{cluster}(N_{cluster}+1)}{L_{field}}.
\end{equation}

These contrasts represent the relative LRG number and luminosity, respectively, of a beam containing a cluster and field galaxies to a beam containing field galaxies alone.  In general, while both richness and total luminosity are strongly correlated, the luminosity contrast is stronger than the number contrast, increasing with greater cluster richness.  As an example, for a richness of $N_{cluster} + 1$ = 12 and $L_{min}$ corresponding to a limiting magnitude of $M_{i} - 5\mathrm{log}_{10}(h) = -21$, $C_{L}$ is 16\% greater than $C_{N}$ .  \citet{reid2009} find the LRG occupation number for $\gtrsim 10^{15} M_{\odot}$ halos to be $\sim 3-5$ LRGs, although their LRG selection was limited to more luminous LRGs than our sample.  We test a range of limiting absolute magnitudes, $-21 \leq M_{i} - 5\mathrm{log}_{10}(h) \leq -22.5$, and find that the contrast for both methods becomes much greater for brighter limiting magnitudes.

For this test, we attempt to be conservative where possible.  The magnitude gap for BCGs, which we ignore, would favor the luminosity method more by increasing the cluster luminosity.  The field early-type LF is based on morphological+spectral PCA criteria, and almost certainly includes more galaxies than would pass our LRG selection criteria.  This effect serves to reduce both contrasts, but given our finding that the luminosity contrast is stronger than the richness contrast, the luminosity contrast will be more strongly affected.

In Table~\ref{tab:nbeams}, we list the $3\farcm5$-radius beams containing $\geq 11$ LRGs that do not overlap with the top 200 luminosity-sorted beams in Table~\ref{tab:beams}.  We present these beams for completeness, as these beams would have been in the top 200 (or had an equal number of LRGs to beams in our top 200) if we had chosen to sort by LRG number counts instead.

\begin{table}
\caption{List of Unique Beams Sorted by N$_{LRG}$ \label{tab:nbeams}}
\begin{ruledtabular}
\begin{tabular}{cccc}
\multirow{2}{*}{RA} &
\multirow{2}{*}{Dec} &
\multirow{2}{*}{N$_{LRG}$} &
log$_{10}$(L$_{i}$)\tablenotemark{a} \\
& & &
($h^{-2}$ L$_{\odot}$)
\\
\tableline
11:09:15.977 &
+09:57:53.352 &
13 &
11.85 \\
13:38:26.535 &
+15:19:27.495 &
13 &
11.84 \\
11:32:31.050 &
+36:50:53.636 &
13 &
11.84 \\
10:05:55.896 &
+47:21:15.407 &
13 &
11.83 \\
08:25:51.875 &
+40:16:59.830 &
13 &
11.82 \\
16:28:13.439 &
+38:24:47.224 &
13 &
11.82 \\
13:49:59.719 &
+39:33:59.337 &
13 &
11.80 \\
10:02:39.789 &
+20:29:22.063 &
12 &
11.86 \\
09:57:38.539 &
+19:38:45.942 &
12 &
11.85 \\
15:11:05.994 &
+67:05:11.986 &
12 &
11.85 \\
00:12:52.314 &
-08:57:47.332 &
12 &
11.85 \\
14:54:20.869 &
+05:55:21.444 &
12 &
11.85 \\
10:35:55.123 &
+59:06:58.574 &
12 &
11.85 \\
16:32:14.530 &
+21:25:22.315 &
12 &
11.85 \\
15:22:35.599 &
+42:34:42.615 &
12 &
11.84 \\
08:53:30.710 &
+23:19:35.213 &
12 &
11.84 \\
09:43:03.652 &
+47:01:14.102 &
12 &
11.84 \\
16:27:27.221 &
+39:41:02.149 &
12 &
11.84 \\
11:27:10.983 &
+20:44:39.185 &
12 &
11.84 \\
10:30:35.164 &
+47:48:39.928 &
12 &
11.83 \\
09:01:55.447 &
+20:54:16.831 &
12 &
11.83 \\
09:08:49.813 &
+61:27:28.658 &
12 &
11.82 \\
15:11:48.439 &
+13:52:02.737 &
12 &
11.82 \\
14:15:18.477 &
+33:44:52.386 &
12 &
11.82 \\
16:17:39.046 &
+42:32:45.143 &
12 &
11.82 \\
11:32:52.075 &
+36:47:13.073 &
12 &
11.81 \\
12:10:48.221 &
+60:17:41.794 &
12 &
11.81 \\
22:43:20.719 &
-09:35:18.906 &
12 &
11.81 \\
12:01:17.199 &
+14:55:49.153 &
12 &
11.81 \\
15:42:25.294 &
+60:03:01.397 &
12 &
11.80 \\
00:44:52.396 &
+07:05:47.833 &
12 &
11.79 \\
09:53:34.737 &
+22:49:09.630 &
12 &
11.79 \\
13:21:22.061 &
+05:59:07.882 &
12 &
11.79 \\
01:38:45.981 &
-10:16:53.809 &
12 &
11.78 \\
08:12:48.436 &
+52:19:06.964 &
12 &
11.78 \\
11:03:04.694 &
+04:19:41.855 &
12 &
11.78 \\
15:17:46.067 &
+04:47:01.250 &
12 &
11.77 \\
15:48:22.764 &
+12:54:52.040 &
12 &
11.77 \\
02:18:14.999 &
-17:04:18.076 &
12 &
11.77 \\
02:39:02.368 &
-01:08:11.545 &
12 &
11.76 \\
10:29:25.019 &
+23:17:29.250 &
12 &
11.76 \\
17:28:37.980 &
+68:14:07.625 &
12 &
11.75 \\
13:00:55.901 &
+22:30:43.953 &
12 &
11.74 \\
11:47:43.158 &
+25:29:07.848 &
12 &
11.68 \\
08:04:54.807 &
+40:26:23.249 &
11 &
11.86 \\
12:31:28.851 &
+17:48:50.446 &
11 &
11.85 \\
00:47:16.514 &
-01:45:44.736 &
11 &
11.85 \\
20:54:49.076 &
-16:20:14.773 &
11 &
11.85 \\
12:02:14.719 &
+61:42:10.414 &
11 &
11.85 \\
14:56:43.695 &
+11:59:50.193 &
11 &
11.85 \\
\end{tabular}
\end{ruledtabular}
\tablenotetext{1}{Total integrated rest-frame $i$-band luminosity in LRGs.}
\end{table}

\addtocounter{table}{-1}
\begin{table}
\caption{Continued.}
\begin{ruledtabular}
\begin{tabular}{cccc}
\multirow{2}{*}{RA} &
\multirow{2}{*}{Dec} &
\multirow{2}{*}{N$_{LRG}$} &
log$_{10}$(L$_{i}$)\tablenotemark{a} \\
& & &
($h^{-2}$ L$_{\odot}$)
\\
\tableline
14:39:20.170 &
+05:44:35.922 &
11 &
11.85 \\
12:22:15.002 &
+42:34:30.585 &
11 &
11.85 \\
09:28:38.426 &
+37:46:55.479 &
11 &
11.85 \\
13:24:06.064 &
+47:40:26.833 &
11 &
11.84 \\
23:11:48.346 &
+03:40:47.616 &
11 &
11.84 \\
12:29:01.509 &
+47:38:55.367 &
11 &
11.84 \\
15:39:40.493 &
+34:25:27.277 &
11 &
11.84 \\
11:06:08.485 &
+33:33:39.682 &
11 &
11.84 \\
12:38:51.513 &
+13:06:30.845 &
11 &
11.84 \\
12:21:43.787 &
+45:25:05.301 &
11 &
11.84 \\
11:59:31.823 &
+49:48:05.792 &
11 &
11.84 \\
09:50:59.745 &
+00:39:43.027 &
11 &
11.84 \\
13:24:11.822 &
+52:18:49.644 &
11 &
11.84 \\
16:14:33.153 &
+62:40:58.110 &
11 &
11.84 \\
21:35:24.074 &
-00:59:50.579 &
11 &
11.83 \\
01:04:27.673 &
+29:03:36.591 &
11 &
11.83 \\
08:53:03.792 &
+52:56:30.040 &
11 &
11.83 \\
15:57:52.855 &
+21:33:41.153 &
11 &
11.83 \\
16:28:17.576 &
+11:09:59.566 &
11 &
11.83 \\
15:37:33.252 &
+28:06:36.831 &
11 &
11.83 \\
09:49:12.774 &
+50:16:09.466 &
11 &
11.83 \\
11:28:15.367 &
+25:49:22.565 &
11 &
11.83 \\
00:06:56.132 &
-00:40:51.829 &
11 &
11.82 \\
14:39:38.927 &
+05:44:00.469 &
11 &
11.82 \\
01:01:40.797 &
+02:36:46.424 &
11 &
11.82 \\
13:44:19.219 &
+39:06:38.980 &
11 &
11.82 \\
11:15:59.327 &
+15:05:20.129 &
11 &
11.82 \\
09:26:08.246 &
+04:25:26.035 &
11 &
11.82 \\
13:59:49.878 &
+49:35:49.136 &
11 &
11.82 \\
00:42:54.313 &
+27:26:47.716 &
11 &
11.82 \\
14:18:21.801 &
+03:03:55.582 &
11 &
11.82 \\
03:17:11.206 &
-07:13:09.000 &
11 &
11.82 \\
08:52:49.102 &
+30:26:33.467 &
11 &
11.81 \\
10:55:13.169 &
+43:57:56.860 &
11 &
11.81 \\
08:34:13.159 &
+45:25:20.803 &
11 &
11.81 \\
08:59:37.436 &
+26:05:36.539 &
11 &
11.81 \\
14:09:04.402 &
+34:45:34.558 &
11 &
11.81 \\
17:30:12.322 &
+55:52:30.416 &
11 &
11.81 \\
23:16:25.189 &
-01:54:32.230 &
11 &
11.81 \\
09:16:25.803 &
+29:52:07.585 &
11 &
11.81 \\
09:27:37.817 &
+56:16:58.192 &
11 &
11.81 \\
02:12:41.969 &
-09:42:27.342 &
11 &
11.81 \\
14:59:34.701 &
+45:14:34.771 &
11 &
11.80 \\
14:40:10.313 &
+14:17:11.818 &
11 &
11.80 \\
14:45:18.655 &
+00:06:31.825 &
11 &
11.80 \\
12:10:39.503 &
+04:29:20.492 &
11 &
11.80 \\
23:05:50.195 &
+00:06:36.279 &
11 &
11.80 \\
10:08:54.896 &
+45:27:58.888 &
11 &
11.80 \\
11:51:54.536 &
+17:50:37.422 &
11 &
11.80 \\
11:43:23.554 &
-00:28:01.790 &
11 &
11.80 \\
\end{tabular}
\end{ruledtabular}
\end{table}

\addtocounter{table}{-1}
\begin{table}
\caption{Continued.}
\begin{ruledtabular}
\begin{tabular}{cccc}
\multirow{2}{*}{RA} &
\multirow{2}{*}{Dec} &
\multirow{2}{*}{N$_{LRG}$} &
log$_{10}$(L$_{i}$)\tablenotemark{a} \\
& & &
($h^{-2}$ L$_{\odot}$)
\\
\tableline
10:21:25.938 &
+12:14:39.260 &
11 &
11.80 \\
02:01:07.278 &
+13:22:24.961 &
11 &
11.80 \\
13:29:24.519 &
+65:27:53.321 &
11 &
11.80 \\
10:10:33.584 &
+08:07:05.982 &
11 &
11.80 \\
13:01:50.667 &
+22:43:34.859 &
11 &
11.79 \\
01:17:33.818 &
+15:45:48.878 &
11 &
11.79 \\
15:43:29.983 &
+35:42:18.202 &
11 &
11.79 \\
12:37:03.247 &
+27:58:19.516 &
11 &
11.79 \\
11:13:14.638 &
+25:59:44.165 &
11 &
11.79 \\
00:11:01.227 &
+29:07:51.757 &
11 &
11.79 \\
11:49:13.307 &
+35:31:54.141 &
11 &
11.79 \\
03:17:53.373 &
-07:05:31.346 &
11 &
11.78 \\
11:12:15.720 &
+23:52:40.054 &
11 &
11.78 \\
00:33:20.715 &
+04:16:19.676 &
11 &
11.78 \\
23:31:26.943 &
+00:35:06.526 &
11 &
11.78 \\
09:00:49.059 &
+38:29:12.213 &
11 &
11.78 \\
22:58:51.623 &
-03:25:22.241 &
11 &
11.78 \\
11:09:31.024 &
+09:55:00.194 &
11 &
11.77 \\
01:02:46.991 &
+11:10:25.567 &
11 &
11.77 \\
12:35:18.759 &
+41:19:56.381 &
11 &
11.77 \\
09:27:22.794 &
+44:10:08.169 &
11 &
11.76 \\
00:00:52.090 &
+28:09:08.009 &
11 &
11.76 \\
00:46:16.731 &
-10:24:23.642 &
11 &
11.76 \\
12:46:24.192 &
+36:26:12.921 &
11 &
11.76 \\
14:27:36.058 &
+55:49:26.812 &
11 &
11.76 \\
01:13:15.628 &
+28:20:05.420 &
11 &
11.75 \\
12:06:09.519 &
+44:41:04.943 &
11 &
11.75 \\
23:04:24.257 &
+02:12:24.002 &
11 &
11.74 \\
09:41:17.904 &
+18:46:46.990 &
11 &
11.74 \\
14:20:58.067 &
+21:08:07.372 &
11 &
11.74 \\
10:31:48.173 &
+47:22:24.694 &
11 &
11.74 \\
13:10:56.248 &
+45:15:35.924 &
11 &
11.74 \\
09:27:59.505 &
+12:45:50.494 &
11 &
11.73 \\
10:33:29.650 &
+37:13:26.586 &
11 &
11.73 \\
14:18:21.330 &
+48:37:36.040 &
11 &
11.73 \\
12:50:52.163 &
+26:49:38.982 &
11 &
11.73 \\
15:20:25.556 &
+06:21:48.889 &
11 &
11.72 \\
12:26:50.804 &
+33:11:20.101 &
11 &
11.72 \\
12:41:02.456 &
+44:12:13.803 &
11 &
11.72 \\
16:07:35.906 &
+16:51:04.155 &
11 &
11.72 \\
02:00:18.811 &
-07:54:08.177 &
11 &
11.71 \\
11:54:30.316 &
+05:24:40.867 &
11 &
11.71 \\
14:03:39.841 &
+10:14:54.200 &
11 &
11.71 \\
11:38:25.361 &
+27:02:21.213 &
11 &
11.70 \\
11:48:32.089 &
+37:48:31.262 &
11 &
11.70 \\
11:47:25.348 &
+44:20:02.282 &
11 &
11.69 \\
14:01:18.524 &
+15:14:08.027 &
11 &
11.68 \\
14:31:57.545 &
+06:37:21.361 &
11 &
11.67 \\
11:09:37.592 &
+38:28:53.623 &
11 &
11.65 \\
\end{tabular}
\end{ruledtabular}
\end{table}

\clearpage

\section{SDSS LRG SELECTION CRITERIA} \label{app:selection}
In this Appendix, we explain our LRG catalog selection criteria.  We base our selection of SDSS DR9 LRGs on the criteria originally defined in \citet{padmanabhan2005}, with some minor modifications.  The LRG selection uses two separate selection criteria (denoted ``Cut I" and ``Cut II" for the $z \lesssim 0.4$ and $z \gtrsim 0.4$ samples, respectively) to define the full sample.  The SQL queries for the two cuts are given in Figures~\ref{fig:sql1} and~\ref{fig:sql2}.  The two cuts are not mutually exclusive, so we remove duplicate objects after combining the two samples into a single catalog.

\begin{figure*}
\begin{center}
\begin{minipage}{0.7\linewidth}
\begin{verbatim}
SELECT
    g.objID,
    g.run,
    g.rerun,
    g.camcol,
    g.field,
    g.fieldID,
    g.obj,
    g.ra,
    g.dec,
    g.b,
    g.dered_g,
    g.dered_r,
    g.dered_i,
    g.extinction_r,
    g.petroMag_r,
    g.psfMag_r,
    g.ModelMag_r,
    g.ModelMagErr_r,
    g.psfMag_i,
    g.ModelMag_i,
    g.ModelMagErr_i,
    g.deVRad_r,
    g.petroR50_r,
    g.flags,
    p.z,
    p.zErr,
    f.quality,
    f.psfWidth_r

FROM
    Galaxy as g, Photoz as p, Field as f

WHERE  p.objID = g.objID
    and f.fieldID = g.fieldID
    and f.quality = 3
    and f.psfWidth_r < 2
    and (g.dered_g - g.dered_r) < 3.0
    and (g.dered_r - g.dered_i) < 1.5
    and g.extinction_r < 0.2
    and g.modelmagerr_r < 0.2
    and (g.petroMag_r + (2.5 * LOG10(2 * pi() * POWER(g.petroR50_r,2.0)))) < 24.2
    and ABS(g.b) > 30
    and ABS((g.dered_r - g.dered_i) - ((g.dered_g - g.dered_r) / 4.0) - 0.18) < 0.2
    and (g.petroMag_r - g.extinction_r) <
         (13.6 + ((0.7 * (g.dered_g - g.dered_r)) + (1.2 * (g.dered_r - g.dered_i - 0.18))) / 0.3)
    and (g.petroMag_r - g.extinction_r) < 19.7
    and (g.psfMag_r - g.ModelMag_r) > 0.3
    and (g.flags & 0x0000000000000002) = 0
    and (g.flags & 0x0000000000000004) = 0
    and (g.flags & 0x0000000000000008) = 0
    and (g.flags & 0x0000000000040000) = 0
    and (g.flags & 0x0000000000100000) = 0
    and (g.flags & 0x0000000002000000) = 0
    and (g.flags & 0x0000000004000000) = 0
    and (g.flags & 0x0000000070000000) != 0
\end{verbatim}
\end{minipage}
\end{center}
\caption{SDSS database SQL query for Cut I.} \label{fig:sql1}
\label{fig:sqlcut1}
\end{figure*}

\begin{figure*}
\begin{center}
\begin{minipage}{0.7\linewidth}
\begin{verbatim}
SELECT
    g.objID,
    g.run,
    g.rerun,
    g.camcol,
    g.field,
    g.fieldID,
    g.obj,
    g.ra,
    g.dec,
    g.b,
    g.dered_g,
    g.dered_r,
    g.dered_i,
    g.extinction_r,
    g.petroMag_r,
    g.psfMag_r,
    g.ModelMag_r,
    g.ModelMagErr_r,
    g.psfMag_i,
    g.ModelMag_i,
    g.ModelMagErr_i,
    g.deVRad_r,
    g.petroR50_r,
    g.flags,
    p.z,
    p.zErr,
    f.quality,
    f.psfWidth_r

FROM
    Galaxy as g, Photoz as p, Field as f

WHERE  P.objID = g.objID
    and f.fieldID = g.fieldID
    and f.quality = 3
    and f.psfWidth_r < 2
    and (g.dered_g - g.dered_r) < 3.0
    and (g.dered_r - g.dered_i) < 1.5
    and g.extinction_r < 0.2
    and g.modelmagerr_r < 0.2
    and (g.petroMag_r + (2.5 * LOG10(2 * pi() * POWER(g.petroR50_r,2.0)))) < 24.2
    and ABS(g.b) > 30
    and ((g.dered_r - g.dered_i) - ((g.dered_g - g.dered_r) / 8)) > 0.55
    and (g.dered_g - g.dered_r) > 1.4
    and g.dered_i < (18.3 + (2 * ((g.dered_r - g.dered_i) - (g.dered_g - g.dered_r) / 8)))
    and g.dered_i < 20
    and (g.psfMag_i - g.ModelMag_i) > 0.2 * (21 - g.dered_i)
    and g.deVRad_r > 0.2
    and (g.flags & 0x0000000000000002) = 0
    and (g.flags & 0x0000000000000004) = 0
    and (g.flags & 0x0000000000000008) = 0
    and (g.flags & 0x0000000000040000) = 0
    and (g.flags & 0x0000000000100000) = 0
    and (g.flags & 0x0000000002000000) = 0
    and (g.flags & 0x0000000004000000) = 0
    and (g.flags & 0x0000000070000000) != 0
\end{verbatim}
\end{minipage}
\end{center}
\caption{SDSS database SQL query for Cut II.} \label{fig:sql2}
\label{fig:sqlcut2}
\end{figure*}

We modify the original LRG selection criteria, which selected LRGs for clustering studies that had different requirements than our applications.  We add cuts to exclude fields with poor quality or that have an effective $r$-band PSF width $\geq 2$\arcsec, indicating poor seeing.  We remove objects with $r$-band extinction $\geq 0.2$ or $r$-band magnitude errors $\geq 0.2$, as we are less confident in these objects' absolute magnitudes (N. Padmanabhan, private communication).  We reduce the galactic latitude criterion from $b \leq 45^{\circ}$ to $b \leq 30^{\circ}$, which strikes a balance between excluding fields with a high-density of stars that could complicate follow-up observations and including as much survey area as possible.  This cut eliminates less than 20\% of the total sky coverage of DR9.  The rest of our color and photometric flag cuts remove most spurious stellar contaminants.  The photometry flags are standard quality cuts to remove spurious objects (A. Ross, private communication).  Based on a visual inspection of subsets of objects in our sample, we also cut objects with the ``subtracted" or ``deblended\_as\_psf" flags set, which removes $< 1$\% of our sample but cuts out objects contaminated by a nearby bright star.  We also cut objects with the ``deblend\_pruned" flag set, which removes a tiny fraction ($< 100$ objects) of our sample with apparent blending issues.

We use similar criteria to get the spectroscopic sample of LRGs, requiring in addition that the ``zwarning" flag is zero and the redshift error is $< 10^{-3}$.  We then match the spectroscopic and photometric samples, replacing the photometric redshift and its error with the spectroscopic redshift and error where applicable.  We apply cuts on the object redshift only after combining the catalogs, as an object with a photometric redshift outside of our target redshift range ($0.1 \leq z \leq 0.7$) may have a spectroscopic redshift within that range.

We then calculate absolute magnitudes for the LRGs, accounting for K-corrections and passive evolution, and apply our absolute $i$-band magnitude cut.  We extend the original \citet{padmanabhan2005} absolute magnitude cut ($-21 < M_{i} - 5\mathrm{log}_{10}(h) < -24$) to include more luminous galaxies up to $-24.7$ based on visual inspection.  While extending this cut does introduce contaminants, it also includes the most luminous LRGs, which contribute the most to the total luminosity of a beam.  We check through visual inspection that these contaminants do not affect our top 200 beams.

\section{KNOWN CLUSTERS IN TOP 200 BEAMS} \label{app:clusters}
We present the list of known clusters in our top 200 beams in Table~\ref{tab:clusters}.  These clusters were identified in a search of the NASA/IPAC Extragalactic Database\footnote{http://ned.ipac.caltech.edu} (NED) as galaxy clusters within $3\farcm5$ of our beam centers\footnote{Z. Wen reports in a private communication that most of our LRG-selected fields contain at least one cluster identified in his larger \citet{wen2012} sample, which was not available through NED at the time of our analysis.}.  The clusters in each beam are ordered by angular offset from the beam center.

\begin{table*}
\caption{List of Known Clusters in SDSS LRG Beams \label{tab:clusters}}
\begin{ruledtabular}
\begin{tabular}{ccc|cc}
Rank &
RA &
Dec &
Cluster\tablenotemark{a} &
Redshift\tablenotemark{b}
\\
\tableline
\multirow{3}{*}{1} &
\multirow{3}{*}{11:23:53.231} &
\multirow{3}{*}{+50:52:53.458} &
WHL J112353.2+505253 &
0.332 \\
& & &
ZwCl 1121.0+5110 &
... \\
& & &
GMBCG J171.03358+50.88460 &
0.416 \\
\tableline
\multirow{4}{*}{2} &
\multirow{4}{*}{12:44:02.713} &
\multirow{4}{*}{+16:51:53.185} &
WHL J124359.7+16511 &
0.5593 \\
& & &
GMBCG J191.00805+16.83941 &
0.516 \\
& & &
GMBCG J191.02734+16.89091 &
0.486 \\
& & &
MS 1241.5+1710 &
0.312 \\
\tableline
\multirow{3}{*}{4} &
\multirow{3}{*}{11:42:24.777} &
\multirow{3}{*}{+58:32:05.333} &
WHL J114224.8+583205 &
0.3109 \\
& & &
Abell 1351 &
0.3224 \\
& & &
GMBCG J175.58575+58.48567 &
0.211 \\
\tableline
\multirow{5}{*}{7} &
\multirow{5}{*}{09:49:58.602} &
\multirow{5}{*}{+17:08:24.240} &
MACS J0949.8+1708 &
0.3826 \\
& & &
GMBCG J147.50544+17.11794 &
0.443 \\
& & &
GMBCG J147.52556+17.13526 &
0.537 \\
& & &
WHL J094951.3+170701 &
0.3638 \\
& & &
GMBCG J147.47303+17.18400 &
0.39 \\
\tableline
\multirow{2}{*}{8} &
\multirow{2}{*}{15:26:13.832} &
\multirow{2}{*}{+04:40:30.109} &
GMBCG J231.56702+04.67892 &
0.499 \\
& & &
WHL J152612.1+043951 &
0.5169 \\
\tableline
\multirow{3}{*}{9} &
\multirow{3}{*}{10:53:30.330} &
\multirow{3}{*}{+56:41:21.327} &
WHL J105325.2+564042 &
0.4571 \\
& & &
GMBCG J163.33806+56.69041 &
0.4 \\
& & &
GMBCG J163.40388+56.66324 &
0.406 \\
\tableline
\multirow{3}{*}{10} &
\multirow{3}{*}{02:18:39.886} &
\multirow{3}{*}{-00:12:31.752} &
SDSS CE J034.664303-00.205654 &
0.31075 \\
& & &
WHL J021843.3-001259 &
0.3483 \\
& & &
WHL J021845.2-001452 &
0.6338 \\
\tableline
\multirow{2}{*}{11} &
\multirow{2}{*}{10:06:11.170} &
\multirow{2}{*}{+08:03:00.487} &
WHL J100601.9+080401 &
0.27275 \\
& & &
GMBCG J151.49575+08.06809 &
0.38 \\
\tableline
\multirow{2}{*}{12} &
\multirow{2}{*}{11:05:34.246} &
\multirow{2}{*}{+17:35:28.209} &
WHL J110538.5+173544 &
0.5038 \\
& & &
WHL J110521.7+173505 &
0.5099 \\
\tableline
\multirow{5}{*}{14} &
\multirow{5}{*}{16:16:11.888} &
\multirow{5}{*}{+06:57:44.695} &
GMBCG J244.03332+06.98599 &
0.34248 \\
& & &
NSC J161620+065841 &
0.3357 \\
& & &
GMBCG J244.07472+06.91786 &
0.408 \\
& & &
WHL J161609.9+065437 &
0.3288 \\
& & &
GMBCG J244.07577+07.00922 &
0.255 \\
\tableline
\multirow{3}{*}{15} &
\multirow{3}{*}{14:17:45.068} &
\multirow{3}{*}{+21:18:27.860} &
GMBCG J214.44457+21.27367 &
0.409 \\
& & &
WHL J141755.5+211840 &
0.4023 \\
& & &
GMBCG J214.48589+21.33680 &
0.429 \\
\tableline
17 &
02:39:53.126 &
--01:34:55.980 &
Abell 370 &
0.375 \\
\tableline
\multirow{2}{*}{18} &
\multirow{2}{*}{08:16:58.695} &
\multirow{2}{*}{+49:33:42.953} &
GMBCG J124.19153+49.54958 &
0.387 \\
& & &
WHL J081644.7+493512 &
0.3349 \\
\tableline
\multirow{3}{*}{20} &
\multirow{3}{*}{08:19:53.827} &
\multirow{3}{*}{+31:59:57.660} &
WHL J081955.9+315904 &
0.5086 \\
& & &
GMBCG J125.00893+31.96433 &
0.318 \\
& & &
GMBCG J125.02582+32.02435 &
0.419 \\
\tableline
\multirow{3}{*}{21} &
\multirow{3}{*}{11:52:08.844} &
\multirow{3}{*}{+31:42:35.278} &
WHL J115203.0+314100 &
0.5353 \\
& & &
WHL J115204.5+314458 &
0.4883 \\
& & &
WHL J115202.8+314516 &
0.349 \\
\tableline
\multirow{4}{*}{23} &
\multirow{4}{*}{10:50:03.460} &
\multirow{4}{*}{+28:29:58.080} &
GMBCG J162.48938+28.51768 &
0.462 \\
& & &
GMBCG J162.50300+28.46881 &
0.389 \\
& & &
WHL J105005.2+283222 &
0.4335 \\
& & &
GMBCG J162.56190+28.50020 &
0.411 \\
\tableline
24 &
12:09:15.992 &
+26:43:33.237 &
WHL J120918.8+264101 &
0.5327 \\
\tableline
25 &
13:40:46.678 &
--02:51:50.541 &
WHL J134046.7-025150 &
0.532 \\
\tableline
\multirow{3}{*}{26} &
\multirow{3}{*}{11:39:26.649} &
\multirow{3}{*}{+47:04:27.608} &
GMBCG J174.81035+47.03759 &
0.44 \\
& & &
WHL J113920.0+470727 &
0.5223 \\
& & &
GMBCG J174.78774+47.10125 &
0.323 \\
\tableline
\multirow{2}{*}{27} &
\multirow{2}{*}{13:51:32.691} &
\multirow{2}{*}{+52:03:33.346} &
GMBCG J207.84562+52.06839 &
0.338 \\
& & &
WHL J135143.9+520407 &
0.3875 \\
\end{tabular}
\end{ruledtabular}
\tablecomments{Galaxy clusters identified from NASA/IPAC Extragalactic Database (NED) within 3\farcm5 of the beam centers.  Clusters are sorted by proximity to the beam center.  References -- WHL = \citet{wen2009}; ZwCl = \citet{zwicky1961}; GMBCG = \citet{hao2010}; MS = \citet{ueda2001}; Abell = \citet{abell1989}; MACS = \citet{ebeling2001}; NSC = \citet{gal2003}; EAD = \citet{estrada2007}; MaxBCG = \citet{koester2007}; NSCS = \citet{lopes2004}; SDSS CE = \citet{goto2002}; SHELS = \citet{geller2005}; AWM = \citet{abate2009}; DDM = \citet{desai2004}}
\tablenotetext{1}{For clusters with multiple designations, we list the first designation given by NED.}
\tablenotetext{2}{Redshift precision to lowest significant non-zero digit.}
\end{table*}

\addtocounter{table}{-1}
\begin{table*}
\caption{Continued.}
\begin{ruledtabular}
\begin{tabular}{ccc|cc}
Rank &
RA &
Dec &
Cluster\tablenotemark{a} &
Redshift\tablenotemark{b}
\\
\tableline
\multirow{3}{*}{28} &
\multirow{3}{*}{14:37:40.295} &
\multirow{3}{*}{+30:12:00.275} &
WHL J143740.3+301200 &
0.3271 \\
& & &
Abell 1943 &
0.182 \\
& & &
NSC J143737+300923 &
0.338 \\
\tableline
\multirow{3}{*}{29} &
\multirow{3}{*}{12:43:08.915} &
\multirow{3}{*}{+20:22:51.768} &
GMBCG J190.77072+20.34803 &
0.418 \\
& & &
GMBCG J190.77001+20.41813 &
0.504 \\
& & &
WHL J124307.6+202537 &
0.5044 \\
\tableline
\multirow{3}{*}{30} &
\multirow{3}{*}{17:43:22.152} &
\multirow{3}{*}{+63:42:57.575} &
WHL J174330.4+634141 &
0.3211 \\
& & &
Abell 2280 &
0.326 \\
& & &
NSC J174255+634144 &
0.3606 \\
\tableline
\multirow{4}{*}{31} &
\multirow{4}{*}{02:02:01.333} &
\multirow{4}{*}{-08:29:02.252} &
WHL J020155.8-082720 &
0.4836 \\
& & &
WHL J020212.0-082849 &
0.3705 \\
& & &
GMBCG J030.46031-08.47258 &
0.305 \\
& & &
GMBCG J030.53418-08.43703 &
0.349 \\
\tableline
\multirow{3}{*}{32} &
\multirow{3}{*}{09:10:41.968} &
\multirow{3}{*}{+38:50:33.710} &
GMBCG J137.68624+38.83708 &
0.511 \\
& & &
WHL J091049.8+385009 &
0.5651 \\
& & &
GMBCG J137.69226+38.79387 &
0.415 \\
\tableline
\multirow{2}{*}{33} &
\multirow{2}{*}{09:42:55.372} &
\multirow{2}{*}{+14:27:20.611} &
WHL J094253.2+142907 &
0.3335 \\
& & &
GMBCG J145.73927+14.50789 &
0.346 \\
\tableline
\multirow{4}{*}{34} &
\multirow{4}{*}{14:33:25.780} &
\multirow{4}{*}{+29:27:45.979} &
Abell 1934 &
0.2194 \\
& & &
GMBCG J218.35869+29.48397 &
0.354 \\
& & &
[EAD2007] 500 &
0.40968 \\
& & &
MaxBCG J218.32159+29.51165 &
0.22145 \\
\tableline
\multirow{5}{*}{35} &
\multirow{5}{*}{13:06:54.628} &
\multirow{5}{*}{+46:30:36.691} &
WHL J130657.3+463206 &
0.2081 \\
& & &
Abell 1682 &
0.2339 \\
& & &
GMBCG J196.70832+46.55927 &
0.245 \\
& & &
NSC J130639+463208 &
0.2508 \\
& & &
GMBCG J196.75262+46.56389 &
0.337 \\
\tableline
\multirow{3}{*}{36} &
\multirow{3}{*}{10:51:34.352} &
\multirow{3}{*}{+42:23:29.626} &
Abell 1110 &
0.194 \\
& & &
NSC J105128+422216 &
0.2982 \\
& & &
GMBCG J162.89536+42.34162 &
0.187 \\
\tableline
\multirow{3}{*}{37} &
\multirow{3}{*}{01:37:26.699} &
\multirow{3}{*}{+07:52:09.305} &
GMBCG J260.55436+32.11438 &
0.304 \\
& & &
GMBCG J260.51908+32.07349 &
0.388 \\
& & &
GMBCG J260.61326+32.13257 &
0.225 \\
\tableline
\multirow{3}{*}{38} &
\multirow{3}{*}{17:22:13.049} &
\multirow{3}{*}{+32:06:51.773} &
GMBCG J260.55436+32.11438 &
0.304 \\
& & &
GMBCG J260.51908+32.07349 &
0.388 \\
& & &
GMBCG J260.61326+32.13257 &
0.225 \\
\tableline
\multirow{4}{*}{39} &
\multirow{4}{*}{12:52:58.597} &
\multirow{4}{*}{+23:42:00.034} &
WHL J125251.8+234206 &
0.3765 \\
& & &
WHL J125300.6+234414 &
0.5187 \\
& & &
ZwCl 1250.6+2401 &
... \\
& & &
GMBCG J193.29269+23.66893 &
0.457 \\
\tableline
40 &
01:19:56.788 &
+12:18:34.735 &
NSCS J011959+121839 &
0.37 \\
\tableline
\multirow{5}{*}{42} &
\multirow{5}{*}{10:35:35.605} &
\multirow{5}{*}{+31:17:47.478} &
NSCS J103537+311759 &
0.36 \\
& & &
GMBCG J158.90316+31.26230 &
0.332 \\
& & &
ZwCl 1032.9+3131 &
0.28085 \\
& & &
WHL J103527.2+312001 &
0.3702 \\
& & &
GMBCG J158.88440+31.35249 &
0.336 \\
\tableline
\multirow{3}{*}{43} &
\multirow{3}{*}{23:34:23.927} &
\multirow{3}{*}{-00:25:00.606} &
SDSS CE J353.605194-00.417486 &
0.37883 \\
& & &
GMBCG J353.58053-00.42458 &
0.462 \\
& & &
WHL J233424.3-002618 &
0.4461 \\
\tableline
45 &
15:26:26.690 &
+04:39:04.707 &
GMBCG J231.56702+04.67892 &
0.499 \\
\tableline
\multirow{2}{*}{46} &
\multirow{2}{*}{00:01:58.481} &
\multirow{2}{*}{+12:03:58.021} &
NSCS J000153+120347 &
0.2033 \\
& & &
Abell 2692 &
0.192 \\
\tableline
\multirow{3}{*}{47} &
\multirow{3}{*}{14:39:56.246} &
\multirow{3}{*}{+54:51:14.221} &
WHL J143958.0+545031 &
0.5435 \\
& & &
MaxBCG J220.01388+54.80918 &
0.26465 \\
& & &
GMBCG J219.97530+54.80381 &
0.398 \\
\tableline
\multirow{5}{*}{48} &
\multirow{5}{*}{09:43:29.460} &
\multirow{5}{*}{+33:18:49.403} &
NSC J094329+331912 &
0.2463 \\
& & &
ZwCl 0940.6+3334 &
... \\
& & &
GMBCG J145.91018+33.31696 &
0.25655 \\
& & &
WHL J094339.7+332037 &
0.4815 \\
& & &
WHL J094316.0+331825 &
0.3216 \\
\tableline
\multirow{3}{*}{50} &
\multirow{3}{*}{22:43:28.058} &
\multirow{3}{*}{-00:25:58.808} &
GMBCG J340.83815-00.43147 &
0.445 \\
& & &
WHL J224326.0-002405 &
0.3495 \\
& & &
GMBCG J340.82973-00.45495 &
0.184 \\
\tableline
51 &
15:01:56.456 &
+33:20:41.059 &
GMBCG J225.48502+33.33266 &
0.304 \\
\tableline
52 &
02:20:56.845 &
+06:52:09.157 &
NSCS J022059+065249 &
0.35 \\
\end{tabular}
\end{ruledtabular}
\end{table*}

\addtocounter{table}{-1}
\begin{table*}
\caption{Continued.}
\begin{ruledtabular}
\begin{tabular}{ccc|cc}
Rank &
RA &
Dec &
Cluster\tablenotemark{a} &
Redshift\tablenotemark{b}
\\
\tableline
\multirow{2}{*}{53} &
\multirow{2}{*}{01:48:08.231} &
\multirow{2}{*}{+00:00:59.692} &
GMBCG J027.04490+00.00106 &
0.38 \\
& & &
SDSS CE J027.042007+00.011633 &
0.34479 \\
\tableline
\multirow{3}{*}{56} &
\multirow{3}{*}{10:39:51.501} &
\multirow{3}{*}{+15:27:25.227} &
WHL J103952.8+152712 &
0.4893 \\
& & &
GMBCG J159.97959+15.45350 &
0.285 \\
& & &
GMBCG J159.90661+15.44490 &
0.496 \\
\tableline
57 &
16:15:59.965 &
+06:55:18.520 &
WHL J161609.9+065437 &
0.3288 \\
\tableline
\multirow{3}{*}{59} &
\multirow{3}{*}{01:59:59.711} &
\multirow{3}{*}{-08:49:39.704} &
WHL J015951.0-084910 &
0.4032 \\
& & &
GMBCG J029.95560-08.83299 &
0.322 \\
& & &
MACS J0159.8-0849 &
0.405 \\
\tableline
\multirow{2}{*}{60} &
\multirow{2}{*}{10:56:14.771} &
\multirow{2}{*}{+28:22:23.064} &
WHL J105614.8+282223 &
0.5731 \\
& & &
GMBCG J164.03905+28.37180 &
0.478 \\
\tableline
\multirow{2}{*}{61} &
\multirow{2}{*}{09:14:23.778} &
\multirow{2}{*}{+21:24:52.512} &
WHL J091424.2+212543 &
0.545 \\
& & &
GMBCG J138.61986+21.41015 &
0.425 \\
\tableline
\multirow{2}{*}{63} &
\multirow{2}{*}{09:02:16.490} &
\multirow{2}{*}{+38:07:07.073} &
GMBCG J135.52728+38.09073 &
0.492 \\
& & &
WHL J090208.9+380450 &
0.4962 \\
\tableline
65 &
16:54:24.482 &
+44:42:10.793 &
WHL J165420.1+444125 &
0.4549 \\
\tableline
\multirow{5}{*}{66} &
\multirow{5}{*}{09:26:35.472} &
\multirow{5}{*}{+29:34:22.128} &
WHL J092635.5+293422 &
0.29165 \\
& & &
ZwCl 0923.6+2946 &
... \\
& & &
WHL J092628.0+293452 &
0.4935 \\
& & &
GMBCG J141.61858+29.59521 &
0.476 \\
& & &
GMBCG J141.68858+29.56935 &
0.191 \\
\tableline
67 &
15:03:01.311 &
+27:57:48.781 &
WHL J150303.7+275519 &
0.5292 \\
\tableline
\multirow{2}{*}{69} &
\multirow{2}{*}{14:33:54.319} &
\multirow{2}{*}{+50:40:45.173} &
ZwCl 1432.0+5054 &
0 \\
& & &
GMBCG J218.43091+50.67096 &
0.38 \\
\tableline
\multirow{2}{*}{70} &
\multirow{2}{*}{08:53:00.399} &
\multirow{2}{*}{+26:22:13.805} &
GMBCG J133.25279+26.36150 &
0.452 \\
& & &
WHL J085254.0+262117 &
0.4549 \\
\tableline
\multirow{2}{*}{72} &
\multirow{2}{*}{11:00:10.270} &
\multirow{2}{*}{+19:16:17.104} &
WHL J110014.2+191622 &
0.2382 \\
& & &
WHL J110003.2+191656 &
0.4442 \\
\tableline
\multirow{2}{*}{74} &
\multirow{2}{*}{13:26:36.060} &
\multirow{2}{*}{+53:53:57.959} &
WHL J132627.0+535348 &
0.375 \\
& & &
GMBCG J201.72979+53.87878 &
0.327 \\
\tableline
\multirow{3}{*}{75} &
\multirow{3}{*}{01:19:34.439} &
\multirow{3}{*}{+14:52:08.957} &
Abell 175 &
0.1292 \\
& & &
WHL J011938.3+145352 &
0.1289 \\
& & &
MaxBCG J019.90953+14.89798 &
0.14315 \\
\tableline
\multirow{2}{*}{77} &
\multirow{2}{*}{11:56:12.252} &
\multirow{2}{*}{-00:21:02.814} &
SDSS CE J179.044357-00.342619 &
0.26537 \\
& & &
GMBCG J179.08749-00.37107 &
0.199 \\
\tableline
\multirow{3}{*}{79} &
\multirow{3}{*}{12:12:08.759} &
\multirow{3}{*}{+27:34:06.919} &
NSCS J121218+273325 &
0 \\
& & &
WHL J121218.5+273255 &
0.3464 \\
& & &
GMBCG J182.99449+27.54719 &
0.447 \\
\tableline
80 &
16:16:27.616 &
+58:12:38.798 &
WHL J161644.8+581118 &
0.2724 \\
\tableline
81 &
12:58:32.002 &
+43:59:47.314 &
WHL J125835.3+440102 &
0.5064 \\
\tableline
82 &
08:07:56.920 &
+65:25:07.350 &
NSC J080809+652543 &
0.1263 \\
\tableline
\multirow{2}{*}{83} &
\multirow{2}{*}{10:50:20.408} &
\multirow{2}{*}{+28:28:04.966} &
GMBCG J162.56190+28.50020 &
0.411 \\
& & &
WHL J105022.4+282439 &
0.4712 \\
\tableline
\multirow{2}{*}{84} &
\multirow{2}{*}{23:19:33.487} &
\multirow{2}{*}{-01:19:26.377} &
NSCS J231933-011656 &
0.3 \\
& & &
ZwCl 2316.8-0135 &
... \\
\tableline
\multirow{2}{*}{85} &
\multirow{2}{*}{09:21:11.999} &
\multirow{2}{*}{+30:29:24.946} &
SHELS J0921.2+3028 &
0.427 \\
& & &
[AWM2009] Cluster 1C &
0.427 \\
\tableline
\multirow{3}{*}{87} &
\multirow{3}{*}{12:06:57.409} &
\multirow{3}{*}{+30:29:22.828} &
NSCS J120702+303056 &
0.21 \\
& & &
WHL J120654.8+303119 &
0.4916 \\
& & &
WHL J120708.2+302741 &
0.4629 \\
\tableline
88 &
15:48:35.149 &
+17:02:22.535 &
WHL J154840.1+170448 &
0.3998 \\
\tableline
\multirow{3}{*}{89} &
\multirow{3}{*}{09:11:06.757} &
\multirow{3}{*}{+61:08:18.085} &
NSC J091110+610734 &
0.2959 \\
& & &
Abell 747 &
... \\
& & &
GMBCG J137.72824+61.09600 &
0.322 \\
\tableline
\multirow{2}{*}{91} &
\multirow{2}{*}{08:22:49.875} &
\multirow{2}{*}{+41:28:12.007} &
WHL J082247.8+412744 &
0.4664 \\
& & &
GMBCG J125.67646+41.45847 &
0.47 \\
\tableline
92 &
23:03:44.474 &
+00:09:38.406 &
GMBCG J345.89121+00.14293 &
0.525 \\
\tableline
\multirow{5}{*}{93} &
\multirow{5}{*}{09:16:14.956} &
\multirow{5}{*}{-00:25:31.237} &
GMBCG J139.05662-00.42820 &
0.217 \\
& & &
GMBCG J139.03846-00.40449 &
0.345 \\
& & &
Abell 776 &
0.33594 \\
& & &
WHL J091605.7-002324 &
0.3208 \\
& & &
GMBCG J139.11397-00.43552 &
0.433 \\
\tableline
\multirow{4}{*}{94} &
\multirow{4}{*}{12:01:25.380} &
\multirow{4}{*}{+23:50:58.316} &
WHL J120122.5+235110 &
0.26735 \\
& & &
NSC J120127+235149 &
0.2471 \\
& & &
MaxBCG J180.32322+23.84515 &
0.22955 \\
& & &
GMBCG J180.35485+23.89116 &
0.363 \\
\end{tabular}
\end{ruledtabular}
\end{table*}

\addtocounter{table}{-1}
\begin{table*}
\caption{Continued.}
\begin{ruledtabular}
\begin{tabular}{ccc|cc}
Rank &
RA &
Dec &
Cluster\tablenotemark{a} &
Redshift\tablenotemark{b}
\\
\tableline
\multirow{2}{*}{96} &
\multirow{2}{*}{12:42:19.077} &
\multirow{2}{*}{+40:23:40.425} &
WHL J124235.9+402255 &
0.4024 \\
& & &
WHL J124201.4+402335 &
0.4462 \\
\tableline
97 &
11:59:04.900 &
+51:11:15.803 &
WHL J115921.9+511238 &
0.388 \\
\tableline
\multirow{3}{*}{98} &
\multirow{3}{*}{11:11:23.230} &
\multirow{3}{*}{+26:01:58.357} &
WHL J111123.2+260158 &
0.3325 \\
& & &
GMBCG J167.84231+26.02512 &
0.324 \\
& & &
GMBCG J167.85188+26.01246 &
0.177 \\
\tableline
99 &
11:23:08.271 &
+54:01:58.928 &
WHL J112247.2+540052 &
0.5005 \\
\tableline
\multirow{2}{*}{100} &
\multirow{2}{*}{14:32:40.352} &
\multirow{2}{*}{+31:41:36.116} &
GMBCG J218.15811+31.64695 &
0.158 \\
& & &
MaxBCG J218.15809+31.64695 &
0.132 \\
\tableline
\multirow{2}{*}{101} &
\multirow{2}{*}{12:45:04.700} &
\multirow{2}{*}{+02:29:08.618} &
GMBCG J191.25782+02.50588 &
0.33 \\
& & &
WHL J124501.2+023155 &
0.4881 \\
\tableline
\multirow{2}{*}{103} &
\multirow{2}{*}{09:15:50.686} &
\multirow{2}{*}{+42:57:08.567} &
WHL J091550.7+425708 &
0.5303 \\
& & &
GMBCG J138.91793+42.95508 &
0.415 \\
\tableline
104 &
13:41:08.628 &
+12:33:45.316 &
WHL J134111.8+123303 &
0.5622 \\
\tableline
105 &
12:25:49.069 &
+08:24:48.700 &
WHL J122555.2+082243 &
0.4821 \\
\tableline
\multirow{3}{*}{106} &
\multirow{3}{*}{08:31:34.886} &
\multirow{3}{*}{+26:52:25.307} &
WHL J083137.5+265054 &
0.4869 \\
& & &
GMBCG J127.93257+26.89906 &
0.26735 \\
& & &
GMBCG J127.85854+26.83155 &
0.453 \\
\tableline
\multirow{3}{*}{107} &
\multirow{3}{*}{10:14:11.602} &
\multirow{3}{*}{+22:31:53.131} &
GMBCG J153.54834+22.53144 &
0.484 \\
& & &
WHL J101407.7+223015 &
0.4663 \\
& & &
GMBCG J153.54087+22.49259 &
0.516 \\
\tableline
\multirow{2}{*}{108} &
\multirow{2}{*}{15:36:44.604} &
\multirow{2}{*}{+02:46:50.702} &
WHL J153646.4+024614 &
0.5121 \\
& & &
GMBCG J234.20097+02.75082 &
0.209 \\
\tableline
109 &
14:54:16.517 &
+04:34:40.491 &
WHL J145416.5+043440 &
0.3354 \\
\tableline
\multirow{3}{*}{110} &
\multirow{3}{*}{15:27:45.828} &
\multirow{3}{*}{+06:52:33.629} &
SDSS J1527+0652 CLUSTER &
0.4 \\
& & &
WHL J152745.8+065233 &
0.3812 \\
& & &
SDSS J1527+0652 &
0.39 \\
\tableline
\multirow{4}{*}{111} &
\multirow{4}{*}{00:06:11.544} &
\multirow{4}{*}{-10:28:19.512} &
WHL J000611.5-102819 &
0.23495 \\
& & &
WHL J000614.3-102820 &
0.5158 \\
& & &
GMBCG J001.53135-10.50724 &
0.221 \\
& & &
GMBCG J001.53243-10.42147 &
0.175 \\
\tableline
\multirow{2}{*}{114} &
\multirow{2}{*}{00:15:23.386} &
\multirow{2}{*}{-09:18:51.103} &
WHL J001513.3-091806 &
0.35 \\
& & &
WHL J001526.4-092207 &
0.4883 \\
\tableline
115 &
11:16:01.248 &
+18:24:23.300 &
WHL J111601.2+182423 &
0.4605 \\
\tableline
118 &
10:42:47.206 &
+33:12:17.845 &
WHL J104253.7+331254 &
0.5057 \\
\tableline
\multirow{2}{*}{119} &
\multirow{2}{*}{15:33:49.313} &
\multirow{2}{*}{+02:38:36.105} &
WHL J153342.2+023744 &
0.5101 \\
& & &
WHL J153354.5+024135 &
0.5622 \\
\tableline
120 &
17:52:27.692 &
+60:10:12.774 &
WHL J175236.4+601045 &
0.4602 \\
\tableline
121 &
12:08:19.794 &
+61:22:03.732 &
WHL J120825.8+612054 &
0.5178 \\
\tableline
\multirow{2}{*}{122} &
\multirow{2}{*}{01:19:07.658} &
\multirow{2}{*}{-09:34:02.693} &
GMBCG J019.79456-09.57832 &
0.348 \\
& & &
WHL J011916.4-093421 &
0.3505 \\
\tableline
\multirow{3}{*}{123} &
\multirow{3}{*}{09:38:11.613} &
\multirow{3}{*}{+27:35:43.705} &
GMBCG J144.56963+27.58991 &
0.334 \\
& & &
WHL J093807.1+273749 &
0.4954 \\
& & &
GMBCG J144.49716+27.62634 &
0.422 \\
\tableline
\multirow{4}{*}{125} &
\multirow{4}{*}{14:52:00.837} &
\multirow{4}{*}{+01:06:56.447} &
SDSS CE J223.007248+01.113963 &
0.37883 \\
& & &
WHL J145211.5+010748 &
0.3927 \\
& & &
SDSS CE J223.038879+01.152872 &
0.41286 \\
& & &
GMBCG J223.04780+01.15035 &
0.392 \\
\tableline
\multirow{3}{*}{126} &
\multirow{3}{*}{12:24:45.458} &
\multirow{3}{*}{-00:39:14.796} &
NSCS J122447-004029 &
0.47 \\
& & &
WHL J122447.3-004056 &
0.4777 \\
& & &
GMBCG J186.15220-00.62346 &
0.403 \\
\tableline
\multirow{2}{*}{127} &
\multirow{2}{*}{15:54:59.348} &
\multirow{2}{*}{+51:37:23.214} &
WHL J155447.4+513757 &
0.48 \\
& & &
GMBCG J238.70078+51.64642 &
0.467 \\
\tableline
\multirow{3}{*}{129} &
\multirow{3}{*}{12:19:21.841} &
\multirow{3}{*}{+50:53:28.236} &
WHL J121917.6+505432 &
0.5327 \\
& & &
GMBCG J184.80005+50.93515 &
0.471 \\
& & &
GMBCG J184.75315+50.88179 &
0.372 \\
\tableline
\multirow{3}{*}{130} &
\multirow{3}{*}{09:51:40.088} &
\multirow{3}{*}{-00:14:20.218} &
GMBCG J147.91704-00.23895 &
0.417 \\
& & &
SDSS CE J147.900848-00.253642 &
0.39017 \\
& & &
WHL J095148.5-001419 &
0.4168 \\
\tableline
131 &
00:51:24.585 &
--10:49:09.758 &
WHL J005121.4-104941 &
0.4847 \\
\tableline
132 &
22:26:27.277 &
+00:53:29.136 &
WHL J222624.0+005405 &
0.2771 \\
\tableline
\multirow{3}{*}{134} &
\multirow{3}{*}{15:50:36.108} &
\multirow{3}{*}{+39:48:56.718} &
WHL J155036.6+394941 &
0.5004 \\
& & &
GMBCG J237.64892+39.78190 &
0.464 \\
& & &
NSCS J155027+394752 &
0.37 \\
\end{tabular}
\end{ruledtabular}
\end{table*}

\addtocounter{table}{-1}
\begin{table*}
\caption{Continued.}
\begin{ruledtabular}
\begin{tabular}{ccc|cc}
Rank &
RA &
Dec &
Cluster\tablenotemark{a} &
Redshift\tablenotemark{b}
\\
\tableline
\multirow{5}{*}{135} &
\multirow{5}{*}{11:33:41.604} &
\multirow{5}{*}{+39:52:25.291} &
GMBCG J173.39894+39.87612 &
0.437 \\
& & &
GMBCG J173.46103+39.87940 &
0.282 \\
& & &
Abell 1310 &
0.2619 \\
& & &
NSCS J113327+395228 &
0.47 \\
& & &
WHL J113353.5+395019 &
0.2755 \\
\tableline
\multirow{2}{*}{136} &
\multirow{2}{*}{13:01:02.878} &
\multirow{2}{*}{+05:35:29.711} &
WHL J130106.9+053411 &
0.4817 \\
& & &
NSCS J130107+053314 &
0.25 \\
\tableline
\multirow{2}{*}{138} &
\multirow{2}{*}{11:33:37.447} &
\multirow{2}{*}{+66:24:44.842} &
Abell 1302 &
0.1165 \\
& & &
WHL J113314.7+662246 &
0.12155 \\
\tableline
139 &
01:57:54.644 &
--00:57:11.347 &
SDSS CE J029.529423-00.979300 &
0.39017 \\
\tableline
\multirow{3}{*}{140} &
\multirow{3}{*}{11:13:46.352} &
\multirow{3}{*}{+56:40:34.462} &
WHL J111344.3+564102 &
0.4754 \\
& & &
GMBCG J168.43311+56.64864 &
0.513 \\
& & &
GMBCG J168.44920+56.72458 &
0.474 \\
\tableline
\multirow{3}{*}{141} &
\multirow{3}{*}{14:44:19.496} &
\multirow{3}{*}{+16:20:12.303} &
WHL J144428.9+162016 &
0.3955 \\
& & &
GMBCG J221.04927+16.35866 &
0.362 \\
& & &
GMBCG J221.08712+16.29277 &
0.338 \\
\tableline
144 &
10:27:02.072 &
+09:16:40.107 &
WHL J102702.1+091640 &
0.5496 \\
\tableline
\multirow{2}{*}{145} &
\multirow{2}{*}{16:48:00.199} &
\multirow{2}{*}{+33:40:03.887} &
WHL J164803.8+334149 &
0.29975 \\
& & &
WHL J164745.7+334127 &
0.5307 \\
\tableline
\multirow{2}{*}{146} &
\multirow{2}{*}{10:22:32.057} &
\multirow{2}{*}{+50:07:07.870} &
Abell 980 &
0.1582 \\
& & &
[EAD2007] 047 &
0.15718 \\
\tableline
\multirow{6}{*}{147} &
\multirow{6}{*}{09:26:51.422} &
\multirow{6}{*}{+04:58:17.559} &
GMBCG J141.70184+04.96670 &
0.401 \\
& & &
NSC J092656+045928 &
0.2845 \\
& & &
ZwCl 0924.4+0511 &
0.2701 \\
& & &
GMBCG J141.69697+05.00110 &
0.486 \\
& & &
GMBCG J141.69759+05.02156 &
0.455 \\
& & &
GMBCG J141.76983+04.97938 &
0.25385 \\
\tableline
\multirow{3}{*}{148} &
\multirow{3}{*}{12:17:31.158} &
\multirow{3}{*}{+36:41:11.240} &
WHL J121731.9+364112 &
0.3916 \\
& & &
SDSS J1217+3641 CLUSTER &
0.364 \\
& & &
GMBCG J184.44333+36.68439 &
0.354 \\
\tableline
\multirow{2}{*}{149} &
\multirow{2}{*}{11:53:05.648} &
\multirow{2}{*}{+41:45:20.510} &
NSCS J115309+414558 &
0.3 \\
& & &
WHL J115312.2+414444 &
0.29165 \\
\tableline
150 &
14:33:05.416 &
+51:03:16.905 &
WHL J143254.9+510154 &
0.4819 \\
\tableline
151 &
09:01:04.594 &
+39:54:49.063 &
GMBCG J135.26913+39.91365 &
0.541 \\
\tableline
152 &
10:50:38.567 &
+35:49:12.425 &
WHL J105038.6+354912 &
0.5021 \\
\tableline
\multirow{3}{*}{154} &
\multirow{3}{*}{08:50:07.915} &
\multirow{3}{*}{+36:04:13.650} &
ZwCl 0847.2+3617 &
0.378 \\
& & &
GMBCG J132.49437+36.10756 &
0.284 \\
& & &
GMBCG J132.52774+36.01979 &
0.241 \\
\tableline
\multirow{4}{*}{156} &
\multirow{4}{*}{08:45:43.999} &
\multirow{4}{*}{+30:10:07.090} &
WHL J084544.0+301007 &
0.4959 \\
& & &
GMBCG J131.39803+30.18358 &
0.428 \\
& & &
NSC J084530+300943 &
0.1253 \\
& & &
WHL J084530.0+300839 &
0.4025 \\
\tableline
\multirow{4}{*}{157} &
\multirow{4}{*}{14:55:07.993} &
\multirow{4}{*}{+38:36:04.879} &
GMBCG J223.76805+38.61049 &
0.23 \\
& & &
ZwCl 1453.3+3849 &
... \\
& & &
NSC J145457+383607 &
0.326 \\
& & &
WHL J145452.1+383716 &
0.3847 \\
\tableline
\multirow{2}{*}{159} &
\multirow{2}{*}{11:07:19.334} &
\multirow{2}{*}{+53:04:17.938} &
GMBCG J166.86887+53.04094 &
0.462 \\
& & &
WHL J110708.1+530129 &
0.4296 \\
\tableline
\multirow{4}{*}{160} &
\multirow{4}{*}{15:38:02.025} &
\multirow{4}{*}{+39:27:39.159} &
SDSS J1537+3926 CLUSTER &
0.444 \\
& & &
NSCS J153814+392905 &
0.23 \\
& & &
NSC J153747+392702 &
0.2532 \\
& & &
WHL J153754.2+392444 &
0.409 \\
\tableline
\multirow{2}{*}{161} &
\multirow{2}{*}{01:03:24.248} &
\multirow{2}{*}{+00:55:37.011} &
SDSS CE J015.847747+00.930720 &
0.2994 \\
& & &
SDSS CE J015.862271+00.873928 &
0.37883 \\
\tableline
\multirow{3}{*}{162} &
\multirow{3}{*}{10:40:17.611} &
\multirow{3}{*}{+54:37:08.607} &
WHL J104016.0+543753 &
0.47 \\
& & &
GMBCG J160.02689+54.62449 &
0.391 \\
& & &
GMBCG J160.01506+54.57856 &
0.478 \\
\tableline
163 &
12:28:58.786 &
+53:37:27.671 &
WHL J122906.7+533551 &
0.5012 \\
\tableline
164 &
15:38:04.005 &
+39:22:32.253 &
WHL J153754.2+392444 &
0.409 \\
\tableline
\multirow{4}{*}{166} &
\multirow{4}{*}{09:50:00.059} &
\multirow{4}{*}{+17:04:27.060} &
GMBCG J147.48329+17.07628 &
0.283 \\
& & &
GMBCG J147.50544+17.11794 &
0.443 \\
& & &
GMBCG J147.46139+17.04589 &
0.354 \\
& & &
WHL J094951.3+170701 &
0.3638 \\
\end{tabular}
\end{ruledtabular}
\end{table*}

\addtocounter{table}{-1}
\begin{table*}
\caption{Continued.}
\begin{ruledtabular}
\begin{tabular}{ccc|cc}
Rank &
RA &
Dec &
Cluster\tablenotemark{a} &
Redshift\tablenotemark{b}
\\
\tableline
\multirow{3}{*}{168} &
\multirow{3}{*}{12:34:49.804} &
\multirow{3}{*}{+23:03:42.109} &
WHL J123446.3+230217 &
0.3238 \\
& & &
NSC J123444+230059 &
0.4125 \\
& & &
ZwCl 1232.1+2319 &
... \\
\tableline
169 &
01:27:10.589 &
+23:14:19.797 &
Abell 196 &
... \\
\tableline
170 &
13:15:23.033 &
--02:50:35.192 &
WHL J131515.4-024840 &
0.4112 \\
\tableline
\multirow{2}{*}{171} &
\multirow{2}{*}{11:39:02.869} &
\multirow{2}{*}{+47:04:43.290} &
GMBCG J174.78774+47.10125 &
0.323 \\
& & &
GMBCG J174.81035+47.03759 &
0.44 \\
\tableline
172 &
14:31:48.010 &
+09:00:15.869 &
NSCS J143159+090154 &
0.38 \\
\tableline
\multirow{3}{*}{173} &
\multirow{3}{*}{09:47:14.189} &
\multirow{3}{*}{+38:10:22.088} &
WHL J094714.2+381022 &
0.4353 \\
& & &
GMBCG J146.78369+38.19401 &
0.328 \\
& & &
GMBCG J146.85598+38.14762 &
0.442 \\
\tableline
174 &
12:17:05.124 &
+26:05:18.445 &
WHL J121705.1+260518 &
0.476 \\
\tableline
175 &
13:26:25.383 &
+53:24:58.472 &
GMBCG J201.55655+53.44045 &
0.504 \\
\tableline
176 &
01:53:42.190 &
+05:35:44.062 &
NSCS J015338+053638 &
0.29 \\
\tableline
\multirow{2}{*}{177} &
\multirow{2}{*}{10:54:40.435} &
\multirow{2}{*}{+55:23:56.307} &
GMBCG J163.66850+55.39898 &
0.487 \\
& & &
WHL J105444.1+552059 &
0.4893 \\
\tableline
\multirow{5}{*}{182} &
\multirow{5}{*}{14:37:17.666} &
\multirow{5}{*}{+34:18:22.187} &
NSC J143709+341851 &
0.2862 \\
& & &
GMBCG J219.34342+34.33407 &
0.393 \\
& & &
WHL J143731.1+341834 &
0.3808 \\
& & &
WHL J143713.7+341530 &
0.5426 \\
& & &
GMBCG J219.31933+34.25157 &
0.541 \\
\tableline
\multirow{3}{*}{183} &
\multirow{3}{*}{14:15:08.392} &
\multirow{3}{*}{-00:29:35.680} &
WHL J141508.4-002935 &
0.1303 \\
& & &
SDSS CE J213.781525-00.487651 &
0.14056 \\
& & &
[DDM2004] J141505.03-002908.1 &
0.141 \\
\tableline
184 &
20:53:55.128 &
--06:34:51.054 &
GMBCG J313.43167-06.55856 &
0.481 \\
\tableline
185 &
12:35:44.353 &
+35:32:47.968 &
WHL J123549.2+353445 &
0.4851 \\
\tableline
\multirow{2}{*}{186} &
\multirow{2}{*}{11:52:35.385} &
\multirow{2}{*}{+37:15:43.111} &
WHL J115235.4+371543 &
0.1475 \\
& & &
NSC J115231+371553 &
0.1175 \\
\tableline
\multirow{2}{*}{187} &
\multirow{2}{*}{14:45:34.036} &
\multirow{2}{*}{+48:00:12.417} &
WHL J144534.0+480012 &
0.5145 \\
& & &
GMBCG J221.47262+47.98388 &
0.422 \\
\tableline
188 &
08:40:08.745 &
+21:56:03.214 &
WHL J084005.8+215315 &
0.4454 \\
\tableline
\multirow{3}{*}{189} &
\multirow{3}{*}{13:48:53.073} &
\multirow{3}{*}{+57:23:46.617} &
WHL J134850.1+572147 &
0.28895 \\
& & &
Abell 1805 &
... \\
& & &
GMBCG J207.24819+57.44672 &
0.312 \\
\tableline
\multirow{6}{*}{190} &
\multirow{6}{*}{13:07:03.631} &
\multirow{6}{*}{+46:33:47.849} &
[EAD2007] 037 &
0.24656 \\
& & &
GMBCG J196.75262+46.56389 &
0.337 \\
& & &
WHL J130657.3+463206 &
0.2081 \\
& & &
GMBCG J196.70832+46.55927 &
0.245 \\
& & &
Abell 1682 &
0.2339 \\
& & &
GMBCG J196.70330+46.60084 &
0.415 \\
\tableline
\multirow{5}{*}{191} &
\multirow{5}{*}{11:40:40.199} &
\multirow{5}{*}{+44:07:40.291} &
MaxBCG J175.15921+44.11784 &
0.14585 \\
& & &
GMBCG J175.15828+44.14893 &
0.487 \\
& & &
NSCS J114032+440607 &
0.42 \\
& & &
WHL J114034.8+440541 &
0.4529 \\
& & &
ZwCl 1137.8+4425 &
... \\
\tableline
\multirow{2}{*}{192} &
\multirow{2}{*}{14:48:20.246} &
\multirow{2}{*}{+20:43:31.168} &
WHL J144820.4+204437 &
0.5046 \\
& & &
GMBCG J222.10454+20.75123 &
0.449 \\
\tableline
\multirow{2}{*}{193} &
\multirow{2}{*}{12:41:56.529} &
\multirow{2}{*}{+03:43:59.760} &
GMBCG J190.52490+03.76859 &
0.357 \\
& & &
WHL J124158.0+034721 &
0.3915 \\
\tableline
\multirow{2}{*}{194} &
\multirow{2}{*}{08:41:23.880} &
\multirow{2}{*}{+25:13:05.204} &
WHL J084123.9+251305 &
0.4655 \\
& & &
GMBCG J130.35004+25.17696 &
0.475 \\
\tableline
196 &
00:24:59.715 &
+08:26:16.778 &
NSCS J002458+082639 &
0.43 \\
\tableline
\multirow{3}{*}{197} &
\multirow{3}{*}{15:50:16.987} &
\multirow{3}{*}{+34:18:33.901} &
WHL J155025.7+341708 &
0.4297 \\
& & &
WHL J155006.3+341917 &
0.4329 \\
& & &
GMBCG J237.60955+34.26717 &
0.468 \\
\tableline
198 &
01:37:18.176 &
+07:55:44.482 &
Abell 220 &
0.33 \\
\tableline
\multirow{3}{*}{199} &
\multirow{3}{*}{11:05:20.978} &
\multirow{3}{*}{+17:37:16.830} &
GMBCG J166.33740+17.62134 &
0.497 \\
& & &
GMBCG J166.30451+17.61325 &
0.375 \\
& & &
WHL J110521.7+173505 &
0.5099 \\
\tableline
200 &
03:33:12.198 &
--06:52:24.614 &
WHL J033303.7-065233 &
0.2918 \\
\end{tabular}
\end{ruledtabular}
\end{table*}

\clearpage

\bibliography{lrgbeams}

\begin{thebibliography}{127}
\expandafter\ifx\csname natexlab\endcsname\relax\def\natexlab#1{#1}\fi

\bibitem[{{Abate} {et~al.}(2009){Abate}, {Wittman}, {Margoniner}, {Bridle},
  {Gee}, {Tyson}, \& {Dell'Antonio}}]{abate2009}
{Abate}, A., {Wittman}, D., {Margoniner}, V.~E., {Bridle}, S.~L., {Gee}, P.,
  {Tyson}, J.~A., \& {Dell'Antonio}, I.~P. 2009, \apj, 702, 603

\bibitem[{{Abell} {et~al.}(1989){Abell}, {Corwin}, \& {Olowin}}]{abell1989}
{Abell}, G.~O., {Corwin}, Jr., H.~G., \& {Olowin}, R.~P. 1989, \apjs, 70, 1

\bibitem[{{Ahn} {et~al.}(2012){Ahn}, {Alexandroff}, {Allende Prieto},
  {Anderson}, {Anderton}, {Andrews}, {Aubourg}, {Bailey}, {Balbinot}, {Barnes},
  \& et~al.}]{ahn2012}
{Ahn}, C.~P., {et~al.} 2012, \apjs, 203, 21

\bibitem[{{Angulo} {et~al.}(2012){Angulo}, {Springel}, {White}, {Jenkins},
  {Baugh}, \& {Frenk}}]{angulo2012}
{Angulo}, R.~E., {Springel}, V., {White}, S.~D.~M., {Jenkins}, A., {Baugh},
  C.~M., \& {Frenk}, C.~S. 2012, \mnras, 426, 2046

\bibitem[{{Auger} {et~al.}(2010){Auger}, {Treu}, {Gavazzi}, {Bolton},
  {Koopmans}, \& {Marshall}}]{auger2010}
{Auger}, M.~W., {Treu}, T., {Gavazzi}, R., {Bolton}, A.~S., {Koopmans},
  L.~V.~E., \& {Marshall}, P.~J. 2010, \apjl, 721, L163

\bibitem[{{Bartelmann} {et~al.}(1998){Bartelmann}, {Huss}, {Colberg},
  {Jenkins}, \& {Pearce}}]{bartelmann1998}
{Bartelmann}, M., {Huss}, A., {Colberg}, J.~M., {Jenkins}, A., \& {Pearce},
  F.~R. 1998, \aap, 330, 1

\bibitem[{{Bartelmann} {et~al.}(1995){Bartelmann}, {Steinmetz}, \&
  {Weiss}}]{bartelmann1995}
{Bartelmann}, M., {Steinmetz}, M., \& {Weiss}, A. 1995, \aap, 297, 1

\bibitem[{{Behroozi} {et~al.}(2010){Behroozi}, {Conroy}, \&
  {Wechsler}}]{behroozi2010}
{Behroozi}, P.~S., {Conroy}, C., \& {Wechsler}, R.~H. 2010, \apj, 717, 379

\bibitem[{{Behroozi} {et~al.}(2012){Behroozi}, {Wechsler}, \&
  {Conroy}}]{behroozi2012}
{Behroozi}, P.~S., {Wechsler}, R.~H., \& {Conroy}, C. 2012, \apj, submitted
  (arXiv:1207.6105)

\bibitem[{{Bell} \& {de Jong}(2001)}]{bell2001}
{Bell}, E.~F., \& {de Jong}, R.~S. 2001, \apj, 550, 212

\bibitem[{{Bernardi} {et~al.}(2003){Bernardi}, {Sheth}, {Annis}, {Burles},
  {Eisenstein}, {Finkbeiner}, {Hogg}, {Lupton}, {Schlegel}, {SubbaRao},
  {Bahcall}, {Blakeslee}, {Brinkmann}, {Castander}, {Connolly}, {Csabai},
  {Doi}, {Fukugita}, {Frieman}, {Heckman}, {Hennessy}, {Ivezi{\'c}}, {Knapp},
  {Lamb}, {McKay}, {Munn}, {Nichol}, {Okamura}, {Schneider}, {Thakar}, \&
  {York}}]{bernardi2003}
{Bernardi}, M., {et~al.} 2003, \aj, 125, 1849

\bibitem[{{B{\"o}hringer} {et~al.}(2000){B{\"o}hringer}, {Voges}, {Huchra},
  {McLean}, {Giacconi}, {Rosati}, {Burg}, {Mader}, {Schuecker}, {Simi{\c c}},
  {Komossa}, {Reiprich}, {Retzlaff}, \& {Tr{\"u}mper}}]{bohringer2000}
{B{\"o}hringer}, H., {et~al.} 2000, \apjs, 129, 435

\bibitem[{{Bonamente} {et~al.}(2008){Bonamente}, {Joy}, {LaRoque}, {Carlstrom},
  {Nagai}, \& {Marrone}}]{bonamente2008}
{Bonamente}, M., {Joy}, M., {LaRoque}, S.~J., {Carlstrom}, J.~E., {Nagai}, D.,
  \& {Marrone}, D.~P. 2008, \apj, 675, 106

\bibitem[{{Bouwens} {et~al.}(2012){Bouwens}, {Bradley}, {Zitrin}, {Coe},
  {Franx}, {Zheng}, {Smit}, {Host}, {Postman}, {Moustakas}, {Labbe},
  {Carrasco}, {Molino}, {Donahue}, {Kelson}, {Meneghetti}, {Jha}, {Benitez},
  {Lemze}, {Umetsu}, {Broadhurst}, {Moustakas}, {Rosati}, {Bartelmann}, {Ford},
  {Graves}, {Grillo}, {Infante}, {Jiminez-Teja}, {Jouvel}, {Lahav}, {Maoz},
  {Medezinski}, {Melchior}, {Merten}, {Nonino}, {Ogaz}, \&
  {Seitz}}]{bouwens2012}
{Bouwens}, R., {et~al.} 2012, \apj, submitted (arXiv:1211.2230)

\bibitem[{{Brada{\v c}} {et~al.}(2012){Brada{\v c}}, {Vanzella}, {Hall},
  {Treu}, {Fontana}, {Gonzalez}, {Clowe}, {Zaritsky}, {Stiavelli}, \&
  {Cl{\'e}ment}}]{bradac2012}
{Brada{\v c}}, M., {et~al.} 2012, \apjl, 755, L7

\bibitem[{{Bradley} {et~al.}(2008){Bradley}, {Bouwens}, {Ford}, {Illingworth},
  {Jee}, {Ben{\'{\i}}tez}, {Broadhurst}, {Franx}, {Frye}, {Infante}, {Motta},
  {Rosati}, {White}, \& {Zheng}}]{bradley2008}
{Bradley}, L.~D., {et~al.} 2008, \apj, 678, 647

\bibitem[{{Bradley} {et~al.}(2012){Bradley}, {Bouwens}, {Zitrin}, {Smit},
  {Coe}, {Ford}, {Zheng}, {Illingworth}, {Ben{\'{\i}}tez}, \&
  {Broadhurst}}]{bradley2012}
---. 2012, \apj, 747, 3

\bibitem[{{Brammer} {et~al.}(2012){Brammer}, {S{\'a}nchez-Janssen},
  {Labb{\'e}}, {da Cunha}, {Erb}, {Franx}, {Fumagalli}, {Lundgren},
  {Marchesini}, {Momcheva}, {Nelson}, {Patel}, {Quadri}, {Rix}, {Skelton},
  {Schmidt}, {van der Wel}, {van Dokkum}, {Wake}, \& {Whitaker}}]{brammer2012}
{Brammer}, G.~B., {et~al.} 2012, \apjl, 758, L17

\bibitem[{{Bruzual} \& {Charlot}(2003)}]{bruzual2003}
{Bruzual}, G., \& {Charlot}, S. 2003, \mnras, 344, 1000

\bibitem[{{Cacciato} {et~al.}(2009){Cacciato}, {van den Bosch}, {More}, {Li},
  {Mo}, \& {Yang}}]{cacciato2009}
{Cacciato}, M., {van den Bosch}, F.~C., {More}, S., {Li}, R., {Mo}, H.~J., \&
  {Yang}, X. 2009, \mnras, 394, 929

\bibitem[{{Cacciato} {et~al.}(2013{\natexlab{a}}){Cacciato}, {van den Bosch},
  {More}, {Mo}, \& {Yang}}]{cacciato2013a}
{Cacciato}, M., {van den Bosch}, F.~C., {More}, S., {Mo}, H., \& {Yang}, X.
  2013{\natexlab{a}}, \mnras, 430, 767

\bibitem[{{Cacciato} {et~al.}(2013{\natexlab{b}}){Cacciato}, {van Uitert}, \&
  {Hoekstra}}]{cacciato2013b}
{Cacciato}, M., {van Uitert}, E., \& {Hoekstra}, H. 2013{\natexlab{b}}, MNRAS,
  submitted (arXiv:1207.6105)

\bibitem[{{Cappellari} {et~al.}(2012){Cappellari}, {McDermid}, {Alatalo},
  {Blitz}, {Bois}, {Bournaud}, {Bureau}, {Crocker}, {Davies}, {Davis}, {de
  Zeeuw}, {Duc}, {Emsellem}, {Khochfar}, {Krajnovi{\'c}}, {Kuntschner},
  {Lablanche}, {Morganti}, {Naab}, {Oosterloo}, {Sarzi}, {Scott}, {Serra},
  {Weijmans}, \& {Young}}]{cappellari2012}
{Cappellari}, M., {et~al.} 2012, \nat, 484, 485

\bibitem[{{Coe} {et~al.}(2012){Coe}, {Umetsu}, {Zitrin}, {Donahue},
  {Medezinski}, {Postman}, {Carrasco}, {Anguita}, {Geller}, {Rines},
  {Diaferio}, {Kurtz}, {Bradley}, {Koekemoer}, {Zheng}, {Nonino}, {Molino},
  {Mahdavi}, {Lemze}, {Infante}, {Ogaz}, {Melchior}, {Host}, {Ford}, {Grillo},
  {Rosati}, {Jim{\'e}nez-Teja}, {Moustakas}, {Broadhurst}, {Ascaso}, {Lahav},
  {Bartelmann}, {Ben{\'{\i}}tez}, {Bouwens}, {Graur}, {Graves}, {Jha},
  {Jouvel}, {Kelson}, {Moustakas}, {Maoz}, {Meneghetti}, {Merten}, {Riess},
  {Rodney}, \& {Seitz}}]{coe2012}
{Coe}, D., {et~al.} 2012, \apj, 757, 22

\bibitem[{{Coe} {et~al.}(2013){Coe}, {Zitrin}, {Carrasco}, {Shu}, {Zheng},
  {Postman}, {Bradley}, {Koekemoer}, {Bouwens}, {Broadhurst}, {Monna}, {Host},
  {Moustakas}, {Ford}, {Moustakas}, {van der Wel}, {Donahue}, {Rodney},
  {Ben{\'{\i}}tez}, {Jouvel}, {Seitz}, {Kelson}, \& {Rosati}}]{coe2013}
---. 2013, \apj, 762, 32

\bibitem[{{Conroy} \& {van Dokkum}(2012)}]{conroy2012}
{Conroy}, C., \& {van Dokkum}, P.~G. 2012, \apj, 760, 71

\bibitem[{{Conroy} \& {Wechsler}(2009)}]{conroy2009}
{Conroy}, C., \& {Wechsler}, R.~H. 2009, \apj, 696, 620

\bibitem[{{Cool} {et~al.}(2008){Cool}, {Eisenstein}, {Fan}, {Fukugita},
  {Jiang}, {Maraston}, {Meiksin}, {Schneider}, \& {Wake}}]{cool2008}
{Cool}, R.~J., {et~al.} 2008, \apj, 682, 919

\bibitem[{{Coziol} {et~al.}(2009){Coziol}, {Andernach}, {Caretta},
  {Alamo-Mart{\'{\i}}nez}, \& {Tago}}]{coziol2009}
{Coziol}, R., {Andernach}, H., {Caretta}, C.~A., {Alamo-Mart{\'{\i}}nez},
  K.~A., \& {Tago}, E. 2009, \aj, 137, 4795

\bibitem[{{Csabai} {et~al.}(2003){Csabai}, {Budav{\'a}ri}, {Connolly},
  {Szalay}, {Gy{\H o}ry}, {Ben{\'{\i}}tez}, {Annis}, {Brinkmann}, {Eisenstein},
  {Fukugita}, {Gunn}, {Kent}, {Lupton}, {Nichol}, \& {Stoughton}}]{csabai2003}
{Csabai}, I., {et~al.} 2003, \aj, 125, 580

\bibitem[{{Dawson} {et~al.}(2013){Dawson}, {Schlegel}, {Ahn}, {Anderson},
  {Aubourg}, {Bailey}, {Barkhouser}, {Bautista}, {Beifiori}, {Berlind},
  {Bhardwaj}, {Bizyaev}, {Blake}, {Blanton}, {Blomqvist}, {Bolton}, {Borde},
  {Bovy}, {Brandt}, {Brewington}, {Brinkmann}, {Brown}, {Brownstein}, {Bundy},
  {Busca}, {Carithers}, {Carnero}, {Carr}, {Chen}, {Comparat}, {Connolly},
  {Cope}, {Croft}, {Cuesta}, {da Costa}, {Davenport}, {Delubac}, {de Putter},
  {Dhital}, {Ealet}, {Ebelke}, {Eisenstein}, {Escoffier}, {Fan}, {Filiz Ak},
  {Finley}, {Font-Ribera}, {G{\'e}nova-Santos}, {Gunn}, {Guo}, {Haggard},
  {Hall}, {Hamilton}, {Harris}, {Harris}, {Ho}, {Hogg}, {Holder}, {Honscheid},
  {Huehnerhoff}, {Jordan}, {Jordan}, {Kauffmann}, {Kazin}, {Kirkby}, {Klaene},
  {Kneib}, {Le Goff}, {Lee}, {Long}, {Loomis}, {Lundgren}, {Lupton}, {Maia},
  {Makler}, {Malanushenko}, {Malanushenko}, {Mandelbaum}, {Manera}, {Maraston},
  {Margala}, {Masters}, {McBride}, {McDonald}, {McGreer}, {McMahon}, {Mena},
  {Miralda-Escud{\'e}}, {Montero-Dorta}, {Montesano}, {Muna}, {Myers},
  {Naugle}, {Nichol}, {Noterdaeme}, {Nuza}, {Olmstead}, {Oravetz}, {Oravetz},
  {Owen}, {Padmanabhan}, {Palanque-Delabrouille}, {Pan}, {Parejko},
  {P{\^a}ris}, {Percival}, {P{\'e}rez-Fournon}, {P{\'e}rez-R{\`a}fols},
  {Petitjean}, {Pfaffenberger}, {Pforr}, {Pieri}, {Prada}, {Price-Whelan},
  {Raddick}, {Rebolo}, {Rich}, {Richards}, {Rockosi}, {Roe}, {Ross}, {Ross},
  {Rossi}, {Rubi{\~n}o-Martin}, {Samushia}, {S{\'a}nchez}, {Sayres}, {Schmidt},
  {Schneider}, {Sc{\'o}ccola}, {Seo}, {Shelden}, {Sheldon}, {Shen}, {Shu},
  {Slosar}, {Smee}, {Snedden}, {Stauffer}, {Steele}, {Strauss}, {Streblyanska},
  {Suzuki}, {Swanson}, {Tal}, {Tanaka}, {Thomas}, {Tinker}, {Tojeiro},
  {Tremonti}, {Vargas Maga{\~n}a}, {Verde}, {Viel}, {Wake}, {Watson}, {Weaver},
  {Weinberg}, {Weiner}, {West}, {White}, {Wood-Vasey}, {Yeche}, {Zehavi},
  {Zhao}, \& {Zheng}}]{dawson2013}
{Dawson}, K.~S., {et~al.} 2013, \aj, 145, 10

\bibitem[{{de Jong} {et~al.}(2013){de Jong}, {Verdoes Kleijn}, {Kuijken}, \&
  {Valentijn}}]{dejong2013}
{de Jong}, J.~T.~A., {Verdoes Kleijn}, G.~A., {Kuijken}, K.~H., \& {Valentijn},
  E.~A. 2013, Experimental Astronomy, 35, 25

\bibitem[{{De Lucia} {et~al.}(2006){De Lucia}, {Springel}, {White}, {Croton},
  \& {Kauffmann}}]{delucia2006}
{De Lucia}, G., {Springel}, V., {White}, S.~D.~M., {Croton}, D., \&
  {Kauffmann}, G. 2006, \mnras, 366, 499

\bibitem[{{Desai} {et~al.}(2004){Desai}, {Dalcanton}, {Mayer}, {Reed}, {Quinn},
  \& {Governato}}]{desai2004}
{Desai}, V., {Dalcanton}, J.~J., {Mayer}, L., {Reed}, D., {Quinn}, T., \&
  {Governato}, F. 2004, \mnras, 351, 265

\bibitem[{{Ebeling} {et~al.}(2001){Ebeling}, {Edge}, \& {Henry}}]{ebeling2001}
{Ebeling}, H., {Edge}, A.~C., \& {Henry}, J.~P. 2001, \apj, 553, 668

\bibitem[{{Eisenstein} {et~al.}(2001){Eisenstein}, {Annis}, {Gunn}, {Szalay},
  {Connolly}, {Nichol}, {Bahcall}, {Bernardi}, {Burles}, {Castander},
  {Fukugita}, {Hogg}, {Ivezi{\'c}}, {Knapp}, {Lupton}, {Narayanan}, {Postman},
  {Reichart}, {Richmond}, {Schneider}, {Schlegel}, {Strauss}, {SubbaRao},
  {Tucker}, {Vanden Berk}, {Vogeley}, {Weinberg}, \& {Yanny}}]{eisenstein2001}
{Eisenstein}, D.~J., {et~al.} 2001, \aj, 122, 2267

\bibitem[{{Eisenstein} {et~al.}(2003){Eisenstein}, {Hogg}, {Fukugita},
  {Nakamura}, {Bernardi}, {Finkbeiner}, {Schlegel}, {Brinkmann}, {Connolly},
  {Csabai}, {Gunn}, {Ivezi{\'c}}, {Lamb}, {Loveday}, {Munn}, {Nichol},
  {Schneider}, {Strauss}, {Szalay}, \& {York}}]{eisenstein2003}
---. 2003, \apj, 585, 694

\bibitem[{{Eisenstein} {et~al.}(2005){Eisenstein}, {Zehavi}, {Hogg},
  {Scoccimarro}, {Blanton}, {Nichol}, {Scranton}, {Seo}, {Tegmark}, {Zheng},
  {Anderson}, {Annis}, {Bahcall}, {Brinkmann}, {Burles}, {Castander},
  {Connolly}, {Csabai}, {Doi}, {Fukugita}, {Frieman}, {Glazebrook}, {Gunn},
  {Hendry}, {Hennessy}, {Ivezi{\'c}}, {Kent}, {Knapp}, {Lin}, {Loh}, {Lupton},
  {Margon}, {McKay}, {Meiksin}, {Munn}, {Pope}, {Richmond}, {Schlegel},
  {Schneider}, {Shimasaku}, {Stoughton}, {Strauss}, {SubbaRao}, {Szalay},
  {Szapudi}, {Tucker}, {Yanny}, \& {York}}]{eisenstein2005}
---. 2005, \apj, 633, 560

\bibitem[{{Estrada} {et~al.}(2007){Estrada}, {Annis}, {Diehl}, {Hall}, {Las},
  {Lin}, {Makler}, {Merritt}, {Scarpine}, {Allam}, \& {Tucker}}]{estrada2007}
{Estrada}, J., {et~al.} 2007, \apj, 660, 1176

\bibitem[{{Ford} {et~al.}(2012){Ford}, {Hildebrandt}, {Van Waerbeke},
  {Leauthaud}, {Capak}, {Finoguenov}, {Tanaka}, {George}, \&
  {Rhodes}}]{ford2012}
{Ford}, J., {et~al.} 2012, \apj, 754, 143

\bibitem[{{Frye} {et~al.}(2012){Frye}, {Hurley}, {Bowen}, {Meurer}, {Sharon},
  {Straughn}, {Coe}, {Broadhurst}, \& {Guhathakurta}}]{frye2012}
{Frye}, B.~L., {et~al.} 2012, \apj, 754, 17

\bibitem[{{Gal} {et~al.}(2003){Gal}, {de Carvalho}, {Lopes}, {Djorgovski},
  {Brunner}, {Mahabal}, \& {Odewahn}}]{gal2003}
{Gal}, R.~R., {de Carvalho}, R.~R., {Lopes}, P.~A.~A., {Djorgovski}, S.~G.,
  {Brunner}, R.~J., {Mahabal}, A., \& {Odewahn}, S.~C. 2003, \aj, 125, 2064

\bibitem[{{Geller} {et~al.}(2005){Geller}, {Dell'Antonio}, {Kurtz}, {Ramella},
  {Fabricant}, {Caldwell}, {Tyson}, \& {Wittman}}]{geller2005}
{Geller}, M.~J., {Dell'Antonio}, I.~P., {Kurtz}, M.~J., {Ramella}, M.,
  {Fabricant}, D.~G., {Caldwell}, N., {Tyson}, J.~A., \& {Wittman}, D. 2005,
  \apjl, 635, L125

\bibitem[{{Gladders} {et~al.}(2003){Gladders}, {Hoekstra}, {Yee}, {Hall}, \&
  {Barrientos}}]{gladders2003}
{Gladders}, M.~D., {Hoekstra}, H., {Yee}, H.~K.~C., {Hall}, P.~B., \&
  {Barrientos}, L.~F. 2003, \apj, 593, 48

\bibitem[{{Gladders} \& {Yee}(2000)}]{gladders2000}
{Gladders}, M.~D., \& {Yee}, H.~K.~C. 2000, \aj, 120, 2148

\bibitem[{{Gonzalez} {et~al.}(2007){Gonzalez}, {Zaritsky}, \&
  {Zabludoff}}]{gonzalez2007}
{Gonzalez}, A.~H., {Zaritsky}, D., \& {Zabludoff}, A.~I. 2007, \apj, 666, 147

\bibitem[{{Gonzalez} {et~al.}(2012){Gonzalez}, {Stanford}, {Brodwin}, {Fedeli},
  {Dey}, {Eisenhardt}, {Mancone}, {Stern}, \& {Zeimann}}]{gonzalez2012}
{Gonzalez}, A.~H., {et~al.} 2012, \apj, 753, 163

\bibitem[{{Goto} {et~al.}(2002){Goto}, {Sekiguchi}, {Nichol}, {Bahcall}, {Kim},
  {Annis}, {Ivezi{\'c}}, {Brinkmann}, {Hennessy}, {Szokoly}, \&
  {Tucker}}]{goto2002}
{Goto}, T., {et~al.} 2002, \aj, 123, 1807

\bibitem[{{Hall} {et~al.}(2012){Hall}, {Brada{\v c}}, {Gonzalez}, {Treu},
  {Clowe}, {Jones}, {Stiavelli}, {Zaritsky}, {Cuby}, \&
  {Cl{\'e}ment}}]{hall2012}
{Hall}, N., {et~al.} 2012, \apj, 745, 155

\bibitem[{{Hao} {et~al.}(2009){Hao}, {Koester}, {Mckay}, {Rykoff}, {Rozo},
  {Evrard}, {Annis}, {Becker}, {Busha}, {Gerdes}, {Johnston}, {Sheldon}, \&
  {Wechsler}}]{hao2009}
{Hao}, J., {et~al.} 2009, \apj, 702, 745

\bibitem[{{Hao} {et~al.}(2010){Hao}, {McKay}, {Koester}, {Rykoff}, {Rozo},
  {Annis}, {Wechsler}, {Evrard}, {Siegel}, {Becker}, {Busha}, {Gerdes},
  {Johnston}, \& {Sheldon}}]{hao2010}
---. 2010, \apjs, 191, 254

\bibitem[{{Hennawi} {et~al.}(2007){Hennawi}, {Dalal}, {Bode}, \&
  {Ostriker}}]{hennawi2007}
{Hennawi}, J.~F., {Dalal}, N., {Bode}, P., \& {Ostriker}, J.~P. 2007, \apj,
  654, 714

\bibitem[{{Hennawi} {et~al.}(2008){Hennawi}, {Gladders}, {Oguri}, {Dalal},
  {Koester}, {Natarajan}, {Strauss}, {Inada}, {Kayo}, {Lin}, {Lampeitl},
  {Annis}, {Bahcall}, \& {Schneider}}]{hennawi2008}
{Hennawi}, J.~F., {et~al.} 2008, \aj, 135, 664

\bibitem[{{Hikage} {et~al.}(2012){Hikage}, {Mandelbaum}, {Takada}, \&
  {Spergel}}]{hikage2012}
{Hikage}, C., {Mandelbaum}, R., {Takada}, M., \& {Spergel}, D.~N. 2012,
  arXiv:1211.1009

\bibitem[{{Hildebrandt} {et~al.}(2011){Hildebrandt}, {Muzzin}, {Erben},
  {Hoekstra}, {Kuijken}, {Surace}, {van Waerbeke}, {Wilson}, \&
  {Yee}}]{hildebrandt2011}
{Hildebrandt}, H., {et~al.} 2011, \apjl, 733, L30

\bibitem[{{Ho} {et~al.}(2009){Ho}, {Lin}, {Spergel}, \& {Hirata}}]{ho2009}
{Ho}, S., {Lin}, Y.-T., {Spergel}, D., \& {Hirata}, C.~M. 2009, \apj, 697, 1358

\bibitem[{{Hoekstra} {et~al.}(2011){Hoekstra}, {Hartlap}, {Hilbert}, \& {van
  Uitert}}]{hoekstra2011}
{Hoekstra}, H., {Hartlap}, J., {Hilbert}, S., \& {van Uitert}, E. 2011, \mnras,
  412, 2095

\bibitem[{{Huang} {et~al.}(2009){Huang}, {Morokuma}, {Fakhouri}, {Aldering},
  {Amanullah}, {Barbary}, {Brodwin}, {Connolly}, {Dawson}, {Doi}, {Faccioli},
  {Fadeyev}, {Fruchter}, {Goldhaber}, {Gladders}, {Hennawi}, {Ihara}, {Jee},
  {Kowalski}, {Konishi}, {Lidman}, {Meyers}, {Moustakas}, {Perlmutter},
  {Rubin}, {Schlegel}, {Spadafora}, {Suzuki}, {Takanashi}, \&
  {Yasuda}}]{huang2009}
{Huang}, X., {et~al.} 2009, \apjl, 707, L12

\bibitem[{{Kaiser}(2004)}]{kaiser2004}
{Kaiser}, N. 2004, in Society of Photo-Optical Instrumentation Engineers (SPIE)
  Conference Series, Vol. 5489, Society of Photo-Optical Instrumentation
  Engineers (SPIE) Conference Series, ed. J.~M. {Oschmann}, Jr., 11--22

\bibitem[{{Kauffmann} {et~al.}(2003){Kauffmann}, {Heckman}, {White}, {Charlot},
  {Tremonti}, {Brinchmann}, {Bruzual}, {Peng}, {Seibert}, {Bernardi},
  {Blanton}, {Brinkmann}, {Castander}, {Cs{\'a}bai}, {Fukugita}, {Ivezic},
  {Munn}, {Nichol}, {Padmanabhan}, {Thakar}, {Weinberg}, \&
  {York}}]{kauffmann2003}
{Kauffmann}, G., {et~al.} 2003, \mnras, 341, 33

\bibitem[{{Kneib} {et~al.}(2004){Kneib}, {Ellis}, {Santos}, \&
  {Richard}}]{kneib2004}
{Kneib}, J., {Ellis}, R.~S., {Santos}, M.~R., \& {Richard}, J. 2004, \apj, 607,
  697

\bibitem[{{Koester} {et~al.}(2007){Koester}, {McKay}, {Annis}, {Wechsler},
  {Evrard}, {Bleem}, {Becker}, {Johnston}, {Sheldon}, {Nichol}, {Miller},
  {Scranton}, {Bahcall}, {Barentine}, {Brewington}, {Brinkmann}, {Harvanek},
  {Kleinman}, {Krzesinski}, {Long}, {Nitta}, {Schneider}, {Sneddin}, {Voges},
  \& {York}}]{koester2007}
{Koester}, B.~P., {et~al.} 2007, \apj, 660, 239

\bibitem[{{Laporte} {et~al.}(2011){Laporte}, {Pell{\'o}}, {Schaerer},
  {Richard}, {Egami}, {Kneib}, {Le Borgne}, {Maizy}, {Boone}, {Hudelot}, \&
  {Mellier}}]{laporte2011}
{Laporte}, N., {et~al.} 2011, \aap, 531, A74+

\bibitem[{{Laureijs} {et~al.}(2011){Laureijs}, {Amiaux}, {Arduini},
  {Augu{\`e}res}, {Brinchmann}, {Cole}, {Cropper}, {Dabin}, {Duvet}, {Ealet},
  \& et~al.}]{laureijs2011}
{Laureijs}, R., {et~al.} 2011, arXiv:1110.3193

\bibitem[{{Leauthaud} {et~al.}(2012){Leauthaud}, {Tinker}, {Bundy}, {Behroozi},
  {Massey}, {Rhodes}, {George}, {Kneib}, {Benson}, {Wechsler}, {Busha},
  {Capak}, {Cort{\^e}s}, {Ilbert}, {Koekemoer}, {Le F{\`e}vre}, {Lilly},
  {McCracken}, {Salvato}, {Schrabback}, {Scoville}, {Smith}, \&
  {Taylor}}]{leauthaud2012}
{Leauthaud}, A., {et~al.} 2012, \apj, 744, 159

\bibitem[{{Li} {et~al.}(2006){Li}, {Kauffmann}, {Jing}, {White}, {B{\"o}rner},
  \& {Cheng}}]{li2006}
{Li}, C., {Kauffmann}, G., {Jing}, Y.~P., {White}, S.~D.~M., {B{\"o}rner}, G.,
  \& {Cheng}, F.~Z. 2006, \mnras, 368, 21

\bibitem[{{Lidman} {et~al.}(2012){Lidman}, {Suherli}, {Muzzin}, {Wilson},
  {Demarco}, {Brough}, {Rettura}, {Cox}, {DeGroot}, {Yee}, {Gilbank},
  {Hoekstra}, {Balogh}, {Ellingson}, {Hicks}, {Nantais}, {Noble}, {Lacy},
  {Surace}, \& {Webb}}]{lidman2012}
{Lidman}, C., {et~al.} 2012, \mnras, 427, 550

\bibitem[{{Lin} \& {Mohr}(2004)}]{lin2004a}
{Lin}, Y.-T., \& {Mohr}, J.~J. 2004, \apj, 617, 879

\bibitem[{{Lin} {et~al.}(2006){Lin}, {Mohr}, {Gonzalez}, \&
  {Stanford}}]{lin2006}
{Lin}, Y.-T., {Mohr}, J.~J., {Gonzalez}, A.~H., \& {Stanford}, S.~A. 2006,
  \apjl, 650, L99

\bibitem[{{Lin} {et~al.}(2003){Lin}, {Mohr}, \& {Stanford}}]{lin2003}
{Lin}, Y.-T., {Mohr}, J.~J., \& {Stanford}, S.~A. 2003, \apj, 591, 749

\bibitem[{{Lin} {et~al.}(2004){Lin}, {Mohr}, \& {Stanford}}]{lin2004b}
---. 2004, \apj, 610, 745

\bibitem[{{Livermore} {et~al.}(2012){Livermore}, {Jones}, {Richard}, {Bower},
  {Ellis}, {Swinbank}, {Rigby}, {Smail}, {Arribas}, {Rodriguez Zaurin},
  {Colina}, {Ebeling}, \& {Crain}}]{livermore2012}
{Livermore}, R.~C., {et~al.} 2012, \mnras, 427, 688

\bibitem[{{Lopes} {et~al.}(2004){Lopes}, {de Carvalho}, {Gal}, {Djorgovski},
  {Odewahn}, {Mahabal}, \& {Brunner}}]{lopes2004}
{Lopes}, P.~A.~A., {de Carvalho}, R.~R., {Gal}, R.~R., {Djorgovski}, S.~G.,
  {Odewahn}, S.~C., {Mahabal}, A.~A., \& {Brunner}, R.~J. 2004, \aj, 128, 1017

\bibitem[{{Mandelbaum} {et~al.}(2006){Mandelbaum}, {Seljak}, {Kauffmann},
  {Hirata}, \& {Brinkmann}}]{mandelbaum2006}
{Mandelbaum}, R., {Seljak}, U., {Kauffmann}, G., {Hirata}, C.~M., \&
  {Brinkmann}, J. 2006, \mnras, 368, 715

\bibitem[{{Mantz} {et~al.}(2010){Mantz}, {Allen}, {Ebeling}, {Rapetti}, \&
  {Drlica-Wagner}}]{mantz2010}
{Mantz}, A., {Allen}, S.~W., {Ebeling}, H., {Rapetti}, D., \& {Drlica-Wagner},
  A. 2010, \mnras, 406, 1773

\bibitem[{{Marriage} {et~al.}(2011){Marriage}, {Acquaviva}, {Ade}, {Aguirre},
  {Amiri}, {Appel}, {Barrientos}, {Battistelli}, {Bond}, {Brown}, {Burger},
  {Chervenak}, {Das}, {Devlin}, {Dicker}, {Bertrand Doriese}, {Dunkley},
  {D{\"u}nner}, {Essinger-Hileman}, {Fisher}, {Fowler}, {Hajian}, {Halpern},
  {Hasselfield}, {Hern{\'a}ndez-Monteagudo}, {Hilton}, {Hilton}, {Hincks},
  {Hlozek}, {Huffenberger}, {Handel Hughes}, {Hughes}, {Infante}, {Irwin},
  {Baptiste Juin}, {Kaul}, {Klein}, {Kosowsky}, {Lau}, {Limon}, {Lin},
  {Lupton}, {Marsden}, {Martocci}, {Mauskopf}, {Menanteau}, {Moodley},
  {Moseley}, {Netterfield}, {Niemack}, {Nolta}, {Page}, {Parker}, {Partridge},
  {Quintana}, {Reese}, {Reid}, {Sehgal}, {Sherwin}, {Sievers}, {Spergel},
  {Staggs}, {Swetz}, {Switzer}, {Thornton}, {Trac}, {Tucker}, {Warne},
  {Wilson}, {Wollack}, \& {Zhao}}]{marriage2011}
{Marriage}, T.~A., {et~al.} 2011, \apj, 737, 61

\bibitem[{{Meneghetti} {et~al.}(2003){Meneghetti}, {Bartelmann}, \&
  {Moscardini}}]{meneghetti2003}
{Meneghetti}, M., {Bartelmann}, M., \& {Moscardini}, L. 2003, \mnras, 340, 105

\bibitem[{{More}(2012)}]{more2012}
{More}, S. 2012, \apj, 761, 127

\bibitem[{{More} {et~al.}(2009){More}, {van den Bosch}, {Cacciato}, {Mo},
  {Yang}, \& {Li}}]{more2009}
{More}, S., {van den Bosch}, F.~C., {Cacciato}, M., {Mo}, H.~J., {Yang}, X., \&
  {Li}, R. 2009, \mnras, 392, 801

\bibitem[{{Moster} {et~al.}(2010){Moster}, {Somerville}, {Maulbetsch}, {van den
  Bosch}, {Macci{\`o}}, {Naab}, \& {Oser}}]{moster2010}
{Moster}, B.~P., {Somerville}, R.~S., {Maulbetsch}, C., {van den Bosch}, F.~C.,
  {Macci{\`o}}, A.~V., {Naab}, T., \& {Oser}, L. 2010, \apj, 710, 903

\bibitem[{{Oke} \& {Gunn}(1983)}]{oke1983}
{Oke}, J.~B., \& {Gunn}, J.~E. 1983, \apj, 266, 713

\bibitem[{{Padmanabhan} {et~al.}(2005){Padmanabhan}, {Budav{\'a}ri},
  {Schlegel}, {Bridges}, {Brinkmann}, {Cannon}, {Connolly}, {Croom}, {Csabai},
  {Drinkwater}, {Eisenstein}, {Hewett}, {Loveday}, {Nichol}, {Pimbblet}, {De
  Propris}, {Schneider}, {Scranton}, {Seljak}, {Shanks}, {Szapudi}, {Szalay},
  \& {Wake}}]{padmanabhan2005}
{Padmanabhan}, N., {et~al.} 2005, \mnras, 359, 237

\bibitem[{{Padmanabhan} {et~al.}(2007){Padmanabhan}, {Schlegel}, {Seljak},
  {Makarov}, {Bahcall}, {Blanton}, {Brinkmann}, {Eisenstein}, {Finkbeiner},
  {Gunn}, {Hogg}, {Ivezi{\'c}}, {Knapp}, {Loveday}, {Lupton}, {Nichol},
  {Schneider}, {Strauss}, {Tegmark}, \& {York}}]{padmanabhan2007}
---. 2007, \mnras, 378, 852

\bibitem[{{Pell{\'o}} {et~al.}(2004){Pell{\'o}}, {Schaerer}, {Richard}, {Le
  Borgne}, \& {Kneib}}]{pello2004}
{Pell{\'o}}, R., {Schaerer}, D., {Richard}, J., {Le Borgne}, J., \& {Kneib}, J.
  2004, \aap, 416, L35

\bibitem[{{P{\'e}rez-Gonz{\'a}lez} {et~al.}(2008){P{\'e}rez-Gonz{\'a}lez},
  {Rieke}, {Villar}, {Barro}, {Blaylock}, {Egami}, {Gallego}, {Gil de Paz},
  {Pascual}, {Zamorano}, \& {Donley}}]{perezgonzalez2008}
{P{\'e}rez-Gonz{\'a}lez}, P.~G., {et~al.} 2008, \apj, 675, 234

\bibitem[{{Popesso} {et~al.}(2005){Popesso}, {B{\"o}hringer}, {Romaniello}, \&
  {Voges}}]{popesso2005}
{Popesso}, P., {B{\"o}hringer}, H., {Romaniello}, M., \& {Voges}, W. 2005,
  \aap, 433, 415

\bibitem[{{Postman} {et~al.}(2012){Postman}, {Coe}, {Ben{\'{\i}}tez},
  {Bradley}, {Broadhurst}, {Donahue}, {Ford}, {Graur}, {Graves}, {Jouvel},
  {Koekemoer}, {Lemze}, {Medezinski}, {Molino}, {Moustakas}, {Ogaz}, {Riess},
  {Rodney}, {Rosati}, {Umetsu}, {Zheng}, {Zitrin}, {Bartelmann}, {Bouwens},
  {Czakon}, {Golwala}, {Host}, {Infante}, {Jha}, {Jimenez-Teja}, {Kelson},
  {Lahav}, {Lazkoz}, {Maoz}, {McCully}, {Melchior}, {Meneghetti}, {Merten},
  {Moustakas}, {Nonino}, {Patel}, {Reg{\"o}s}, {Sayers}, {Seitz}, \& {Van der
  Wel}}]{postman2012}
{Postman}, M., {et~al.} 2012, \apjs, 199, 25

\bibitem[{{Reid} \& {Spergel}(2009)}]{reid2009}
{Reid}, B.~A., \& {Spergel}, D.~N. 2009, \apj, 698, 143

\bibitem[{{Richard} {et~al.}(2006){Richard}, {Pell{\'o}}, {Schaerer}, {Le
  Borgne}, \& {Kneib}}]{richard2006}
{Richard}, J., {Pell{\'o}}, R., {Schaerer}, D., {Le Borgne}, J., \& {Kneib}, J.
  2006, \aap, 456, 861

\bibitem[{{Richard} {et~al.}(2008){Richard}, {Stark}, {Ellis}, {George},
  {Egami}, {Kneib}, \& {Smith}}]{richard2008}
{Richard}, J., {Stark}, D.~P., {Ellis}, R.~S., {George}, M.~R., {Egami}, E.,
  {Kneib}, J., \& {Smith}, G.~P. 2008, \apj, 685, 705

\bibitem[{{Ross} {et~al.}(2011){Ross}, {Ho}, {Cuesta}, {Tojeiro}, {Percival},
  {Wake}, {Masters}, {Nichol}, {Myers}, {de Simoni}, {Seo},
  {Hern{\'a}ndez-Monteagudo}, {Crittenden}, {Blanton}, {Brinkmann}, {da Costa},
  {Guo}, {Kazin}, {Maia}, {Maraston}, {Padmanabhan}, {Prada}, {Ramos},
  {Sanchez}, {Schlafly}, {Schlegel}, {Schneider}, {Skibba}, {Thomas}, {Weaver},
  {White}, \& {Zehavi}}]{ross2011}
{Ross}, A.~J., {et~al.} 2011, \mnras, 417, 1350

\bibitem[{{Salpeter}(1955)}]{salpeter1955}
{Salpeter}, E.~E. 1955, \apj, 121, 161

\bibitem[{{Schaerer} \& {Pell{\'o}}(2005)}]{schaerer2005}
{Schaerer}, D., \& {Pell{\'o}}, R. 2005, \mnras, 362, 1054

\bibitem[{{Sharon} {et~al.}(2012){Sharon}, {Gladders}, {Rigby}, {Wuyts},
  {Koester}, {Bayliss}, \& {Barrientos}}]{sharon2012}
{Sharon}, K., {Gladders}, M.~D., {Rigby}, J.~R., {Wuyts}, E., {Koester}, B.~P.,
  {Bayliss}, M.~B., \& {Barrientos}, L.~F. 2012, \apj, 746, 161

\bibitem[{{Skibba} {et~al.}(2011){Skibba}, {van den Bosch}, {Yang}, {More},
  {Mo}, \& {Fontanot}}]{skibba2011}
{Skibba}, R.~A., {van den Bosch}, F.~C., {Yang}, X., {More}, S., {Mo}, H., \&
  {Fontanot}, F. 2011, \mnras, 410, 417

\bibitem[{{Spiniello} {et~al.}(2012){Spiniello}, {Trager}, {Koopmans}, \&
  {Chen}}]{spiniello2012}
{Spiniello}, C., {Trager}, S.~C., {Koopmans}, L.~V.~E., \& {Chen}, Y.~P. 2012,
  \apjl, 753, L32

\bibitem[{{Springel} {et~al.}(2005){Springel}, {White}, {Jenkins}, {Frenk},
  {Yoshida}, {Gao}, {Navarro}, {Thacker}, {Croton}, {Helly}, {Peacock}, {Cole},
  {Thomas}, {Couchman}, {Evrard}, {Colberg}, \& {Pearce}}]{springel2005}
{Springel}, V., {et~al.} 2005, \nat, 435, 629

\bibitem[{{Stark} {et~al.}(2007){Stark}, {Ellis}, {Richard}, {Kneib}, {Smith},
  \& {Santos}}]{stark2007}
{Stark}, D.~P., {Ellis}, R.~S., {Richard}, J., {Kneib}, J.-P., {Smith}, G.~P.,
  \& {Santos}, M.~R. 2007, \apj, 663, 10

\bibitem[{{Sunyaev} \& {Zeldovich}(1972)}]{sunyaev1972}
{Sunyaev}, R.~A., \& {Zeldovich}, Y.~B. 1972, Comments on Astrophysics and
  Space Physics, 4, 173

\bibitem[{{Tinker} {et~al.}(2008){Tinker}, {Kravtsov}, {Klypin}, {Abazajian},
  {Warren}, {Yepes}, {Gottl{\"o}ber}, \& {Holz}}]{tinker2008}
{Tinker}, J., {Kravtsov}, A.~V., {Klypin}, A., {Abazajian}, K., {Warren}, M.,
  {Yepes}, G., {Gottl{\"o}ber}, S., \& {Holz}, D.~E. 2008, \apj, 688, 709

\bibitem[{{Tinker} {et~al.}(2005){Tinker}, {Weinberg}, {Zheng}, \&
  {Zehavi}}]{tinker2005}
{Tinker}, J.~L., {Weinberg}, D.~H., {Zheng}, Z., \& {Zehavi}, I. 2005, \apj,
  631, 41

\bibitem[{{Treu} {et~al.}(2010){Treu}, {Auger}, {Koopmans}, {Gavazzi},
  {Marshall}, \& {Bolton}}]{treu2010}
{Treu}, T., {Auger}, M.~W., {Koopmans}, L.~V.~E., {Gavazzi}, R., {Marshall},
  P.~J., \& {Bolton}, A.~S. 2010, \apj, 709, 1195

\bibitem[{{Treu} {et~al.}(2005{\natexlab{a}}){Treu}, {Ellis}, {Liao}, \& {van
  Dokkum}}]{treu2005a}
{Treu}, T., {Ellis}, R.~S., {Liao}, T.~X., \& {van Dokkum}, P.~G.
  2005{\natexlab{a}}, \apjl, 622, L5

\bibitem[{{Treu} {et~al.}(2005{\natexlab{b}}){Treu}, {Ellis}, {Liao}, {van
  Dokkum}, {Tozzi}, {Coil}, {Newman}, {Cooper}, \& {Davis}}]{treu2005b}
{Treu}, T., {et~al.} 2005{\natexlab{b}}, \apj, 633, 174

\bibitem[{{Ueda} {et~al.}(2001){Ueda}, {Ishisaki}, {Takahashi}, {Makishima}, \&
  {Ohashi}}]{ueda2001}
{Ueda}, Y., {Ishisaki}, Y., {Takahashi}, T., {Makishima}, K., \& {Ohashi}, T.
  2001, \apjs, 133, 1

\bibitem[{{van Dokkum} \& {Conroy}(2012)}]{vandokkum2012}
{van Dokkum}, P.~G., \& {Conroy}, C. 2012, \apj, 760, 70

\bibitem[{{van Dokkum} \& {Stanford}(2003)}]{vandokkum2003}
{van Dokkum}, P.~G., \& {Stanford}, S.~A. 2003, \apj, 585, 78

\bibitem[{{Van Waerbeke} {et~al.}(2010){Van Waerbeke}, {Hildebrandt}, {Ford},
  \& {Milkeraitis}}]{vanwaerbeke2010}
{Van Waerbeke}, L., {Hildebrandt}, H., {Ford}, J., \& {Milkeraitis}, M. 2010,
  \apjl, 723, L13

\bibitem[{{Vanderlinde} {et~al.}(2010){Vanderlinde}, {Crawford}, {de Haan},
  {Dudley}, {Shaw}, {Ade}, {Aird}, {Benson}, {Bleem}, {Brodwin}, {Carlstrom},
  {Chang}, {Crites}, {Desai}, {Dobbs}, {Foley}, {George}, {Gladders}, {Hall},
  {Halverson}, {High}, {Holder}, {Holzapfel}, {Hrubes}, {Joy}, {Keisler},
  {Knox}, {Lee}, {Leitch}, {Loehr}, {Lueker}, {Marrone}, {McMahon}, {Mehl},
  {Meyer}, {Mohr}, {Montroy}, {Ngeow}, {Padin}, {Plagge}, {Pryke}, {Reichardt},
  {Rest}, {Ruel}, {Ruhl}, {Schaffer}, {Shirokoff}, {Song}, {Spieler},
  {Stalder}, {Staniszewski}, {Stark}, {Stubbs}, {van Engelen}, {Vieira},
  {Williamson}, {Yang}, {Zahn}, \& {Zenteno}}]{vanderlinde2010}
{Vanderlinde}, K., {et~al.} 2010, \apj, 722, 1180

\bibitem[{{Vikhlinin} {et~al.}(2009){Vikhlinin}, {Burenin}, {Ebeling},
  {Forman}, {Hornstrup}, {Jones}, {Kravtsov}, {Murray}, {Nagai}, {Quintana}, \&
  {Voevodkin}}]{vikhlinin2009}
{Vikhlinin}, A., {et~al.} 2009, \apj, 692, 1033

\bibitem[{{von der Linden} {et~al.}(2007){von der Linden}, {Best}, {Kauffmann},
  \& {White}}]{vonderlinden2007}
{von der Linden}, A., {Best}, P.~N., {Kauffmann}, G., \& {White}, S.~D.~M.
  2007, \mnras, 379, 867

\bibitem[{{Wen} {et~al.}(2009){Wen}, {Han}, \& {Liu}}]{wen2009}
{Wen}, Z.~L., {Han}, J.~L., \& {Liu}, F.~S. 2009, \apjs, 183, 197

\bibitem[{{Wen} {et~al.}(2012){Wen}, {Han}, \& {Liu}}]{wen2012}
---. 2012, \apjs, 199, 34

\bibitem[{{White} {et~al.}(2011){White}, {Blanton}, {Bolton}, {Schlegel},
  {Tinker}, {Berlind}, {da Costa}, {Kazin}, {Lin}, {Maia}, {McBride},
  {Padmanabhan}, {Parejko}, {Percival}, {Prada}, {Ramos}, {Sheldon}, {de
  Simoni}, {Skibba}, {Thomas}, {Wake}, {Zehavi}, {Zheng}, {Nichol},
  {Schneider}, {Strauss}, {Weaver}, \& {Weinberg}}]{white2011}
{White}, M., {et~al.} 2011, \apj, 728, 126

\bibitem[{{Williamson} {et~al.}(2011){Williamson}, {Benson}, {High},
  {Vanderlinde}, {Ade}, {Aird}, {Andersson}, {Armstrong}, {Ashby}, {Bautz},
  {Bazin}, {Bertin}, {Bleem}, {Bonamente}, {Brodwin}, {Carlstrom}, {Chang},
  {Chapman}, {Clocchiatti}, {Crawford}, {Crites}, {de Haan}, {Desai}, {Dobbs},
  {Dudley}, {Fazio}, {Foley}, {Forman}, {Garmire}, {George}, {Gladders},
  {Gonzalez}, {Halverson}, {Holder}, {Holzapfel}, {Hoover}, {Hrubes}, {Jones},
  {Joy}, {Keisler}, {Knox}, {Lee}, {Leitch}, {Lueker}, {Luong-Van}, {Marrone},
  {McMahon}, {Mehl}, {Meyer}, {Mohr}, {Montroy}, {Murray}, {Padin}, {Plagge},
  {Pryke}, {Reichardt}, {Rest}, {Ruel}, {Ruhl}, {Saliwanchik}, {Saro},
  {Schaffer}, {Shaw}, {Shirokoff}, {Song}, {Spieler}, {Stalder}, {Stanford},
  {Staniszewski}, {Stark}, {Story}, {Stubbs}, {Vieira}, {Vikhlinin}, \&
  {Zenteno}}]{williamson2011}
{Williamson}, R., {et~al.} 2011, \apj, 738, 139

\bibitem[{{Wong} {et~al.}(2012){Wong}, {Ammons}, {Keeton}, \&
  {Zabludoff}}]{wong2012}
{Wong}, K.~C., {Ammons}, S.~M., {Keeton}, C.~R., \& {Zabludoff}, A.~I. 2012,
  \apj, 752, 104

\bibitem[{{Yang} {et~al.}(2008){Yang}, {Mo}, \& {van den Bosch}}]{yang2008}
{Yang}, X., {Mo}, H.~J., \& {van den Bosch}, F.~C. 2008, \apj, 676, 248

\bibitem[{{Yuan} {et~al.}(2012){Yuan}, {Kewley}, {Swinbank}, \&
  {Richard}}]{yuan2012}
{Yuan}, T.-T., {Kewley}, L.~J., {Swinbank}, A.~M., \& {Richard}, J. 2012, \apj,
  759, 66

\bibitem[{{Zaritsky} {et~al.}(2006){Zaritsky}, {Gonzalez}, \&
  {Zabludoff}}]{zaritsky2006}
{Zaritsky}, D., {Gonzalez}, A.~H., \& {Zabludoff}, A.~I. 2006, \apj, 638, 725

\bibitem[{{Zehavi} {et~al.}(2005){Zehavi}, {Eisenstein}, {Nichol}, {Blanton},
  {Hogg}, {Brinkmann}, {Loveday}, {Meiksin}, {Schneider}, \&
  {Tegmark}}]{zehavi2005}
{Zehavi}, I., {et~al.} 2005, \apj, 621, 22

\bibitem[{{Zheng} {et~al.}(2009){Zheng}, {Bradley}, {Bouwens}, {Ford},
  {Illingworth}, {Ben{\'{\i}}tez}, {Broadhurst}, {Frye}, {Infante}, {Jee},
  {Motta}, {Shu}, \& {Zitrin}}]{zheng2009}
{Zheng}, W., {et~al.} 2009, \apj, 697, 1907

\bibitem[{{Zheng} {et~al.}(2012){Zheng}, {Postman}, {Zitrin}, {Moustakas},
  {Shu}, {Jouvel}, {H{\o}st}, {Molino}, {Bradley}, {Coe}, {Moustakas},
  {Carrasco}, {Ford}, {Ben{\'{\i}}tez}, {Lauer}, {Seitz}, {Bouwens},
  {Koekemoer}, {Medezinski}, {Bartelmann}, {Broadhurst}, {Donahue}, {Grillo},
  {Infante}, {Jha}, {Kelson}, {Lahav}, {Lemze}, {Melchior}, {Meneghetti},
  {Merten}, {Nonino}, {Ogaz}, {Rosati}, {Umetsu}, \& {van der Wel}}]{zheng2012}
---. 2012, \nat, 489, 406

\bibitem[{{Zitrin} {et~al.}(2012){Zitrin}, {Broadhurst}, {Bartelmann},
  {Rephaeli}, {Oguri}, {Ben{\'{\i}}tez}, {Hao}, \& {Umetsu}}]{zitrin2012}
{Zitrin}, A., {Broadhurst}, T., {Bartelmann}, M., {Rephaeli}, Y., {Oguri}, M.,
  {Ben{\'{\i}}tez}, N., {Hao}, J., \& {Umetsu}, K. 2012, \mnras, 423, 2308

\bibitem[{{Zitrin} {et~al.}(2011){Zitrin}, {Broadhurst}, {Coe}, {Umetsu},
  {Postman}, {Ben{\'{\i}}tez}, {Meneghetti}, {Medezinski}, {Jouvel}, {Bradley},
  {Koekemoer}, {Zheng}, {Ford}, {Merten}, {Kelson}, {Lahav}, {Lemze}, {Molino},
  {Nonino}, {Donahue}, {Rosati}, {Van der Wel}, {Bartelmann}, {Bouwens},
  {Graur}, {Graves}, {Host}, {Infante}, {Jha}, {Jimenez-Teja}, {Lazkoz},
  {Maoz}, {McCully}, {Melchior}, {Moustakas}, {Ogaz}, {Patel}, {Regoes},
  {Riess}, {Rodney}, \& {Seitz}}]{zitrin2011}
{Zitrin}, A., {et~al.} 2011, \apj, 742, 117

\bibitem[{{Zwicky} {et~al.}(1961){Zwicky}, {Herzog}, {Wild}, {Karpowicz}, \&
  {Kowal}}]{zwicky1961}
{Zwicky}, F., {Herzog}, E., {Wild}, P., {Karpowicz}, M., \& {Kowal}, C.~T.
  1961, {Catalogue of galaxies and of clusters of galaxies, Vol. I}

\end{thebibliography}

\end{document}